\documentclass[fleqn,usenatbib]{mnras}

\usepackage[T1]{fontenc}

\usepackage{multirow}
\usepackage{tablefootnote}
\DeclareRobustCommand{\VAN}[3]{#2}
\let\VANthebibliography\thebibliography
\def\thebibliography{\DeclareRobustCommand{\VAN}[3]{##3}\VANthebibliography}

\usepackage{multicol}
\usepackage{graphicx}	
\usepackage{amsmath}	
\usepackage{amssymb}	

\renewcommand{\d}[1]{\ensuremath{\operatorname{d}\!{#1}}}
\usepackage{newtxtext,newtxmath}

\title[Tidal Migration of Hot Jupiters]{Tidal Migration of Hot Jupiters: Introducing the Impact of Gravity Wave Dissipation}

\author[Y. A. Lazovik]{
Yaroslav A. Lazovik$^{1,2}$\thanks{E-mail: yaroslav.lazovik@gmail.com}
\\
$^{1}$Lomonosov Moscow State University, Faculty of Physics, 1 Leninskie Gory, bldg.2, Moscow, 119991, Russia\\
$^{2}$Sternberg Astronomical Institute, Lomonosov Moscow State University, Universitetsky pr. 13, Moscow, 119234, Russia
}

\date{Accepted XXX. Received YYY; in original form ZZZ}

\pubyear{2021}

\begin{document}
\label{firstpage}
\pagerange{\pageref{firstpage}--\pageref{lastpage}}
\maketitle

\begin{abstract}
We study the migration of hot Jupiters orbiting solar-type pre-main sequence and main sequence stars under the effect of tidal dissipation. The explored range of stellar mass extends from 0.6 to 1.3 $M_{\odot}$. We apply recently developed prescriptions which allow us to explore the orbital evolution over the wide parameter space. Three types of tides are considered: equilibrium tide, inertial waves, and gravity waves. We combine the results of our simulations with the observed distribution of stellar and planetary parameters to evaluate the infall rate of hot Jupiters in the Milky Way Galaxy. In particular, we find that, for 11 -- 21\% of the initial hot Jupiter population, coalescence occurs before the host star's main sequence termination. If the planet is massive enough, such an event can potentially be accompanied by a powerful transient detectable with new facilities. Orbital decay by itself can be observed through transit-timing variation. However, the obtained coalescence rate in the Galaxy is too low (340 -- 650 events per million years) to make positive predictions about the observational possibility. Potentially identifiable decaying systems formed by a star corresponding to a given mass interval might be too rare to be detected with the modern space telescopes, like TESS, within a 10-year baseline. At the same time, the forthcoming missions, like PLATO, look more promising in this regard.
\end{abstract}



\section{Introduction}

The discovery of the first hot Jupiter 51 Peg b (\citealt{Mayor}) has opened the doors for the conceptually new branch of studies. Over 25 years of research significantly expanded our understanding of the nature of exoplanet populations. According to current estimates, most stars are likely to host planets (e.g., \citealt{Mulders}). The growth in observational data motivated fundamental work, which ultimately led to the development of the planet population synthesis method, designed to reconcile theory and observations (\citealt{IdaLin,Mordasini1,Mordasini2, Benz}).

Recent realizations of the population synthesis consider more complicated migration models (\citealt{Alessi}), pebble accretion (\citealt{Chambers}), planet-planet interaction (\citealt{Alibert2}), and alternative scenarios for planetary formation through disc instability (\citealt{Forgan, Forgan2, Muller}). Another crucial improvement is to investigate the effect of stellar (\citealt{Alibert1}) and disc (\citealt{Mordasini3}) properties on planetary statistics and follow the long-term evolution (\citealt{Emsenhuber1, Emsenhuber2,Schlecker, Burn}). The above steps are aimed to provide an opportunity to directly compare observable and synthetic exoplanet populations and make predictions for the growth of the number of detected exoplanets.

However, close-in planets are likely to be a subject of significant migration after the dissipation of a protoplanetary disc, which raises the need to take into account star-planet and planet-planet interaction to reproduce the observed population of exoplanets correctly (\citealt{Juric1,Emsenhuber2,Ahuir}). Various studies approached this idea from different points of view. The relative motion between the planet and the magnetized ambient wind induces star-planet magnetic interaction. Depending on the configuration of a star-planet system and the magnetic field strengths (stellar and planetary), the interaction may be either unipolar (\citealt{Laine1, Laine2}) or dipolar (\citealt{Strugarek1,Strugarek2,Strugarek3}). The first attempt to consider both regimes of magnetic interaction within one model was made by \cite{Strugarek4}. Aside from magnetic interaction, many different mechanisms responsible for planetary migration have been proposed, including planetary obliquity tides (\citealt{Millholland}), Kozai migration (\citealt{Naoz,Attia}), the gravitational interaction between hot Jupiter and its atmospheric outflows (\citealt{Kurbatov}), migration inside a disc cavity (\citealt{Debras}), secular planet-planet interaction (\citealt{Wu, Laskar,BolmontRaymond, PuLai, Becker}), and many others. 

Among various mechanisms, tidal interaction plays a critical role in its impact on the dynamics of the most massive close-in planets, hot Jupiters (e.g. \citealt{Ahuir}). The presence of a gravitational perturber causes large-scale equilibrium flows on the stellar surface and dynamical interior oscillations. The energy of such deformations dissipates under the effect of various factors, which leads to a redistribution of angular momentum in a star-planet system followed by planetary migration. The tidal response can thus be decomposed into two components: an equilibrium (non-wavelike) tide and a dynamical (wavelike) tide (\citealt{Zahn1,Zahn2,Zahn3}). The dynamical waves are able to propagate both in radiation and in convection zones, depending on the type of a wave. The corresponding tides are called the gravity waves and the inertial waves, respectively.

Numerous studies provided prescriptions for tidal dissipation rates. Equilibrium tides have been extensively studied by \cite{Zahn2,Zahn3, Hansen1, Hansen2}. Inertial wave dissipation can be estimated using a frequency-averaged formalism developed by \cite{Ogilvie1}. The processes governing gravity wave dissipation depend on the amplitude of the ``primary'' waves. Three different regimes are possible, namely linear (\citealt{Goodman}), nonlinear (\citealt{Kumar,BarkerOgilvie1,Weinberg,Essick}), and strongly nonlinear (\citealt{Goodman,Ogilvie00,BarkerOgilvie, Barker1, Barker}). It is worth mentioning that the strength of tidal interaction can also be expressed in terms of so-called overlap integrals (\citealt{IP,PI, IPch,ChPI,ChIP}). The corresponding dissipation rates satisfy and complement the results obtained with other techniques.

The aforementioned methodological approaches must be constrained by the empirical estimates. There are several ways to derive the information about star-planet interaction from the observations. The most straightforward one is the detection of the orbital decay rate through transit-timing variation. However, WASP-12b remains the only planet to date for which the orbital decay has been confirmed (\citealt{Maciejewski,Yee,Turner}). Secondly, magnetic and tidal interaction affects the outer layers of a star. The atmospheric changes induce stellar activity, which can be observed in extreme cases (\citealt{Cuntz}). Thirdly, angular momentum exchange caused by the planetary migration can enhance stellar rotation rates (\citealt{BolmontMathis,Penev,Arevalo}), leading to biases in gyrochronological ages (\citealt{Gallet}) and anomalies in the rotation period distribution of young stellar clusters (\citealt{GalletBolmont}). Planetary engulfment influences stellar rotation even more dramatically. For example, \cite{Qureshi} attributes the observed bimodality in spin-period distributions of young stellar clusters to a fraction of stars consuming a planet at the early times. Finally, the planetary merger can be accompanied by a bright optical or/and UV/X-ray transient, as shown by \cite{Metzger}. Such events can be detected even at Mpc distances, making them a potentially useful tool to test star-planet interaction models.

In the present study, we follow the orbital evolution of hot Jupiters around solar-type pre-main sequence (hereafter PMS) and main sequence (MS) stars. To do so, we apply tidal dissipation prescriptions from \cite{Barker}, hereafter B20, to the stellar evolutionary models computed with the MESA code (\citealt{MESA1,MESA2,MESA3,MESA4,MESA5}). In our model, we consider equilibrium tide, inertial waves, and gravity waves. The corresponding tidal dissipation rates are subsequently used to obtain the orbital tracks for hot Jupiters with different masses and initial positions. The orbital tracks are eventually converted into the infall diagram plotted for a star with a given mass, initial rotation period, and metallicity. Based on the initial distribution of stellar and planetary parameters, hot Jupiter occurrence rate, and star formation history, we derive the statistics of planetary mergers within the Galactic thin disc. In particular, we estimate the rates of events that may be observed through the bright optical transients and transit-timing variation. The present paper updates the results obtained for hot Jupiter population in the previous works by \cite{Metzger} and \cite{Popkov}. Our model advances the studies by \cite{GalletBolmont1,BolmontGallet,Rao}, as we include gravity wave dissipation which plays the key role in shaping the distribution of planets engulfed by their host stars. At the same time, we confined ourselves to considering tidal interaction only. In future work, we will improve our simulations by taking into account other processes, like magnetic interaction (\citealt{Ahuir}) and photo-evaporation (\citealt{Rao1}).

This paper is structured as follows. In Sec.~\ref{sec:model}, we emphasize the main features of our model. In Sec.~\ref{sec:orbit}, we investigate the impact of the initial conditions (stellar and planetary mass, semi-major axis, stellar rotation period, and metallicity) on the orbital evolution. Our modeling of the Milky Way hot Jupiter population is described in Sec.~\ref{sec:populaton}. The results are presented in Sec.~\ref{sec:results} and discussed in Sec.~\ref{sec:discussion}. Finally, we summarize our work in Sec.~\ref{sec:conclusion}.

\section{Model description} \label{sec:model}

In the present section, we describe the methods we use to simulate the orbital evolution of hot Jupiters. We start with the stellar model and then dwell on the prescriptions for orbital migration and infall scenarios.

\subsection{Stellar model} \label{subsec:stellar}
\label{sec:star} 

\begin{table}
	\centering
	\caption{Parameters used in the present wind braking model\label{tab1}}
	
	\begin{tabular}{cc} 
		\hline
		Parameter & Value\\
		\hline
		$K$ & $1.5 \times 10^{30}$ erg \\
        $m$ & $0.22$ \\
        $p$ & $2.3$ \\
        $\chi$ & $14$ \\
        $\mathrm{Ro_{sat}}$ & $0.14$ \\
        $\alpha_\mathrm{MLT}$ & $1.82$ \\
		\hline
	\end{tabular}
\end{table}

In the present work, the star is assumed as a spherically-symmetric uniformly rotating body. We study orbital evolution of planets around stars with masses in the range of 0.6 to 1.3 $M_{\odot}$ during the MS phase at ages less than the Galactic age (14 Gyr). We use a step of 0.1 $M_{\odot}$ in the range [0.6; 0.9] ${M_{\odot}}$ and a step of 0.05 $M_{\odot}$ in the range [0.9; 1.3] ${M_{\odot}}$. Our grid contains 8 initial rotation periods (${P_\mathrm{rot,init}}$ = 2, 2.5, 3, 3.5, 4.5, 5.5, 8, and 12 days). We compute our stellar models with the evolution code MESA r11701 and prescriptions from \cite{Gossage}.  Their study provides an opportunity to extend the MIST prescriptions (\citealt{Dotter,Choi}), allowing to implement various magnetic braking laws. Few changes are made with respect to the original framework. Firstly, we disable Type 2 opacity to achieve a more realistic age of the MS termination for the solar model. Secondly, we adopt the expression for protoplanetary disc dissipation timescale from \cite{GalletBolmont}:
\begin{equation}
    \tau_\mathrm{disc} = 10^{0.17} \; \left(\frac{P_{\rm rot,init}}{ {\rm days}}\right)^{0.86}  \; \left(\frac{M_{*}}{M_{\odot}}\right)^{1.55}\, {\rm Myr}, 
	\label{eq:disc}
\end{equation}
where $M_{*}$ is the stellar mass and $P_{\rm rot,init}$ --- the initial stellar rotation period. The symbol $\odot$ denotes the solar value.

\begin{figure}
	\includegraphics[width=\columnwidth]{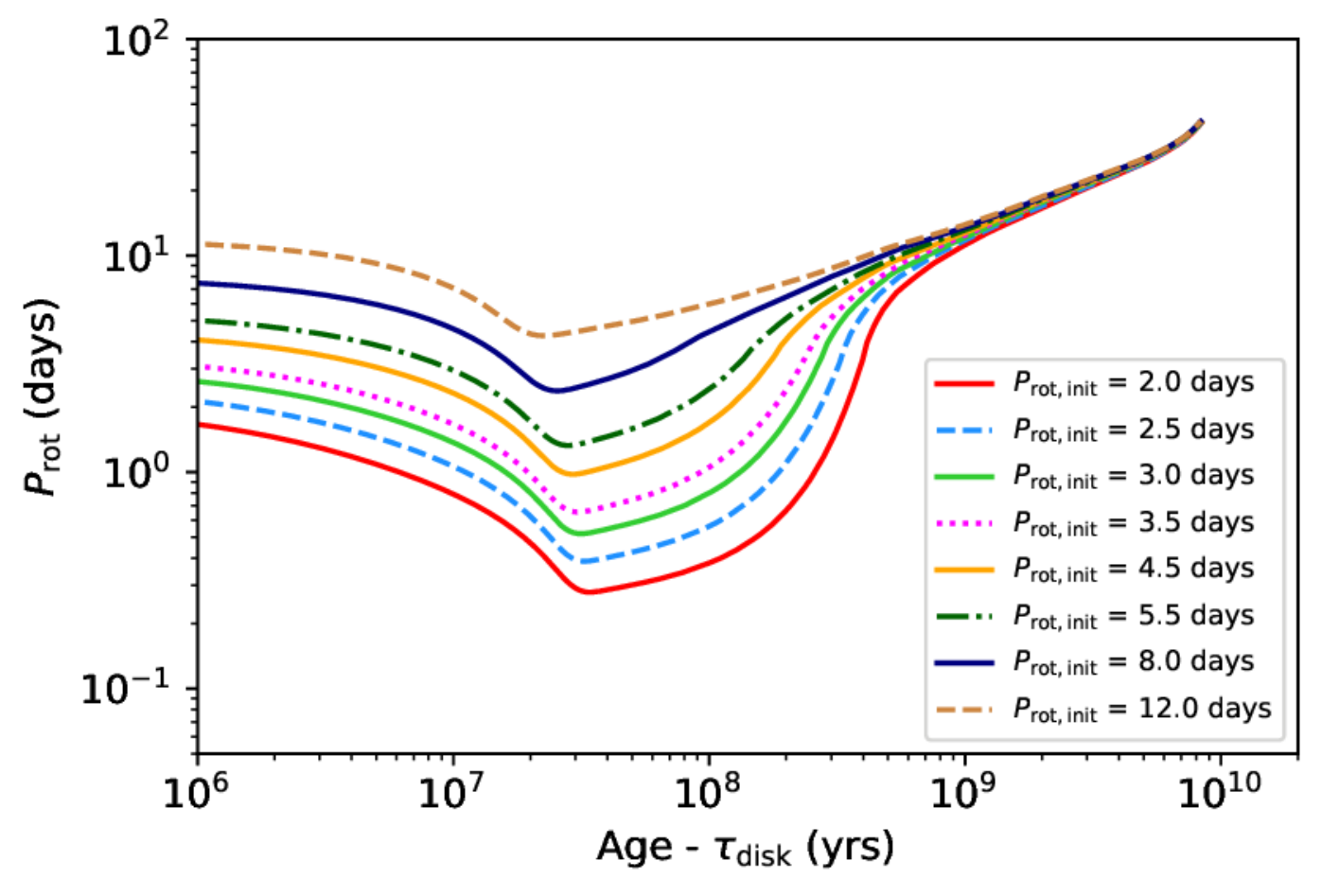}
    \caption{Evolution of the rotation period as a function of time after disc dissipation for all solar mass models with [Fe/H] = 0.0 dex.}
    \label{fig1_F}
\end{figure}

Following \cite{Rebull}, stellar rotation is held constant within the disc lifetime, as accretion and contraction compensate the magnetic wind braking and other braking processes. After disc dissipation, the total angular momentum of an isolated star decreases due to the wind torque parametrized according to the braking model from \cite{Matt, Amard}:

\begin{equation}
    \Gamma_\mathrm{wind} = - \Gamma_{0}  \left(\frac{\tau_\mathrm{cz}}{\tau_\mathrm{cz\odot}}\right)^{p} \; \left(\frac{\Omega_{*}}{\Omega_{\odot}}\right)^{p+1},\; \mathrm{Ro} > \mathrm{Ro}_\mathrm{sat},
	\label{eq:braking1}
\end{equation}
\begin{equation}
    \Gamma_\mathrm{wind} = - \Gamma_{0}  \chi^{p} \left(\frac{\Omega_{*}}{\Omega_{\odot}}\right), \; \mathrm{Ro} < \mathrm{Ro}_\mathrm{sat},
	\label{eq:braking2}
\end{equation}
where
\begin{equation}
    \Gamma_{0} = K \left(\frac{R_{*}}{R_{\odot}}\right)^{3.1}\; \left(\frac{M{*}}{M_{\odot}}\right)^{0.5} \; \gamma^{-2m},
    \label{eq:braking3}
\end{equation}
\begin{equation}
\gamma = \sqrt {1 + 193\left(\frac{\Omega_{*}}{\omega_\mathrm{crit}}\right)^2},
    \label{eq:braking4}
\end{equation}
$\omega_\mathrm{crit} = \sqrt{\frac{GM_*}{R_*^3}}$; $R_{*}$ and $\Omega_{*}$ are the stellar radius and angular rotation rate, respectively. $\mathrm{Ro}$ is the Rossby number, defined as
\begin{equation}
    \mathrm{Ro} = \frac{2\pi}{\Omega_{*}\tau_\mathrm{cz}},
    \label{eq:braking5}
\end{equation}
with $\tau_\mathrm{cz}$ the convective turnover timescale at one half a pressure scale height above the bottom of the outermost convection zone. This quantity is derived in the same manner as in \cite{Gossage}:
\begin{equation}
    \tau_\mathrm{cz}(r) = \alpha_\mathrm{MLT}H_\mathrm{P}(r)/v_\mathrm{c}(r),
    \label{eq:braking6}
\end{equation}
where $\alpha_\mathrm{MLT}$ is the convective mixing length parameter, $H_\mathrm{P}(r)$ is the scale height, and $v_\mathrm{c}(r)$ is the convective velocity. The adopted parameters of the wind braking model are given in Table~\ref{tab1}.

Fig.~\ref{fig1_F} illustrates the spin evolution of solar mass stars with different initial periods of rotation. The rotation periods of stars with the same mass converge into the single $P_\mathrm{rot}(\mathrm{age})$ dependence, allowing to define the gyrochronological ages of the isolated MS stars based on their angular frequency measurements.

\subsection{Migration model} \label{subsec:migration}

Our simulations are based on the assumption that the only way a star-planet system can lose its angular momentum is due to the magnetic wind braking. Thus we begin with the following expression:

\begin{equation}
    \frac{\d L_{*}}{\d t} + \frac{\d L_\mathrm{pl}}{\d t} = \Gamma_\mathrm{wind},
	\label{eq:orbit1}
\end{equation}
where $L_{*}$ and $L_\mathrm{pl}$ are the stellar angular momentum and the angular momentum of the planetary orbit, respectively:
\begin{equation}
    L_{*}=  \Omega_{*} I_{*},
	\label{eq:orbit2}
\end{equation}
\begin{equation}
    L_\mathrm{pl}= n \,M_\mathrm{pl}\, a^2.   
	\label{eq:orbit3}
\end{equation}
\begin{equation}
    n= \sqrt{\frac{G M_{*}}{a^3}}.   
	\label{eq:orbit4}
\end{equation}
\noindent
$I_{*}$ is the stellar moment of inertia, $n$ is the orbital angular frequency, $a$ is the semi-major axis, and $M_\mathrm{pl}$ is the planetary mass.

In the present study, we focus on the dynamics of a star-planet system composed of a spherically-symmetric uniformly rotating star and a point-mass planet on a circular equatorial orbit. In eq.(\ref{eq:orbit3}), we neglect planetary spin. This is a reasonable assumption as the planet's amount of spin angular momentum is negligible compared with the orbital angular momentum. We note, though, that the planetary spin can affect close-in systems through the obliquity-driven tides (\citealt{Millholland}). However, this possibility is beyond the scope of the present research. Given that hot Jupiter rotation is typically synchronized within a relatively short timescale (\citealt{Guillot}), we expect that only stellar tides are significant in the context of the orbital migration of hot Jupiters (the same conclusion was reached in \cite{Matsumura}). Angular momentum transfer establishes the relationship between orbital migration and spin evolution in the star-planet system.  Combining eqs.(\ref{eq:orbit1})~--~(\ref{eq:orbit4}), we obtain the expression for the derivative of the stellar angular rotation rate:
\begin{equation}
    \frac{\d \Omega_{*}}{\d t} = \frac{1}{I_{*}} \left(\Gamma_\mathrm{wind} - \Omega_{*} \frac{\d I_{*}}{\d t} - \frac{1}{2} M_\mathrm{pl} \sqrt{\frac{G M_{*}}{a}} \frac{\d a}{\d t} \right).   
	\label{eq:orbit5}
\end{equation}

Tidally induced planetary migration impacts the host's spin, which, in turn, modifies its internal structure through rotationally-driven mixing and centrifugal deformation. However, given that this effect is relatively small, we assume no feedback on the stellar structure caused by the angular momentum exchange. 

The planetary migration rate is determined by the tidal quality factor $Q'$, which characterizes the efficiency of tidal energy dissipation. The tidal quality factor is proportional to the ratio of the maximum energy stored in the tide to the amount of energy dissipated over one tidal period. The value of $Q'$ depends on the tidal forcing frequency $\omega_\mathrm{t} = 2|n - \Omega_{*}|$ and stellar properties, including rotation. The evolution of the semi-major axis of the planet in the case of circular and aligned orbits is given by (B20):
\begin{equation}
    \frac{1}{a}\frac{\d a}{\d t} = \frac{\Omega_{*} - n}{|\Omega_{*} - n|}\frac{9n}{2} \left(\frac{M_\mathrm{pl}}{M_{*}} \right) \left(\frac{R_{*}}{a} \right)^5 \frac{1}{Q'}.  
	\label{eq:orbit7}
\end{equation}
Eq.(\ref{eq:orbit7}) outlines the region of the inward migration located inside the corotation radius $a_\mathrm{cor}$ (where $n = \Omega_{*}$). If the planet is located outside $a_\mathrm{cor}$, it migrates away from the host star. Eq.(\ref{eq:orbit7}) also shows that the migration rate anticorrelates with the tidal quality factor $Q'$. High stellar angular rotation rate provides rapid migration, especially for close-in planets.

Eqs.(\ref{eq:orbit5}) and (\ref{eq:orbit7}) govern our star-planet simulations. The computation is interrupted after a hot Jupiter merges with the host star. We assume that the coalescence occurs when the orbital frequency raises above $\omega_\mathrm{crit}$. We independently treat the phase of the orbital evolution when the planet gets captured by the $n = 2 \Omega_{*}$ limit during the phase of stellar contraction. As noted in subsection~\ref{subsec:tide}, inertial waves are no longer active inside the corresponding orbital radius, which significantly slows down the migration in contrast to the outer region. Taking into account that gravity wave dissipation does not operate during PMS (which is shown in Fig.~\ref{fig2_F}), the planet located under the $n = 2 \Omega_{*}$ limit is not able to approach the stellar surface within the short timescale. As the star contracts, the $n = 2 \Omega_{*}$ limit crosses the orbit enabling inertial wave excitation and thus promoting rapid planetary migration until the planet returns below the limit. The hot Jupiter stays on the edge of the inertial wave excitation region and migrates inwards as the star spins up. This scenario is reviewed in \cite{BolmontGallet,Rao}.

In order to speed up the computation in the situation described above, we substitute $n = 2  \Omega_{*}$ into eqs. (\ref{eq:orbit4}) and (\ref{eq:orbit5}). The differential equation describing the evolution of stellar rotation takes the form:
\begin{equation}
    \frac{\d \Omega_{*}}{\d t} = \frac{1}{I_{*}} \left( \frac{\Gamma_\mathrm{wind} - \Omega_{*} \frac{\d I_{*}}{\d t}}{1-\frac{2}{3}\frac{I_\mathrm{pl}}{I_{*}}} \right),  
	\label{eq:orbit8}
\end{equation}
with $I_\mathrm{pl} = M_\mathrm{pl} \; a^2$ and $a = \left( \frac{GM_{*}}{(2\Omega_{*})^2} \right)^{1/3}$. Eq.(\ref{eq:orbit8}) reduces the number of differential equations to one, which significantly simplifies calculations. When the planet is able to unbind from the $n = 2 \Omega_{*}$ limit, we return to the standard system of differential equations, represented by eqs.(\ref{eq:orbit5}) and (\ref{eq:orbit7}).

\subsection{Tidal dissipation} \label{subsec:tide}

In the following subsection, we briefly outline the prescriptions from B20 that we use to derive the tidal quality factor $Q'$. The common approach to study tidal interactions is to decompose the potential of the perturbing body into a Fourier series and separately explore the tidal response to every mode of perturbation. Similarly to B20, we focus on the $l = m = 2$ mode related to the most important tidal component, particularly for circular and aligned orbits considered in this research.

The tidal quality factor representing the dissipation of equilibrium tide is given by the expression:
\begin{equation}
    \frac{1}{Q'_\mathrm{eq}} = \frac{16\pi G}{3(2l+1)R_{*}^{2l+1}|A|^{2}}\frac{D_\mathrm{v}}{|\omega_\mathrm{t}|},
    \label{eq:tide_eq1}
\end{equation}
with $A^2 = 6\pi/5$ and $D_\mathrm{v}$ the viscous dissipation of equilibrium tide calculated by integrating over the radial extent of each convective region:
\begin{equation}
    D_\mathrm{v} = \frac{1}{2}\omega_\mathrm{t}^2 \int r^{2}\mu(r) D_\mathrm{l}(r)\d r,
    \label{eq:tide_eq2}
\end{equation}
where $\mu(r) = \rho(r)\nu_\mathrm{E}(r)$. Here $\rho(r)$ is the density and $\nu_\mathrm{E}(r)$ is the effective turbulent viscosity. \cite{Duguid1,Duguid2,VidalBarker1,VidalBarker} reported the existence of three different regimes of tidal dissipation with frequency-independent, intermediate, and quadratic scaling laws for $\nu_\mathrm{E}(r)$, respectively. The quadratic law is relevant to high-frequency tidal forcing valid for close-in planets, which is why we assume the only scaling law for the effective turbulent viscosity:

\begin{equation}
    \nu_{E}(r) = \frac{25}{\sqrt{20}}\left(\frac{\omega_\mathrm{c}}{\omega_\mathrm{t}}\right)^2 u_\mathrm{c} l_\mathrm{c},
    \label{eq:tide_eq3}
\end{equation}
with $u_\mathrm{c}$ the convective velocity, $l_\mathrm{c}$ the mixing-length, and $\omega_\mathrm{c} = u_\mathrm{c}/l_\mathrm{c}$ the convective frequency. Such approximation allows one to make the integrand in eq.~(\ref{eq:tide_eq2}) $\omega_\mathrm{t}$-independent, thus speeding up the computation and providing an opportunity to conduct a large number of simulations.

The function $D_\mathrm{l}(r)$ is defined by the components of equilibrium tidal displacement vector \mbox{$\xi$} (eqs. (21), (22) in B20) which is obtained by solving the ordinary differential equation with boundary conditions given by eqs. (16)--(18) in B20.

Inertial wave dissipation is estimated within the framework of the frequency-averaged formalism described in \cite{Ogilvie1}. The above-mentioned work provided the commonly accepted prescription which was subsequently applied in \cite{Mathis} to calculate the dissipation rates based on a simplified homogeneous two-layer stellar model with radiative core and convective envelope. Using the same approach, B20 derived relations for a more complex heterogeneous model adopted in the present paper. Following B20, the tidal quality factor for inertial wave dissipation can be evaluated from the equation:
\begin{equation}
    \frac{1}{Q'_\mathrm{iw}} = \frac{32\pi^2 G}{3(2l+1)R_{*}^{2l+1}|A|^{2}}(E_\mathrm{l} + E_\mathrm{l-1} + E_\mathrm{l+1}),
    \label{eq:tide_iw1}
\end{equation}
with the parameters $E_\mathrm{l}$,$E_\mathrm{l-1}$, and $E_\mathrm{l+1}$ specified by eqs.~(31)--(33) in B20. We note that inertial waves can only be excited when the tidal frequency $\omega_\mathrm{t}$ is in the range $\mathrm{[-2\Omega_{*},2\Omega_{*}]}$. The quantities $E_\mathrm{l}$,$E_\mathrm{l-1}$, and $E_\mathrm{l+1}$ are proportional to the squared angular velocity $\Omega_{*}^2$, which explains fast planetary migration around rapidly rotating stars, see Sec.~\ref{sec:orbit}.

In this work, we dwell on a strongly nonlinear gravity wave dissipation regime. Gravity waves are launched near radiative/convective boundary, increase their amplitude while propagating toward stellar center due to geometric focusing, overturn the background stratification, and break. Such interaction results in effective energy release, as all the energy stored in the primary waves is absorbed by stellar interior, providing the dominant mechanism for planetary migration. The corresponding tidal quality factor is estimated from:

\begin{equation}
    \frac{1}{Q'_\mathrm{gw}} = \frac{2 [\Gamma(\frac{1}{3})]^2}{3^{\frac{1}{3}}(2l+1)(l(l+1))^{\frac{4}{3}}} \frac{R_{*}}{G M_{*}^2} \mathcal{G} |\omega_\mathrm{t}|^{\frac{8}{3}},
    \label{eq:tide_gw1}
\end{equation}
where
\begin{equation}
\mathcal{G} = \sigma_\mathrm{c}^2 \rho_\mathrm{c} r_\mathrm{c}^5 \bigg|\frac{\d \,N^2}{\d \,\ln\,r}\bigg|_{r=r_\mathrm{c}}^{-\frac{1}{3}}.
    \label{eq:tide_gw2}
\end{equation}
Subscript $c$ refers to radiative/convective interface, $N$ is the buoyancy frequency, and the parameter $\sigma_\mathrm{c}$ depends on the derivative of the dynamical tide radial displacement (see eq.(43) in B20).

The feasibility of a strongly nonlinear regime is determined by the initial amplitude of primary waves, which correlates with the tidal frequency and the planetary mass. For stars with radiative core, B20 proposed wave-breaking criterion (a similar criterion is derived in \cite{Ogilvie00,BarkerOgilvie}):

\begin{equation}
A_\mathrm{nl}^2 = \frac{3^{2/3}54\sqrt{6}[\Gamma(\frac{1}{3})]^2}{25\pi(l(l+1))^{\frac{4}{3}}} \frac{\mathcal{G} C^5}{\rho_0} \left(\frac{M_\mathrm{pl}}{M_{*}}\right)^2 \left(\frac{R_{*}}{a}\right)^6 |\omega_\mathrm{t}|^{-13/3} \gtrsim 1,
    \label{eq:tide_gw3}
\end{equation}
with $\rho_0$ the central density and $C$ the slope of the buoyancy frequency profile near center of the star. For the typical hot Jupiter masses, condition (\ref{eq:tide_gw3}) is satisfied at late ages of the MS. Given that the star has already lost most of its angular momentum due to magnetic wind braking, it is useful to assume a non-rotating configuration while calculating gravity wave dissipation ($\omega_\mathrm{t} = 2n$). The latter justifies neglecting the rotational effects on the tidal quality factor $Q'_\mathrm{gw}$, investigated in \cite{IPch}. Finally, the wave-damping criterion can be transformed into the expression for either the critical semi-major axis $a \gtrsim a_\mathrm{crit}(M_{*},M_\mathrm{pl},t)$, or the critical planetary mass $M_\mathrm{pl} \gtrsim M_\mathrm{crit}(M_{*},a,t)$ ($a_\mathrm{crit}$ is not related to $\omega_\mathrm{crit}$ from subsection~\ref{subsec:stellar}). When the condition (\ref{eq:tide_gw3}) is not satisfied, we assume no gravity wave dissipation. In reality, dissipation below the critical amplitude is still expected due to linear and weakly nonlinear damping. However, the efficiency of the relevant processes is markedly reduced compared with the fully damped regime implemented in this work.

For stars possessing convective core, B20 used WKB theory to derive the criterion for wave breaking to occur (see eq. (53) in B20). However, when applied to close-in objects, eq. (53) of B20 gives the critical mass value exceeding $\mathrm{13\;M_\mathrm{J}}$ before the terminal-age main sequence (TAMS). Thus we do not consider gravity wave dissipation in stars with convective cores. Our assumption is confirmed by the results presented in B20. The absence of gravity wave dissipation is one explanation for the survival of the most massive hot Jupiters, which are typically found around F-stars with convective cores (\citealt{BarkerOgilvie0, BarkerOgilvie}).
\begin{figure}
	\includegraphics[width=\columnwidth]{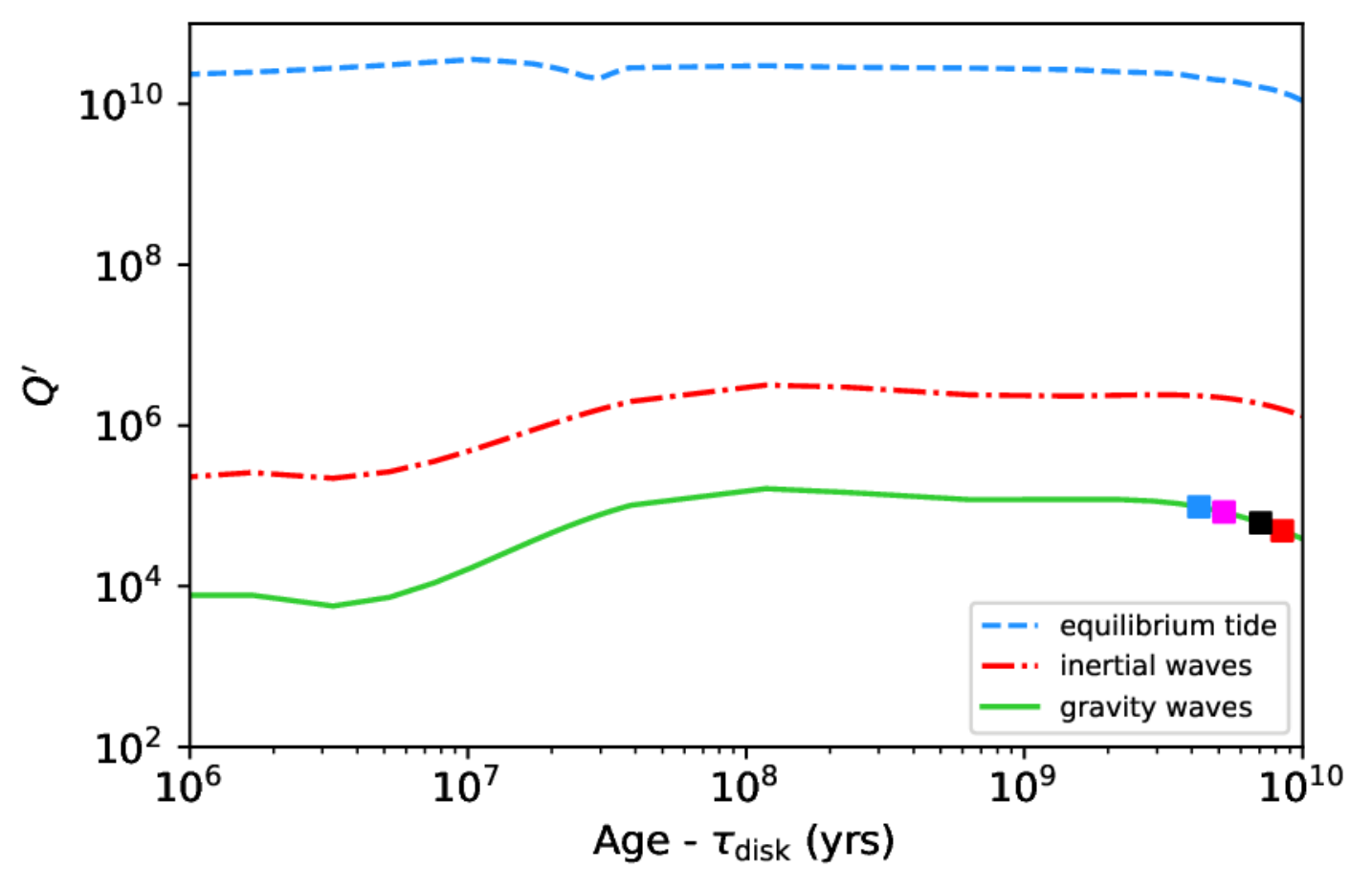}
    \caption{Tidal quality factor as a function of time after disc dissipation ($M_{*}$ = 1.0$ \; M_{\odot}$, [Fe/H] = +0.2 dex, $\tau_\mathrm{disc}$ = 6.4 Myr, $P_\mathrm{orb}$ = 1 day, $P_\mathrm{rot}$ = 5.5 days). Stellar rotation and planetary orbital separation are held constant. Lines, from top to bottom, represent equilibrium tide, inertial waves, and gravity waves, respectively. Red, black, magenta, and blue squares (from left to right) correspond to the initiation of gravity wave dissipation for a planet with $M_\mathrm{pl}$ = 0.3 $M_\mathrm{J}$, 1 $M_\mathrm{J}$, 3 $M_\mathrm{J}$, and 10 $M_\mathrm{J}$, respectively.}
    \label{fig2_F}
\end{figure}

Fig.~\ref{fig2_F} compares the tidal quality factors due to each tidal mechanism as a function of time after disc dissipation. From now on, our reference model corresponds to [Fe/H] = +0.2 dex since hot Jupiters are found to orbit metal-rich stars preferentially. The mean metallicity of the hot Jupiter hosts is enhanced by around 0.2 dex relative to the field star population (\citealt{Petigura}). Chemical abundances are scaled according to the solar chemical mixture by \cite{Asplund}.

It is clear that, when enabled, the dynamical tide dissipation makes a leading contribution. Equilibrium tide dissipation, in turn, is much weaker. However, its occurrence does not depend on the fulfillment of any conditions. Another important note concerns the dependence of the beginning of gravity wave dissipation on planetary mass, represented by the colored squares in Fig.~\ref{fig2_F}. More massive planets induce gravity waves with higher amplitude, which is why they are able to overturn the stratification and break a few billion years earlier. As a result, the duration of the rapid phase of migration is longer, with more distant planets merging with the host star before the TAMS. The dissipation of inertial waves is more intense during the PMS and remains almost constant on the MS.

If both inertial and gravity waves dissipate, the resulting tidal quality factor is expressed by:

\begin{equation}
    \frac{1}{Q'} = \frac{1}{Q'_\mathrm{eq}} + \frac{1}{Q'_\mathrm{iw}} + \frac{1}{Q'_\mathrm{gw}}
    \label{eq:tide}
\end{equation}

\subsection{Types of mergers} \label{subsec:merger}
We follow the classification by \cite{Metzger}, who divided planet-star mergers into three categories, depending on the ratio between planetary and stellar mean densities $\rho_\mathrm{pl}/\rho_{*}$.

If the relative density is high ($\rho_\mathrm{pl}/\rho_{*}>5$), interaction occurs without tidal disruption or Roche lobe overflow, leading to a very bright transient. This case is called direct impact. The maximum luminosity of the corresponding event is:
\begin{equation}
    L_\mathrm{peak, di} = (4\pi\sigma)^{\frac{1}{3}} \; T_\mathrm{rec}^{\frac{4}{3}} \; \;\frac{GM_\mathrm{*}}{R_\mathrm{*}}^{\frac{1}{3}}\; \left(\frac{M_\mathrm{pl}}{10m_\mathrm{p}}\mathrm{Ry}\right)^{1.14}\; {\rm erg}\; {\rm s^{-1}},
    \label{eq:luminosity0}
\end{equation}
with $\sigma$ Stefan-Boltzmann constant, $T_\mathrm{rec} = 6000 K$ the recombination temperature of hydrogen, $m_\mathrm{p}$ proton mass, and $\mathrm{Ry} = 13.6 eV$.

In the case of average relative densities ($1<\rho_\mathrm{pl}/\rho_{*}<5$), tidal forces destroy the planet, which is called tidal disruption. Such events are accompanied by the bright transients lasting for weeks or months with the peak luminosity (\citealt{Popkov}):
\begin{equation}
    L_\mathrm{peak, td} = 9.5 \times 10^{36} \; \frac{M_\mathrm{pl}}{M_\mathrm{J}}\; {\rm erg}\; {\rm s^{-1}},
    \label{eq:luminosity}
\end{equation}
where $M_\mathrm{J}$ is the Jupiter mass.

Finally, if the planet has a lower density than the star ($\rho_\mathrm{pl}/\rho_{*}<1$), the corresponding interaction is called stable accretion. Stable accretion proceeds on a long timescale with no significant bright events (\citealt{Valsecchi,Jackson}), but might result in spin-up and chemical enrichment of stellar outer layers.

\subsection{Density of hot Jupiters} \label{subsec:density}

\begin{figure}
	\includegraphics[width=\columnwidth]{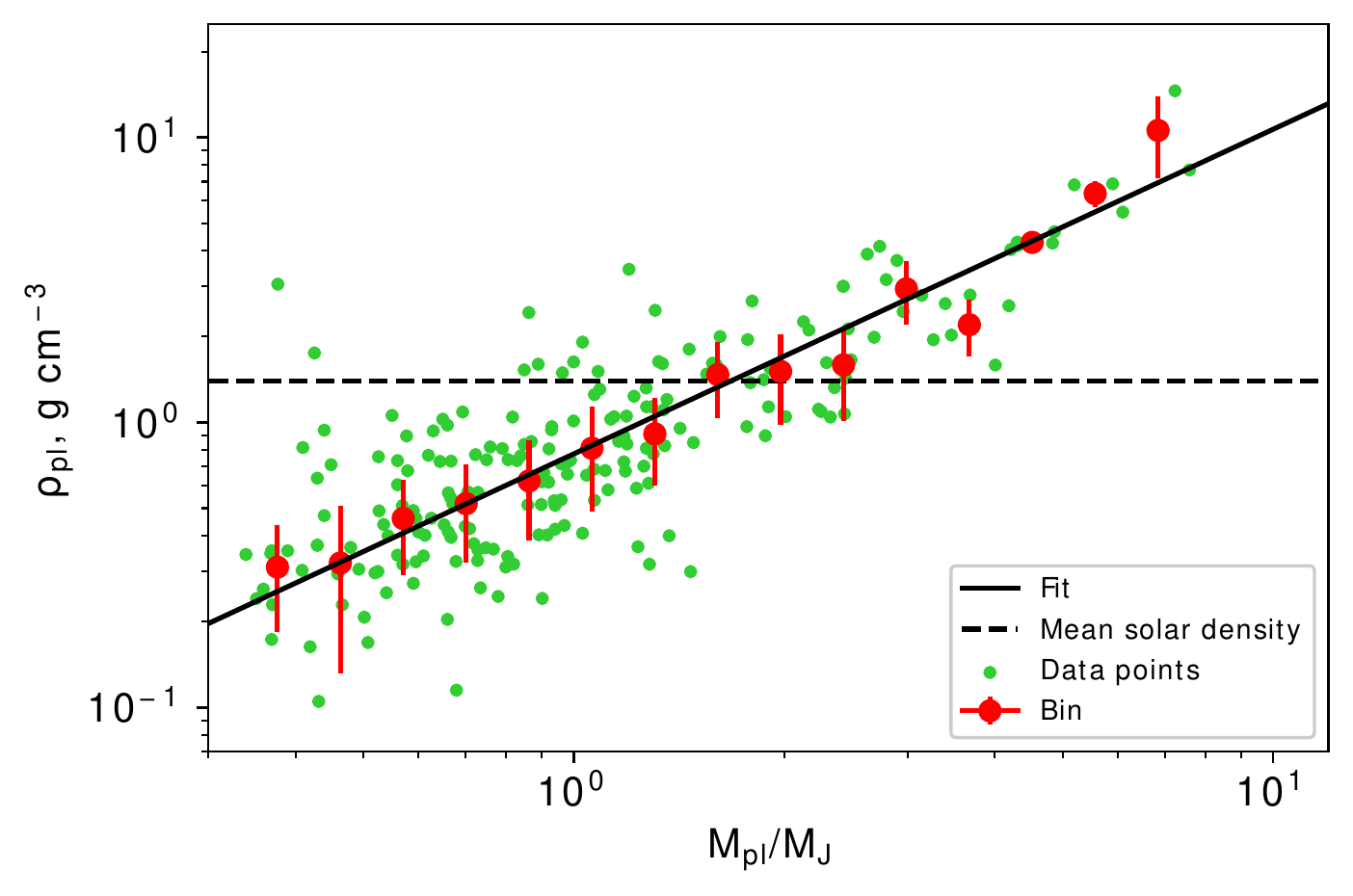}
    \caption{Mass--density diagram. The planetary mass is given in Earth masses. Green points represent the initial sample taken from \url{https://exoplanetarchive.ipac.caltech.edu/}. Red points with error bars represent bins. Black solid line corresponds to the best fit, expressed by eq.(\ref{eq:density}).}
    \label{fig3_F}
\end{figure}

In subsection~\ref{subsec:merger}, we discuss the impact of hot Jupiter density on the outcome of a star-planet interaction. To obtain density as a function of a planetary mass, we select from the NASA Exoplanet Archive all close systems ($\mathrm{1\;{\rm day} < P_\mathrm{orb} < 10\;{\rm days}}$) with a hot Jupiter ($\mathrm{0.3\;M_\mathrm{J} < M_\mathrm{pl} < 10\;M_\mathrm{J}}$) orbiting a solar-type star ($\mathrm{0.6\; M_{\odot} < M_{*} < 1.3 \;M_{\odot},\; 4500\; K < T_\mathrm{eff} < 7000\; {\rm K},\; log\; g > 4.0}$), for which both planetary mass and radii estimates are available. The computed density ($\rho_\mathrm{pl} = \frac{M_\mathrm{pl}}{4/3\pi R_\mathrm{pl}^3)}$) is plotted against mass of the planet in Fig.~\ref{fig3_F}. The data is binned and cleaned as we remove the points lying outside $\mathrm{3\sigma}$ range with respect to the corresponding bin. The mass--density dependence is subsequently fitted with the following expression:
\begin{equation}
    \rho_\mathrm{pl} = 0.78 \; \left(\frac{M_\mathrm{pl}}{M_\mathrm{J}}\right)^{1.14}\; {\rm g}\; {\rm cm^{-3}}.
    \label{eq:density}
\end{equation}

Fig.~\ref{fig3_F} shows that the majority of hot Jupiters has a lower density than the mean density of the Sun ($\mathrm{\rho_{\odot} = 1.4 \;  {\rm g}\; {\rm cm^{-3}}}$), thus most of the mergers do not produce bright transients.

\section{Orbital evolution} \label{sec:orbit}
In this section, we investigate the effect of various factors on the orbital evolution of hot Jupiters. We start with varying the parameters associated with the planets and then move on to stellar properties. 
\subsection{Impact of the initial semi-major axis} \label{subsec:axis}
\begin{figure}
	\includegraphics[width=\columnwidth]{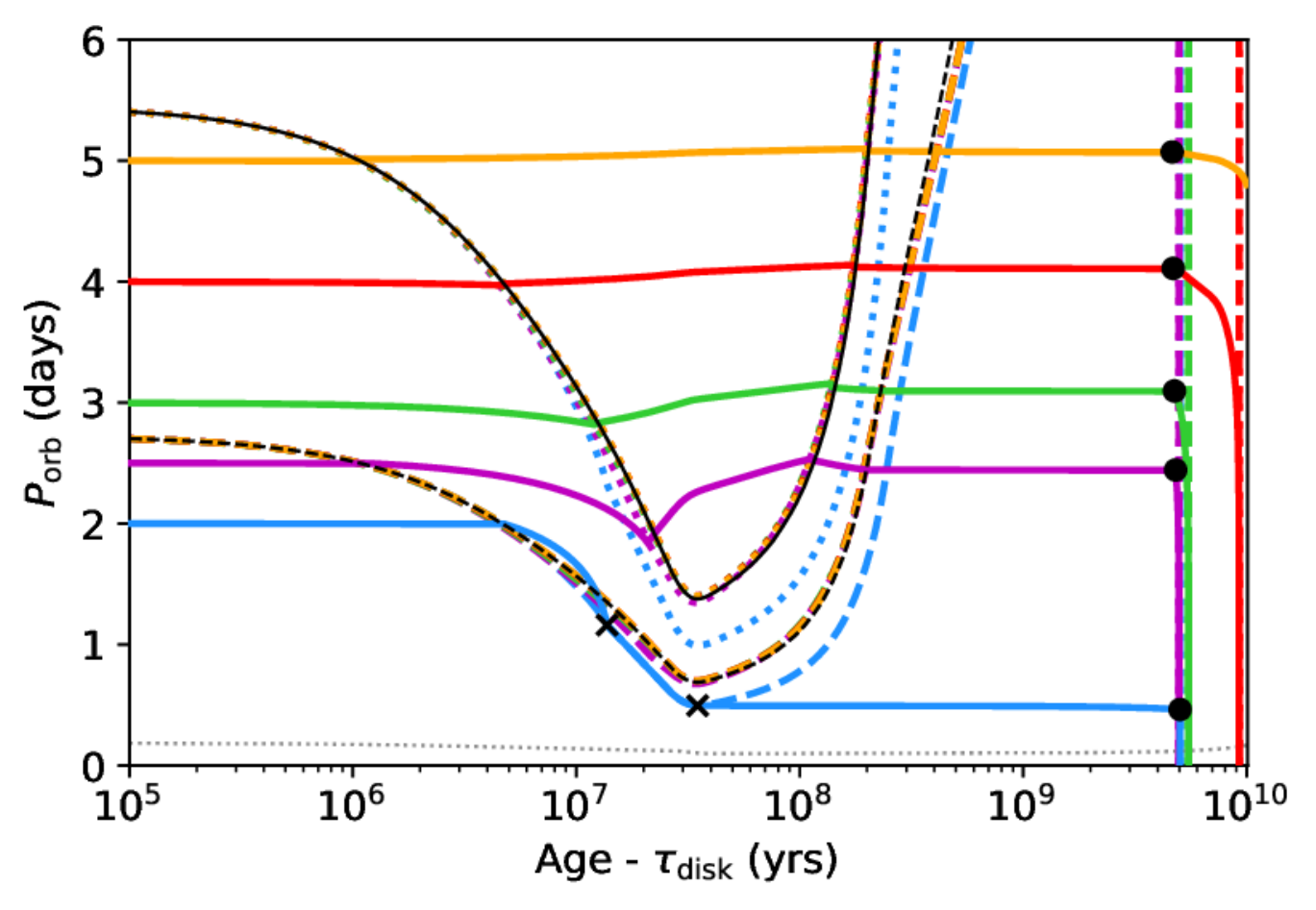}
    \caption{Secular evolution of hot Jupiters with $M_\mathrm{pl} = 3\;M_\mathrm{J}$ around a solar-mass star with $P_\mathrm{rot,init}$ = 5.5 days and [Fe/H] = +0.2 dex. Black solid and dashed lines indicate the corotation radius and the $n = 2 \Omega_{*}$ limit of an isolated star, respectively. Gray dotted line represents stellar surface. Colored solid lines correspond to the orbital period evolution of the planets. From top to bottom: $P_\mathrm{orb,init}$ = 5, 4, 3, 2.5, 2 days. Colored dotted and dashed lines represent the corotation radius and $n = 2 \Omega_{*}$ limit of a star in the presence of a corresponding planet, respectively. Black circles indicate the starting time of gravity wave dissipation, black crosses restrict the phase of the orbital evolution when hot Jupiter is captured on the $n = 2 \Omega_{*}$ limit.}
    \label{fig4_F}
\end{figure}

Fig.~\ref{fig4_F} illustrates the secular evolution of a star-planet system composed of a solar-mass star with a median initial rotation and a hot Jupiter with $M_\mathrm{pl} = 3\;M_\mathrm{J}$. We consider five initial orbital periods: $P_\mathrm{orb,init}$ = 2, 2.5, 3, 4, and 5 days. Colored solid lines represent the planetary migration, while colored dashed and dotted lines correspond to the dynamics of the corotation radius and the $n = 2 \Omega_{*}$ limit, respectively. In all five cases, the planet is initially located in the region of inward migration. Three of the most distant planets are able to excite the inertial waves inside the host star. As the star spins up, these planets cross the corotation radius, reversing migration from inward to outward. Hot Jupiters with $P_\mathrm{orb,init}$ = 2 and 2.5 days, in turn, are initially located in the region of slow migration where only equilibrium tide operates. Subsequently, they cross the $n = 2 \Omega_{*}$ limit, significantly enhancing the migration rates. However, only the planet with $P_\mathrm{orb,init}$ = 2 days is located close enough to be captured by the $n = 2 \Omega_{*}$ limit. In all five cases, the planet gets into the equilibrium tide region after the star has lost most of its angular momentum due to the wind braking. Unlike the models from \cite{BolmontGallet}, our stellar evolutionary tracks do not reveal a rapid decrease of the stellar radius at 20 Myr. Besides, the equilibrium tide dissipation computed using the prescriptions from B20 is too weak to result in the engulfment of the closest hot Jupiter within the first billion years, in contrast to Fig. 8 of \cite{BolmontGallet}.

The initiation of gravity wave dissipation shown by the black circles serves as a turnaround point in the context of the orbital evolution for all hot Jupiters, except for the furthest one, eventually leading to the coalescence with the host star. One can see that the wave-breaking criterion becomes satisfied at almost the same age for all star-planet systems with the same planetary mass. This is due to the robust dependence of $A_\mathrm{nl}^2$ on $C$ expressed by eq.(\ref{eq:tide_gw3}). The quantity $C$, which characterizes the strength of the stratification at the center of a star, increases throughout stellar evolution on the MS. The latter is shown in Fig. 10 of B20. Five billion years of gravity wave dissipation is enough for a planet represented by red color to merge with the star. Hot Jupiters shown in green and magenta require roughly a billion years to do so. The planet represented by blue color falls onto the star within the first thousand years since the beginning of gravity wave breaking.

\subsection{Impact of planetary mass} \label{subsec:plmass}
\begin{figure*}
\begin{multicols}{2}
    \includegraphics[width=\linewidth]{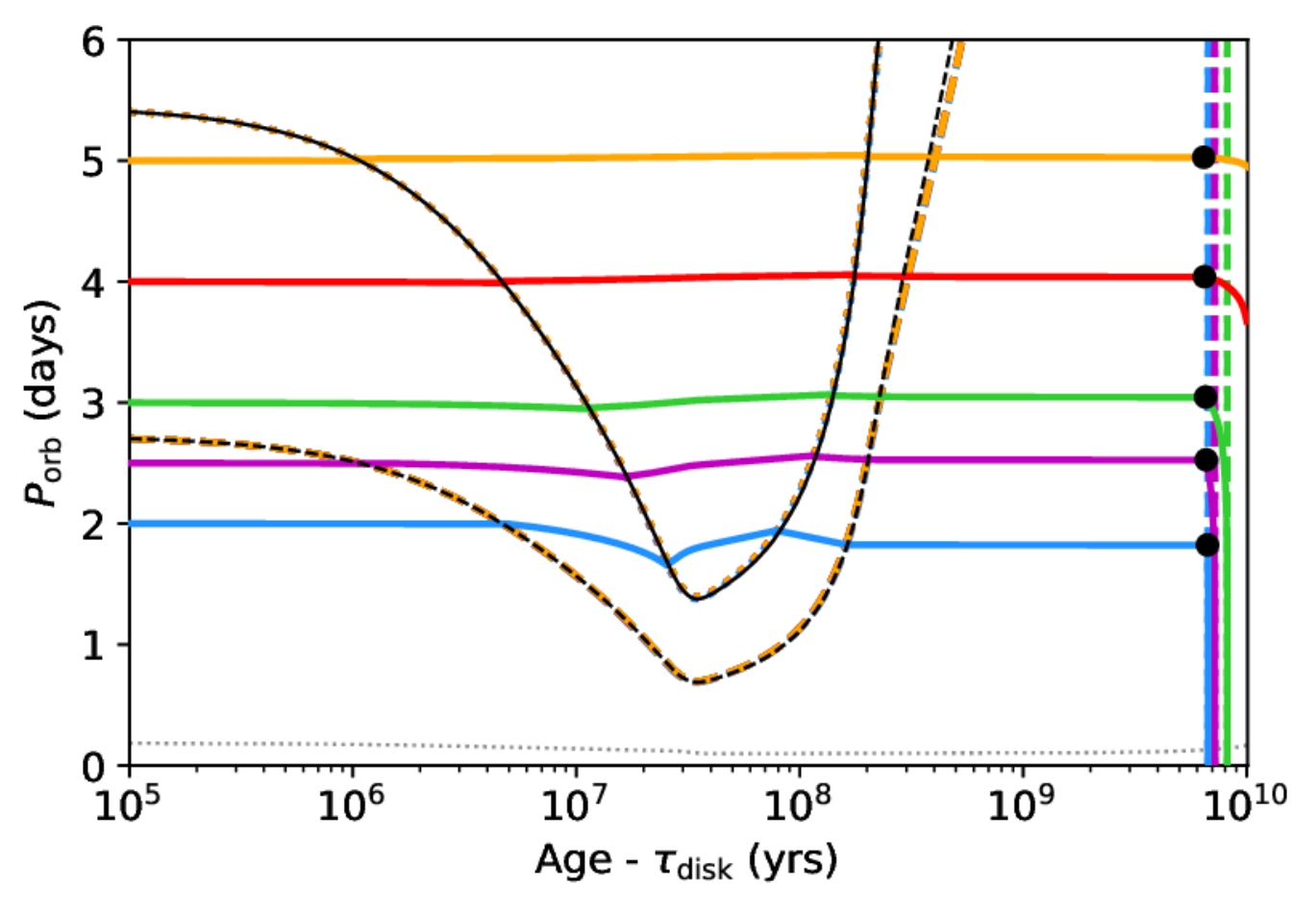}\par 
    \includegraphics[width=\linewidth]{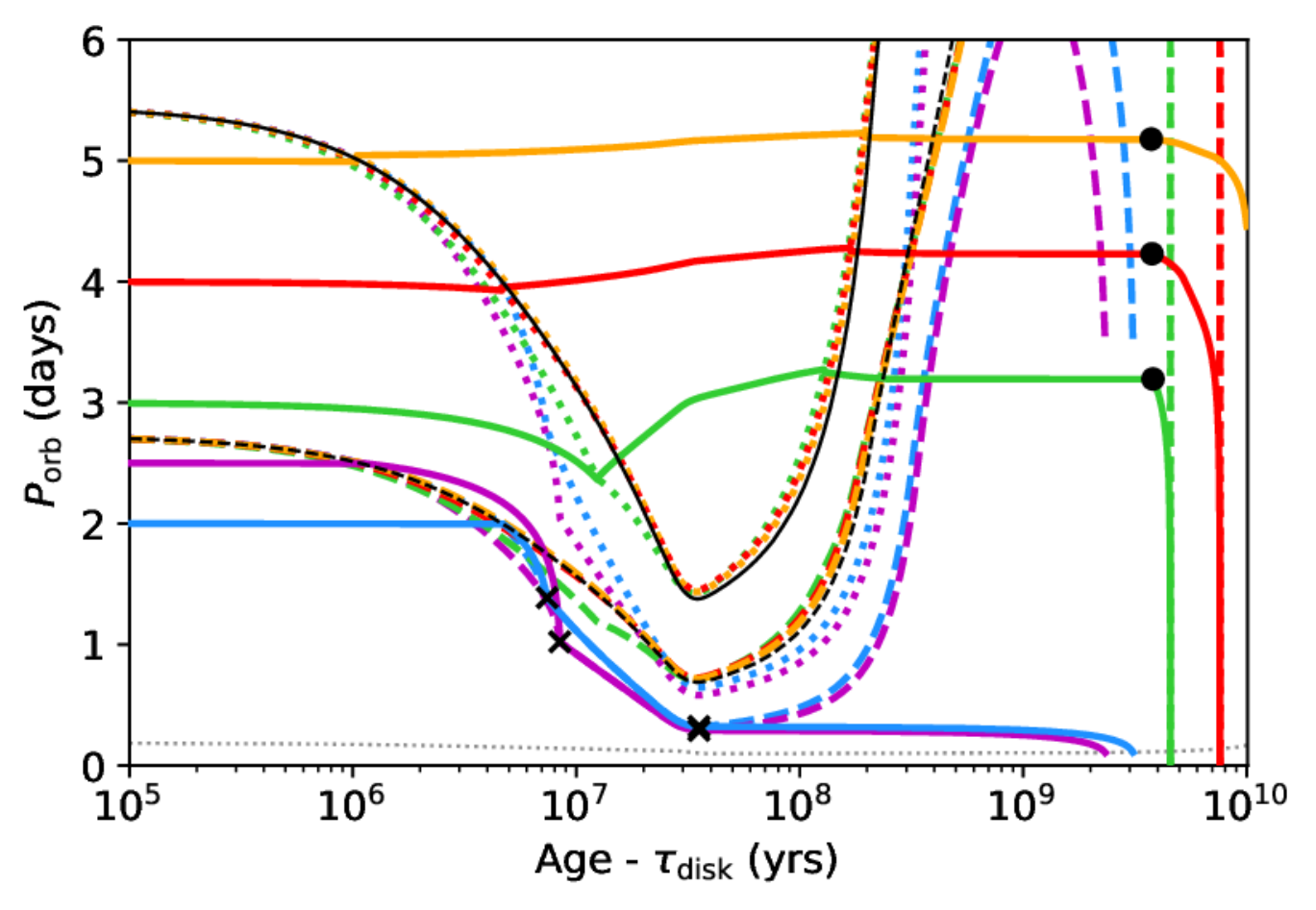}\par 
    \end{multicols}
\caption{Secular evolution of hot Jupiters with different masses. Designations and stellar properties are the same as those in Fig.~\ref{fig4_F}. Left panel corresponds to $M_\mathrm{pl} = 1\;M_\mathrm{J}$, right panel corresponds to $M_\mathrm{pl} = 7\;M_\mathrm{J}$.}
\label{fig5_F}
\end{figure*}

 Fig.~\ref{fig5_F} demonstrates the effect of planetary mass on the secular evolution of a star-planet system formed by a star with the same properties as in subsection~\ref{subsec:axis}. The left panel represents the population of hot Jupiters with $M_\mathrm{pl} = 1\;M_\mathrm{J}$. As opposed to the case of massive planets with $M_\mathrm{pl} = 7\;M_\mathrm{J}$, shown on the right, the inertial waves do not induce rapid migration. Less massive hot Jupiters are not captured on the $n = 2 \Omega_{*}$ limit and undergo slow tidal evolution. As a result, they are located close to the initial orbit at the beginning of gravity wave dissipation. 
 
 On the contrary, the right panel of Fig.~\ref{fig5_F} indicates that inertial waves strongly influence the dynamics of close planets with $M_\mathrm{pl} = 7\;M_\mathrm{J}$. The planets shown in blue and magenta attach to the $n = 2 \Omega_{*}$ limit and migrate effectively before zero-age main sequence (ZAMS). More massive hot Jupiters are able to spin up the host star to a higher angular velocity. Consequently, these planets leave the $n = 2 \Omega_{*}$ limit closer to the stellar surface, leading to early engulfment without the impact of gravity waves. Note that initially the closest planet merges with the host star later than the second closest hot Jupiter. Such change of order happens because the distant planet is able to deposit more angular momentum into the stellar spin, forcing the $n = 2 \Omega_{*}$ limit to move lower. Another essential feature concerns the dissipation of gravity waves. According to Fig.~\ref{fig5_F}, gravity waves provide the infall of more distant planets compared to low-mass hot Jupiters. First of all, this is due to the fact that the migration rate, expressed by eq.(\ref{eq:orbit7}), correlates linearly with planetary mass. Secondly, as reported in subsection \ref{subsec:tide}, the dissipation of gravity waves begins earlier for massive hot Jupiters, meaning that such objects have more time to merge before the TAMS.

\begin{figure}
	\includegraphics[width=\columnwidth]{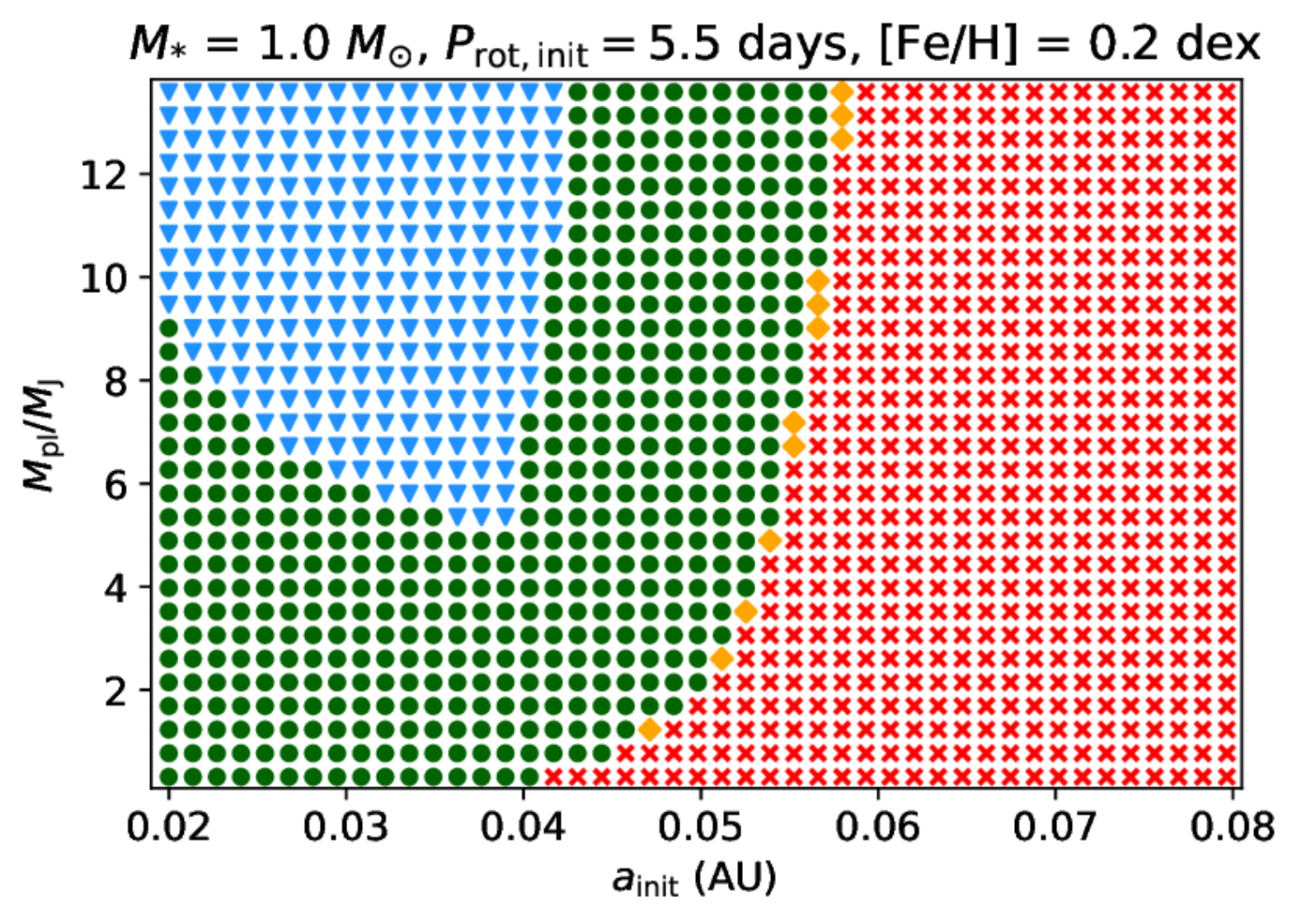}
    \caption{Infall diagram for a solar-mass star with $P_\mathrm{rot,init}$ = 5.5 days and [Fe/H] = +0.2 dex. The depicted mergers are prior to the end of the main sequence of the host.} Red crosses correspond to the no-infall region. Orange diamonds indicate the intermediate outcome (no infall, but the planet reduces the semi-major axis by more than 20 \% during the gravity wave dissipation phase). Green circles mark the infall due to gravity waves. Blue triangles indicate the infall proceeding before the initiation of gravity wave breaking. 
    \label{fig6_F}
\end{figure}

To investigate the fate of hot Jupiter depending on its location in the parameter space, we perform our simulations with 30 planetary masses uniformly spaced in the range [0.3,13.6] $M_\mathrm{J}$ and 60 initial positions uniformly spaced in the range [0.02,0.10] AU. For every engulfment, we determine the type, age of the system, luminosity of event (for bright transients), and luminosity of the star. The results are plotted in Fig.~\ref{fig6_F}. This so-called infall diagram reproduces the main information reported in the first two subsections of the present section. The overall infall region, shown by blue and green markers, widens with increasing planetary mass. The mergers occurring without gravity waves, represented by blue triangles, correspond to the planets with $M_\mathrm{pl} \gtrsim 5\;M_\mathrm{J}$ initially located inside the [0.02,0.04] AU interval. Most of the infall region in the parameter space is occupied by coalescences accompanied by gravity wave dissipation, represented by green circles.  

\subsection{Impact of the initial stellar rotation rate} \label{subsec:rotation}
\begin{figure*}
\begin{multicols}{2}
    \includegraphics[width=\linewidth]{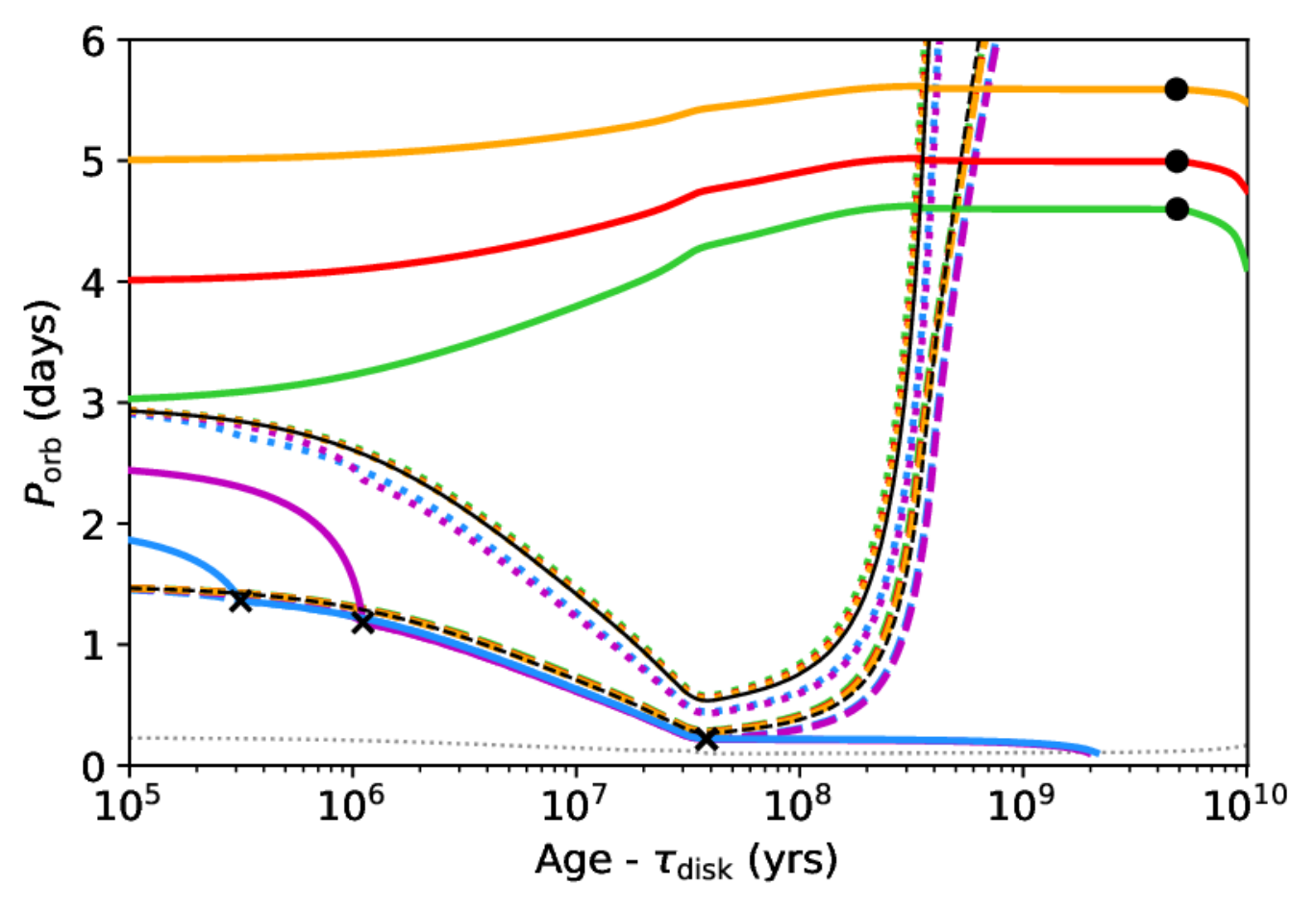}\par 
    \includegraphics[width=\linewidth]{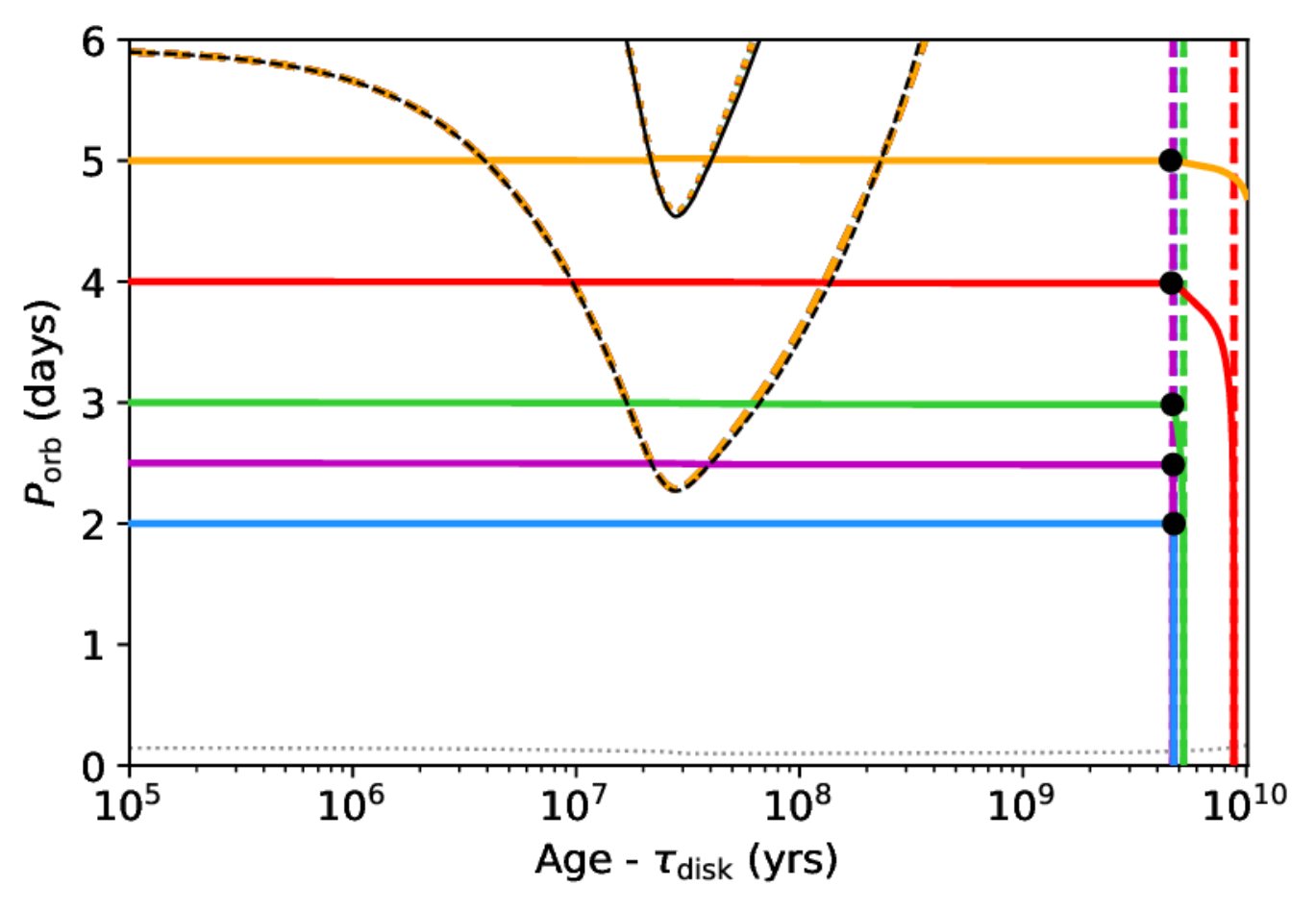}\par 
    \end{multicols}
\caption{Secular evolution of hot Jupiters with $M_\mathrm{pl} = 3\;M_\mathrm{J}$ around solar-mass stars with different initial spin. Designations are the same as those in Fig.~\ref{fig4_F}. Left panel corresponds to $P_\mathrm{rot,init}$ = 3 days, right panel corresponds to $P_\mathrm{rot,init}$ = 12 days.}
\label{fig7_F}
\end{figure*}

Stellar spin is another significant factor governing the tidal evolution of a star-planet system. Fig.~\ref{fig7_F} demonstrates the migration around a solar-mass star with the initial rotation period of 3 and 12 days. The planets orbiting fast rotators, shown in the left panel, undergo rapid migration under the dissipation of inertial waves. Those located outside the corotation radius move far enough from the host star to remain stable against tidal inspiral by the TAMS. Hot Jupiters initially lying inside the region of inward migration merge with the star through the dissipation of equilibrium tide. Thus, the contribution of gravity waves to secular evolution around fast rotators is substantially reduced in favor of dominating inertial waves. Conversely, the coalescence with a slow rotator is a direct consequence of gravity wave dissipation. The right panel reveals no migration around the star with $P_\mathrm{rot,init}$ = 12 days until the wave-breaking criterion is satisfied, which highlights the unimportance of equilibrium tides when modeling the migration around slow rotators. In this way, star-planet mergers occur toward the end of the evolution on the MS, in contrast to the case of a rapid rotator.

To investigate the impact of stellar rotation in detail, we refer the reader to Fig.~\ref{ap1} in Appendix A, which contains the infall diagrams obtained for every solar-mass model. Note that extremely fast initial rotation leads to the coalescence of the most massive hot Jupiters with the PMS stars. In the absence of gravity waves, the infall region is outlined by the initial location of the corotation radius. The importance of gravity waves in the context of orbital decay gradually increases with decreasing initial spin. Finally, the overall infall region reaches the maximum for the models with slow rotation ($P_\mathrm{rot,init}$ = 8, 12 days).
\subsection{Impact of stellar mass} \label{subsec:stmass}
\begin{figure}
	\includegraphics[width=\columnwidth]{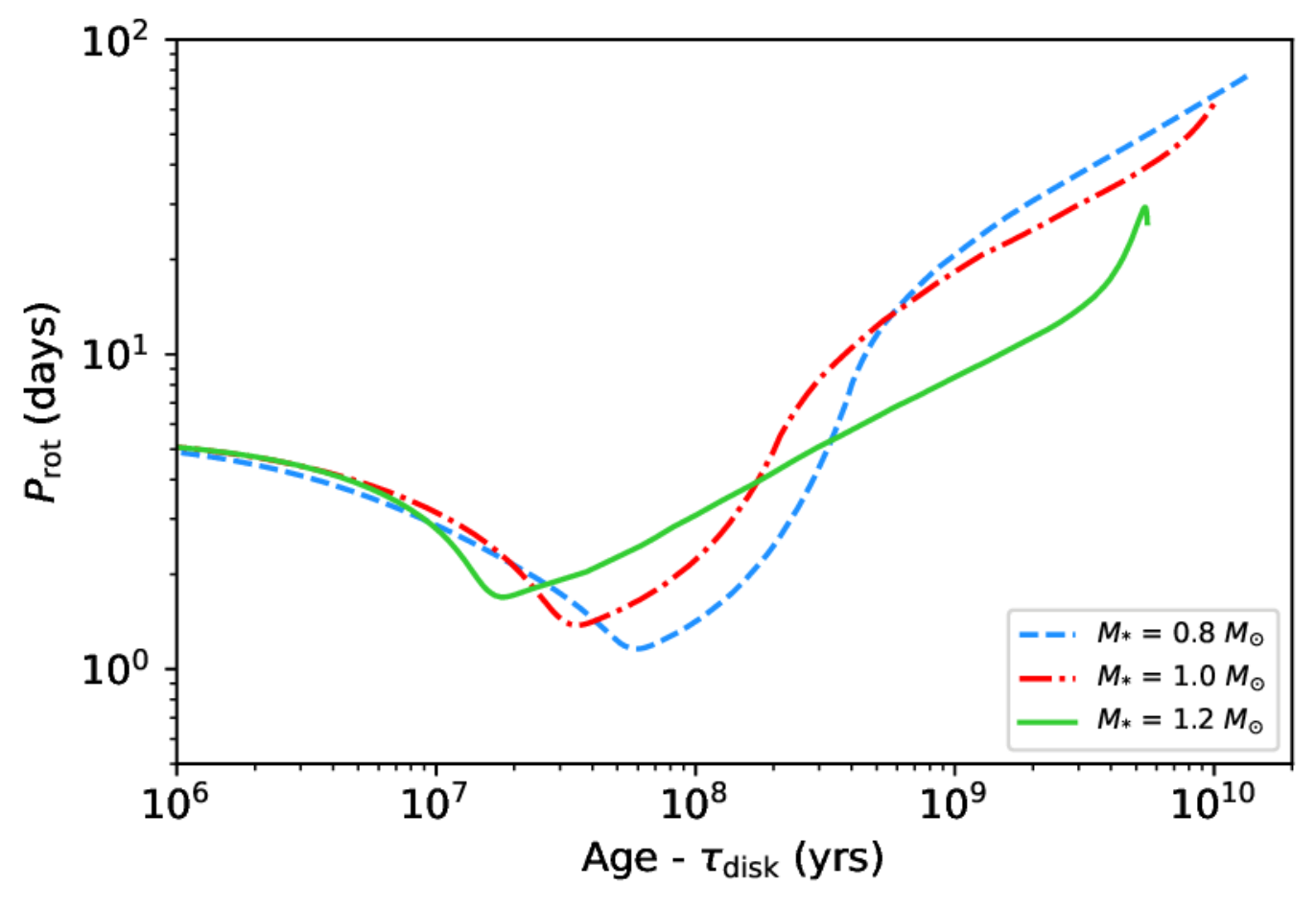}
    \caption{Evolution of the rotation period as a function of time after disc dissipation for isolated stars with $M_\mathrm{*} = 0.8, 1.0, 1.2 \; M_\mathrm{\odot}$. Stellar models correspond to [Fe/H] = +0.2 dex.}
    \label{fig8_F}
\end{figure}
For stars possessing radiative core, the contribution of gravity waves depends on stellar mass. Fig.~\ref{ap2} shows infall diagrams for a 0.8 $M_\mathrm{\odot}$ star. The distribution of planets unstable against orbital decay around fast rotators is slightly different from what is illustrated in Fig.~\ref{ap1} for a 1.0 $M_\mathrm{\odot}$ star. The corresponding variation is related to the initial position of the corotation radius for a given orbital period. The fate of planets orbiting median and slow rotators is more sensitive to mass of the host. First of all, one can see the absence of dissipation of gravity waves produced by low-mass hot Jupiters in a 0.8 $M_\mathrm{\odot}$ star. The critical planetary mass required for gravity wave damping within the first 14 Gyr increases sharply with decreasing stellar mass. Accordingly, we do not expect gravity wave breaking in stars with  $M_* \leq 0.6 \; M_\mathrm{\odot}$. Even if the planet is above the critical mass limit, it does not have enough time to merge from the initial separation of more than 0.05 AU as the phase of active migration begins too late. Secondly, as demonstrated in Fig.~\ref{fig8_F}, low-mass stars spin up to a higher angular velocity before ZAMS. Thus, inertial waves dissipate more effectively and prevail over gravity waves. It means that hot Jupiters initially located outside the corotation radius migrate far enough to remain stable while the closest hot Jupiters manage to merge before the starting time of gravity wave breaking. This feature is observed in Fig.~\ref{ap2}, as the fraction of planets unaffected by gravity waves is higher than in the diagrams computed for the solar-mass model.

In stars with convective cores, the whole picture changes dramatically, which is revealed for a 1.2 $M_\mathrm{\odot}$ star in Fig.~\ref{ap3}. As noted in subsection~\ref{subsec:tide}, these stars do not exhibit gravity wave dissipation during the MS stage. The latter is the main factor coming into play. The upper two diagrams on the right in Fig.~\ref{ap3} demonstrate the offset in the distribution of decaying planets orbiting median rotators toward massive hot Jupiters. The infall region corresponding to slow rotators is separated into two distinct parts. The part related to a smaller semi-major axis represents hot Jupiters migrating under the dissipation of equilibrium tide only. The right part refers to a hot Jupiter population initially located above the $n = 2 \Omega_{*}$ limit, hence being able to excite inertial waves. Rapid migration driven by the dynamical tide allows the planets to get closer to the stellar surface before ZAMS. Consequently, these hot Jupiters manage to fall onto the host star due to the dissipation of equilibrium tide.

\subsection{Impact of metallicity} 
\label{subsec:metallicity}
Stellar chemical composition strongly affects the population of planetary systems. Metal-rich stars possess protoplanetary discs with high dust to gas ratio. For such discs, solid accretion is enhanced, which is the necessary condition for the formation of close-in giant planets. Thereby, hot Jupiters are more common around stars with high abundance. The latter has confirmation both at theoretical and observational levels. For instance, applying population synthesis calculations, \cite{Mordasini3} confirmed the excess in the metallicity of hot Jupiter hosts. This result was later reproduced by \cite{Alessi}. Moreover, \cite{Mordasini3} showed that, at low metallicities, Jovian planets cannot form inside the ice line. The above findings agree well with observational trends (\citealt{Santos,Fischer,Adibekyan,Petigura}). Another important factor affected by metallicity is the efficiency of Type I migration which plays a critical role in shaping the initial distribution of planets in the mass--separation diagram (\citealt{Mordasini2}). 

\begin{figure}
	\includegraphics[width=\columnwidth]{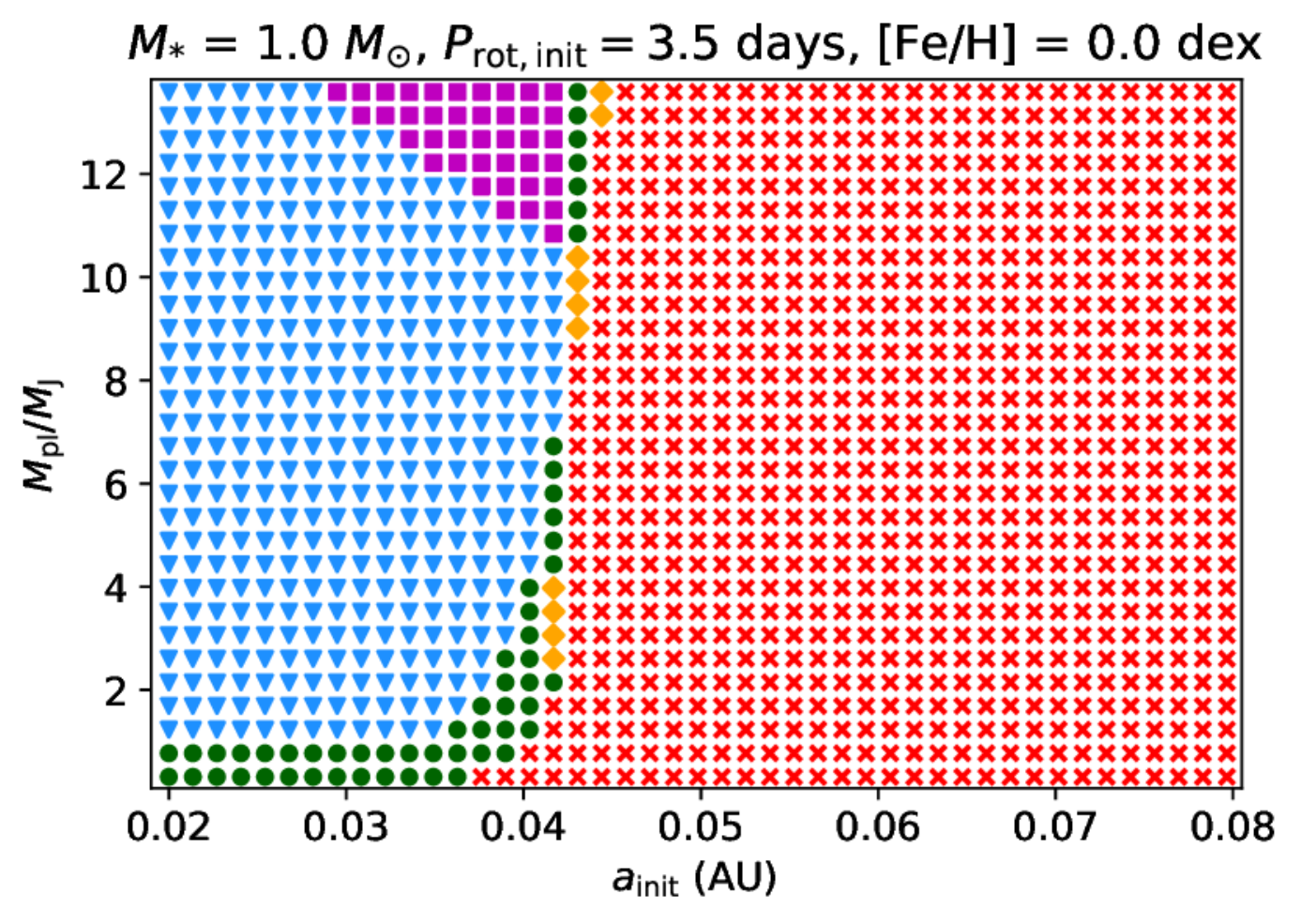}
	\includegraphics[width=\columnwidth]{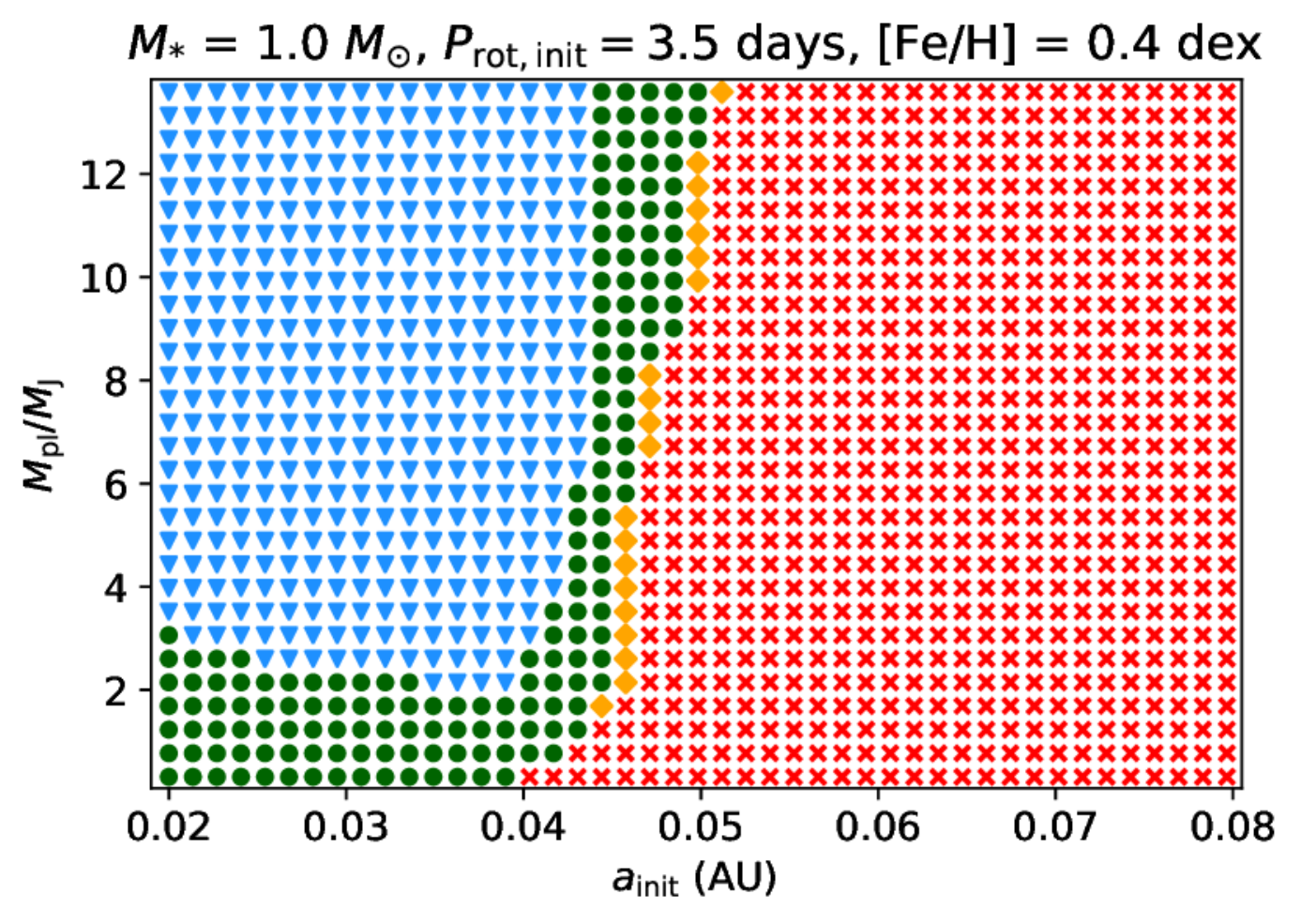}
    \caption{Infall diagrams for solar-mass stars of two different metallicites. Top panel refers to [Fe/H] = 0.0 dex, bottom panel refers to [Fe/H] = +0.4 dex. $P_\mathrm{rot,init}$ = 3.5 days for both panels. Designations are the same as those in Fig.~\ref{fig6_F}. Magenta squares correspond to the infall before ZAMS.}
    \label{fig9_F}
\end{figure}

In the present study, we focus on the metallicity effect on hot Jupiter migration after the disc dissipation. In particular, we find that the engulfment is more likely in metal-rich systems.
This feature is especially pronounced for median rotators with $P_\mathrm{rot,init}$ between 3.5 and 5.5 days. As reported in subsection~\ref{subsec:rotation}, planets orbiting rapid rotators undergo the most intensive migration under the dissipation of inertial waves at early ages. In contrast, systems with slow rotators decay through gravity wave dissipation at late ages of the MS. The transition between these two specific regimes of star-planet dynamics is sensitive to the chemical composition of the host star, which is demonstrated in Fig.~\ref{fig9_F}. One can see that the solar-metallicity model, shown on the top, is marginally affected by gravity waves, compared with the metal-rich model depicted on the bottom. At the same time, metal-poor stars engulf their planets earlier, which opens up an opportunity for some of the most massive planets, depicted in magenta on the top, to merge with the host star before ZAMS.

\begin{figure}
	\includegraphics[width=\columnwidth]{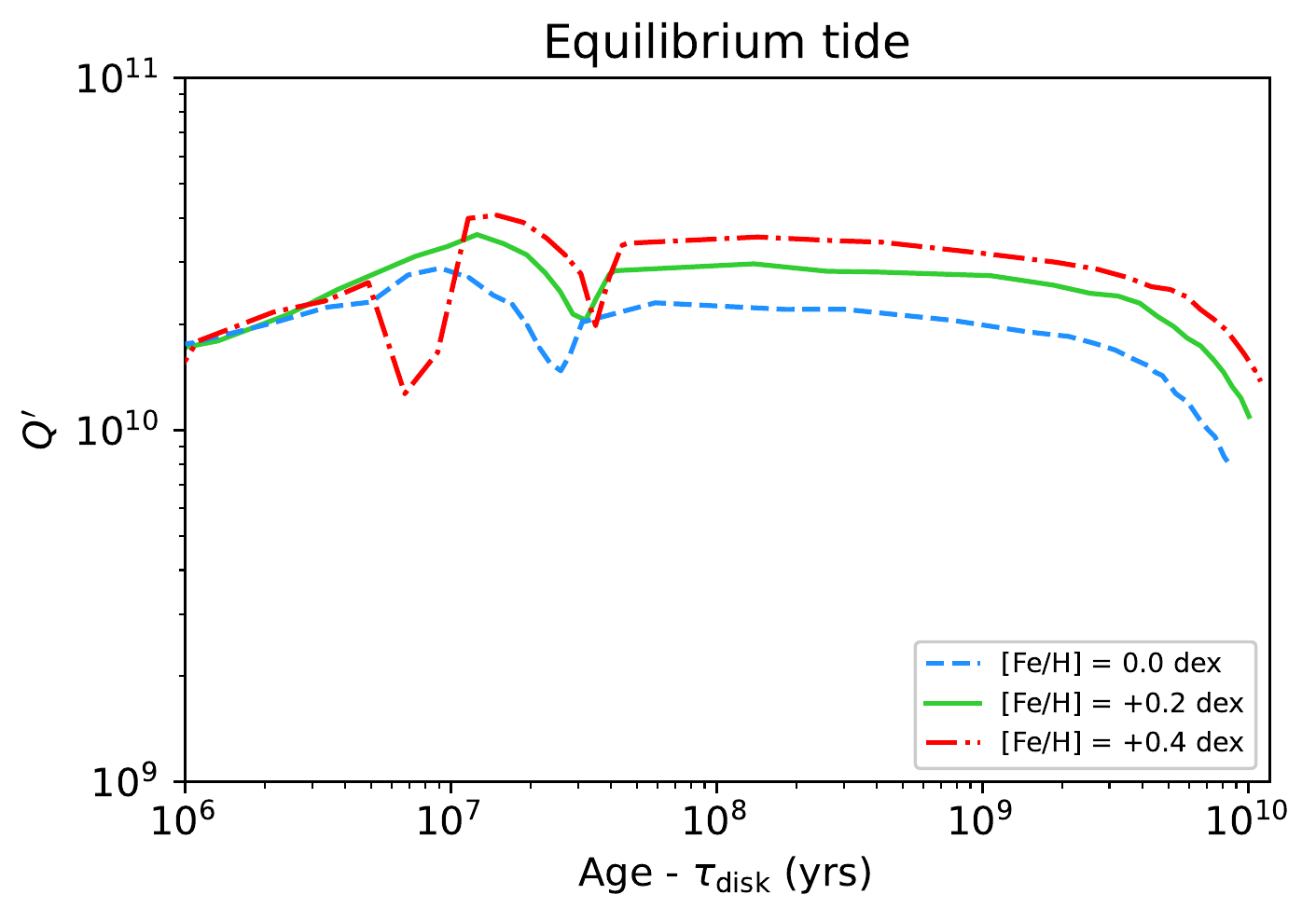}
	\includegraphics[width=\columnwidth]{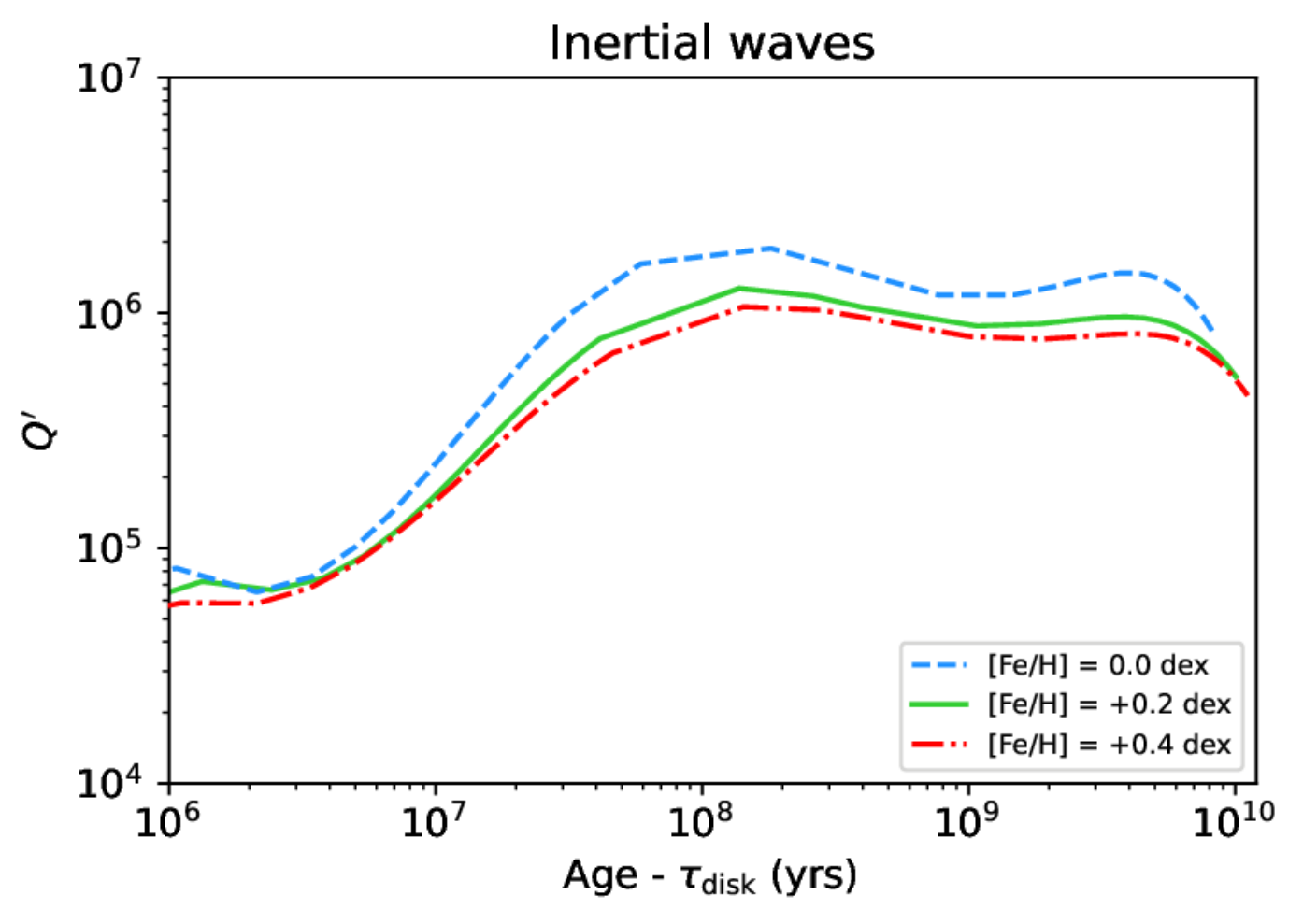}
	\includegraphics[width=\columnwidth]{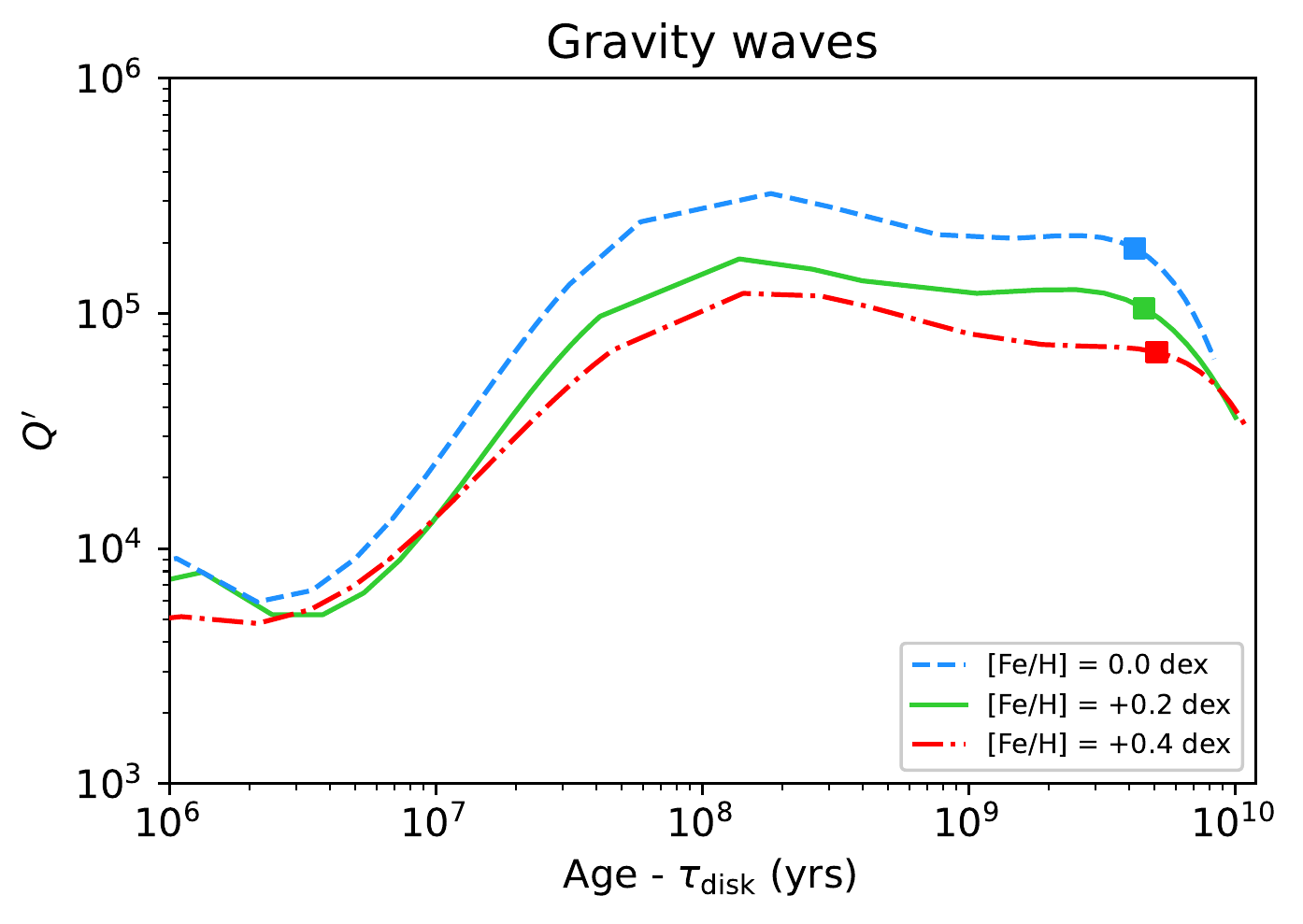}
    \caption{Tidal quality factor as a function of time for solar-mass models of three metallicities. $\tau_\mathrm{disc}$ = 4.3 Myr. We have assumed the orbital period  $P_\mathrm{orb}$ = 1 day and the rotation period $P_\mathrm{rot}$ = 3.5 days (both quantities are held constant). From top to bottom: tidal quality factors for equilibrium tide, inertial waves, and gravity waves. Blue, green, and red lines correspond to [Fe/H] = 0.0, +0.2, +0.4 dex, respectively. Squares correspond to the initiation of gravity wave dissipation for a planet with $M_\mathrm{pl}$ = 5 $M_\mathrm{J}$}
    \label{fig10_F}
\end{figure}

In order to explain in detail the reasons for this variation, we plotted the tidal quality factors for solar-mass models corresponding to three different metallicities in Fig.~\ref{fig10_F}. First of all, it is worth mentioning that, for a given age, the tidal dissipation rates lie within an order of magnitude for all three models spanning the metallicity range of typical hot Jupiter hosts. The dissipation of equilibrium tide is more effective in stars with low abundance, which is why some of the most massive hot Jupiters are able to merge with the solar-metallicity star before ZAMS. The dynamical tides are generally stronger in metal-rich stars. However, the difference in tidal quality factor for inertial waves is negligible during the phase of early migration. Consequently, metallicity does not seem to impact star-planet systems formed by a rapid rotator significantly. The enhancement of gravity wave dissipation, in turn, has a stronger effect on the infall diagrams computed for the median and slow rotators. The influence of metallicity is manifested through the quantity $\mathcal{G}$, expressed by Eq.(\ref{eq:tide_gw2}). This quantity characterizes the buoyancy frequency profile at the interface between the radiative and convective regions where gravity waves are launched. The conditions at the interface depend on the parameter $f_\mathrm{ov}$ ($f_\mathrm{ov}$ determines the efficiency of overshoot mixing, see \cite{Choi}) which is linked to metallicity. According to the adopted prescriptions from \cite{Gossage}, $f_\mathrm{ov}$ is set to 0.016, 0.011, and 0.010 in the models with [Fe/H] = 0.0, +0.2, and +0.4 dex, respectively.

Probably, the biggest contribution arises from the fact that the MS duration is longer for metal-rich stars, which is reflected in a planet having more time to merge under the dissipation of gravity waves. In a star-planet system composed of a solar-mass star and hot Jupiter with $M_\mathrm{pl}$ = 5 $M_\mathrm{J}$, the associated phase of intensive planetary migration proceeds 4.2, 5.5, and 6.0 Gyr in a model with [Fe/H] = 0.0, +0.2, and +0.4 dex, respectively. The increase in time results in the extension of the infall region observable in the bottom panel of Fig.~\ref{fig9_F}.

\section{Modeling hot Jupiter population} \label{sec:populaton}
In order to derive statistics of planet-star mergers, one needs to set the model of the initial hot Jupiter population. Here we describe the initial distributions of stellar and planetary properties we use to obtain total coalescence rates in the Galaxy. We consider the single value of stellar metallicity [Fe/H] = +0.2 dex in accordance with the mean metallicity of the hot Jupiter hosts (\citealt{Petigura}). 
\subsection{Stellar initial mass function, initial spin distribution} \label{subsec:imf}
The distribution of stellar mass is based on stellar initial mass function (IMF) from \cite{Kroupa}:
\begin{equation}
    P(M < M_{*} < M + \d M) \propto \begin{cases}
   M^{-1.3}, & M \leq 0.5 M_{\odot}\\
    M^{-2.3}, & M > 0.5 M_{\odot}
 \end{cases}.
    \label{eq:st_mass}
\end{equation}
We select stars lying in the mass range of 0.6 to 1.3 ${M_{\odot}}$ to create a synthetic population of hot Jupiter hosts.

\begin{figure}
	\includegraphics[width=\columnwidth]{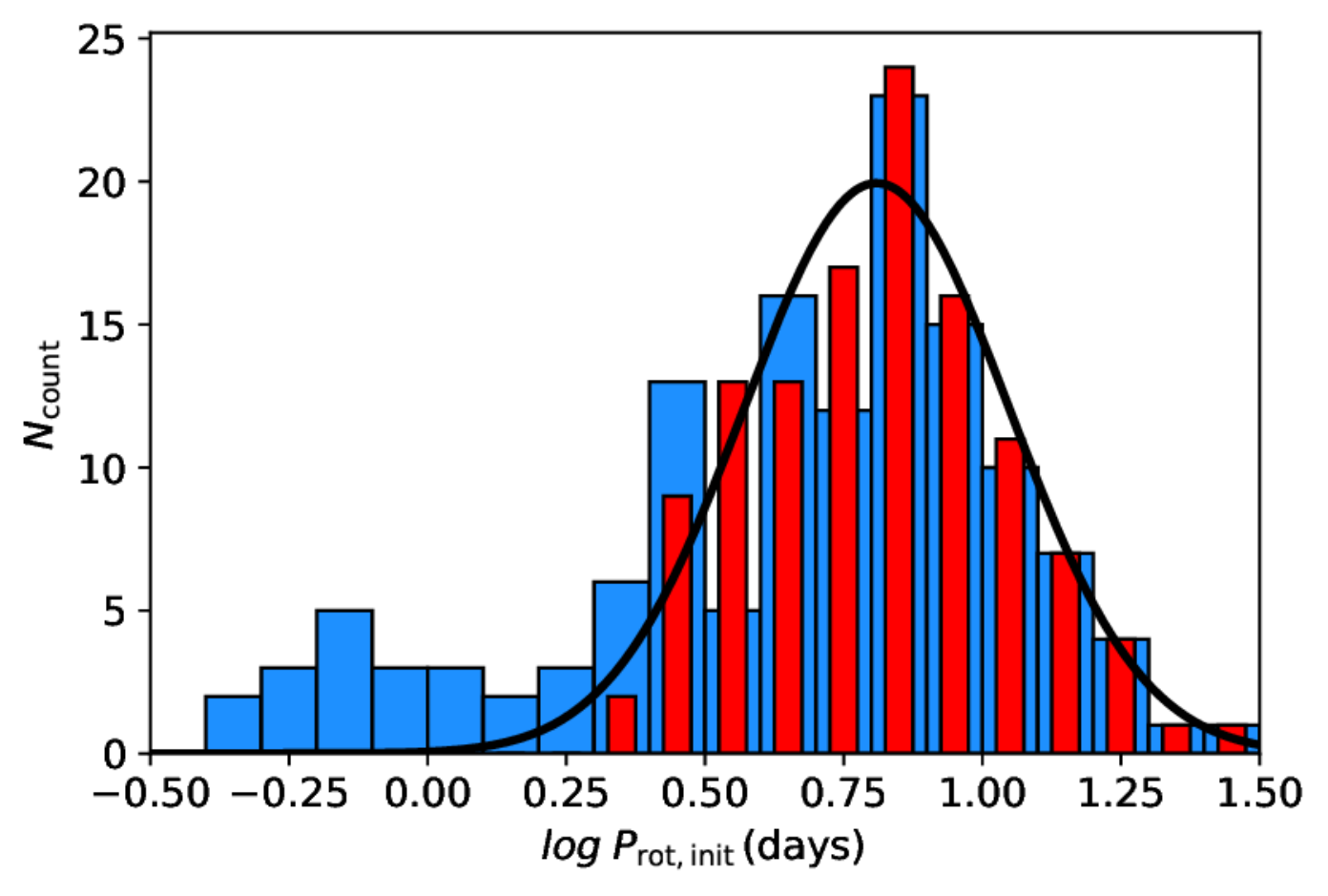}
    \caption{Distribution of the initial rotation period based on the NGC 2362 sample. Blue histogram: raw data from~\protect\cite{Irwin}. Red histogram: distribution obtained after calculating the initial period of stars that passed disc dissipation lifetime. Black solid line corresponds to the fit.}
    \label{fig11_F}
\end{figure}

To parametrize the initial spin distribution, we address the NGC 2362 sample by \cite{Irwin} providing the rotation period measurements. Applying eq. (\ref{eq:disc}), we found that out of 134 objects within the mass range specified above, 64 have disc dissipation timescale below 5 Myr, the cluster's age according to \cite{Moitinho}. The initial rotation period of these objects is recalculated using a set of our MESA models of solar metallicity stars. We exclude the extremely rapid rotators with the initial rotation period lower than two days since our grid does not allow definite estimates of their initial spin by interpolating between the grid $\log \; P_\mathrm{rot, init}$ values. The remaining 118 stars are selected to obtain the initial spin distribution plotted in red in Fig.~\ref{fig11_F}. The raw data from \cite{Irwin} is shown in blue. As expected, recalculation of the initial period leads to a slight offset toward higher periods. The presented distribution is fitted by a Gaussian function:

\begin{equation}
  p(\log \; P_\mathrm{rot,init})    \propto \exp\left(-\frac{(\log \;P_\mathrm{rot,init} - \zeta_*)^2}{2\sigma_*^2}\right),
    \label{eq:pdistribution}
\end{equation}
with $\zeta_* = 0.81$, $\sigma_* = 0.24$. Again, when simulating the initial rotation period, we exclude stars with the fastest rotation rates ($P_\mathrm{rot,init} < 2$ days). As demonstrated in subsection~\ref{subsec:rotation}, these objects do not make a significant contribution to the total number of mergers. 

At the same time, we consider stars with extremely low rotation rates ($P_\mathrm{rot,init} > 12$ days). Such hosts represent the case of virtually zero rotation described by our models with $P_\mathrm{rot,init} = 12$  days. This approximation is justified since low stellar spin practically does not affect the orbital evolution of hot Jupiters. Indeed,  the migration driven by the inertial wave dissipation is negligibly small in the planetary systems formed by a slow rotator. In contrast, the dissipation of equilibrium tide and gravity waves depends on the tidal forcing frequency determined by the orbital period rather than the rotation period, when $P_\mathrm{orb} \ll P_\mathrm{rot}$.

\subsection{Star formation history and spatial distribution of star-planet systems} \label{subsec:sfh}

The star formation rate (SFR) history of the inner disc is taken from \cite{Haywood}:
\begin{equation}
     SFR(\tau) = \begin{cases}
   3\;M_{\odot}/ {\rm yr}, & \tau \leq 7 \times 10^9\;{\rm yr;}\\
   0\;M_{\odot}/{\rm yr}, & 7 \times 10^9\;{\rm yr} < \tau \leq 9.5 \times 10^9\;{\rm yr};\\
    10\;M_{\odot}/{\rm yr}, & 9.5 \times 10^9\;{\rm yr} < \tau \leq 12.5 \times 10^9\;{\rm yr},
 \end{cases},
    \label{eq:sfr}
\end{equation}
where $\tau$ is the lookback Galactic time ($\tau = 0$ corresponds to the present epoch).

Eq. (\ref{eq:sfr}) indicates the presence of two distinct epochs of star formation activity related to the Galactic thin and thick disc, separated by the quenching phase. In the present study, we consider the thin disc population only, as the planet-metallicity correlation for Jovian-mass planets ($\Gamma_\mathrm{z} \propto 10^{3.4 \;\mathrm{ [Fe/H]}}$ taken from \cite{Petigura}) applied for the typical thick disc metallicity ([Fe/H] = -0.7 dex, see \cite{Gilmore}) reduces hot Jupiter occurrence rate within the Galactic thick disc by a factor of 200. For the same reason, taking into account a significant offset between the inner and outer thin disc abundances revealed in \cite{Snaith}, we decided to neglect hot Jupiter population of the outer thin disc. Thereby we adopt only the first line of eq. (\ref{eq:sfr}) to define the total number of simulated stars and their ages. 

Simulated star-planet systems are distributed inside the inner ($R < 10$ kpc) thin disc according to the exponential law derived by \cite{Juric}:
\begin{equation}
p(R,Z) \propto \exp \left(-\frac{R}{L} - \frac{Z}{H} \right)
    \label{eq:loc}
\end{equation}
with $H = 300$ pc, $L = 2600$ pc. For every system, we derive the distance adopting $R_{\odot} = 8 $ kpc, $H_{\odot} = 25$  pc. To calculate the apparent magnitude of bright transients and stars in decaying systems we assume the extinction of 1.8 mag per kpc (\citealt{Whittet}). 

\subsection{Initial distribution of hot Jupiter mass and orbital period} 
\label{subsec:planet}

\begin{figure}
	\includegraphics[width=\columnwidth]{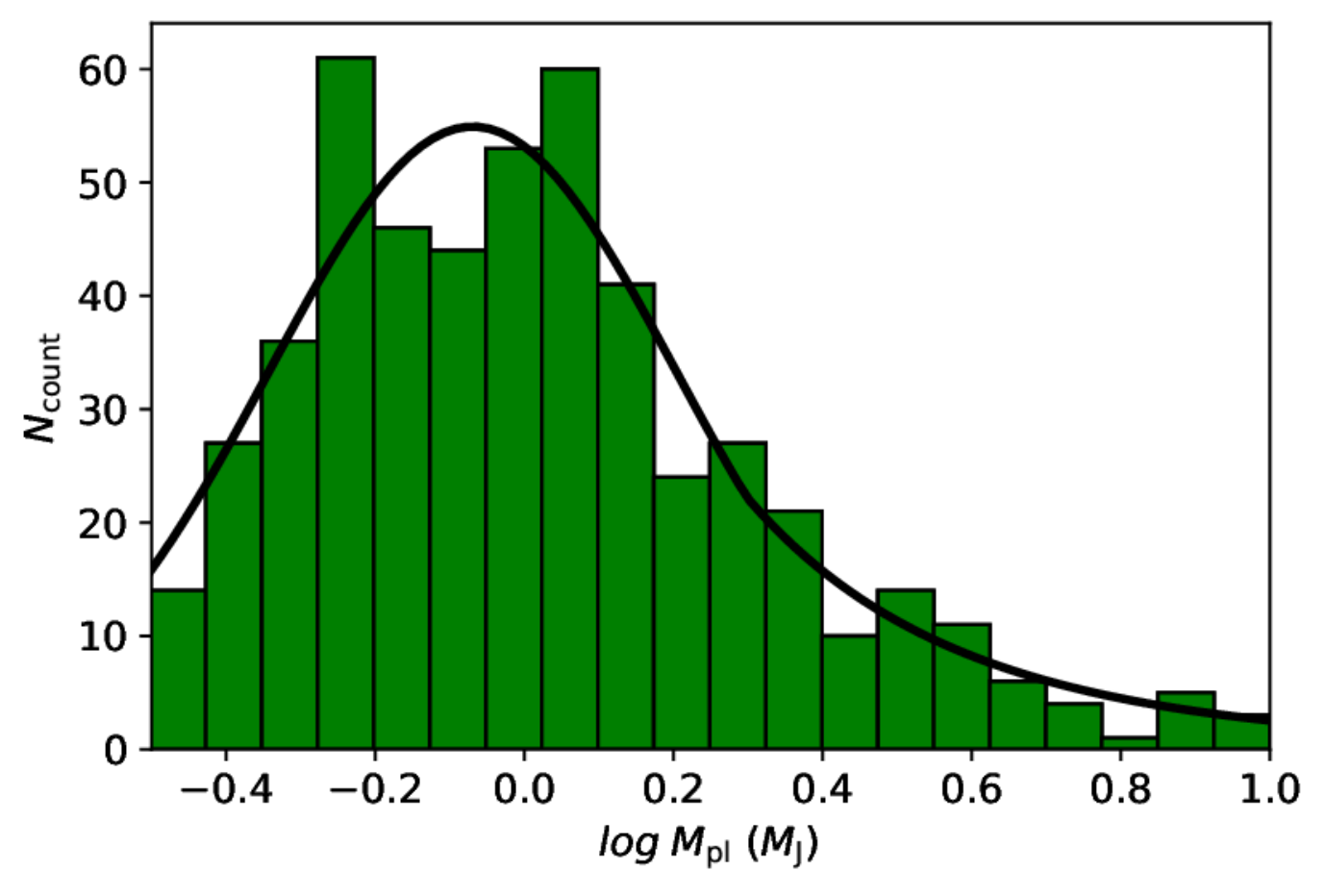}
    \caption{Distribution of hot Jupiter mass. Green histogram represents the observed planetary sample from \url{https://exoplanetarchive.ipac.caltech.edu/} Black solid line corresponds to the fit.}
    \label{fig12_F}
\end{figure}

The distribution of planetary mass, shown in Fig.~\ref{fig12_F}, is based on the hot Jupiter sample taken from the NASA Exoplanet Archive imposing the same restrictions as in subsection~\ref{subsec:density}, except we do not require the availability of planetary radii measurements. The established range of hot Jupiter mass extends from $0.3 \; M_\mathrm{J}$ to $10 \; M_\mathrm{J}$. We refer to \cite{Schlaufman}, who found evidence supporting the idea that more massive objects are formed via gravitational instability rather than core accretion, which underlines their belonging to the population of brown dwarfs. To keep the fraction of hot Jupiters with $M_\mathrm{pl} > 3\;M_\mathrm{J}$ close to the observed value, we approximate the distribution with the following piecewise continuous function:
\begin{equation}
   p\left(\log \frac{M_\mathrm{pl}}{M_\mathrm{J}}\right) \propto \begin{cases}
   \exp\left(-\frac{(\log \frac{M_\mathrm{pl}}{M_\mathrm{J}} - \zeta_\mathrm{pl})^2}{2\sigma_\mathrm{pl}^2}\right), & \log\; \frac{M_\mathrm{pl}}{M_\mathrm{J}} \leq 0.3\\
    \left(\log \frac{M_\mathrm{pl}}{M_\mathrm{J}}\right)^{\beta}, & \log \frac{M_\mathrm{pl}}{M_\mathrm{J}} > 0.3
 \end{cases},
    \label{eq:pl_mass}
\end{equation}
where $\zeta_\mathrm{pl} = -0.068$, $\sigma_\mathrm{pl} = 0.28$ , and $\beta = -1.5$.

We considered two different distribution of the initial orbital period. The first one (hereafter D1) is from \cite{Petigura}:
\begin{equation}
p(\log \;P_\mathrm{orb, init}) \propto P_\mathrm{orb, init}^{\alpha},
    \label{eq:orb1}
\end{equation}
with $\alpha = 0.9$. The above relation is sharper than the power law from \cite{Cumming} (with $\alpha = 0.26$). We find it more reliable since its calibration involved more diverse planetary sample. We assume that the distribution of Jovian-mass planets is spaced in orbital period within the interval [1 day, 10 days].

Another distribution (hereafter D2) of the initial hot Jupiter position is uniform with respect to the logarithm of the initial orbital period. Such distribution is commonly used by various authors. For instance,  it was applied by \cite{Collier} to recover the sharp upper-left boundary in the mass--separation diagram within the framework of a model with a constant tidal quality factor. The second distribution is meant to provide an upper bound for our statistical estimates. 

The fraction of FGK stars hosting a hot Jupiter is fixed at a 1\% level. We neglect the variation of hot Jupiter occurrence rate with stellar mass. In future studies, we will take into account the chemical evolution in the Galaxy and the associated variation of the hot Jupiter occurrence rate with the lookback Galactic time. In total, we simulate 52 million star-planet systems each of which is represented by a set of parameters, namely stellar mass, initial period of stellar rotation, planetary mass, initial orbital period, age, and Galactocentric cylindrical coordinates. The outcome of the orbital evolution is obtained applying the precomputed grid of models. In particular, we determine the closest grid values of stellar and planetary mass, the logarithm of the initial period of stellar rotation, and the initial orbital period with respect to the parameters of the simulated star-planet system. The fate of the corresponding grid model is subsequently defined as the outcome of the performed simulation. 

\section{Results} \label{sec:results}
\begin{figure}
	\includegraphics[width=\columnwidth]{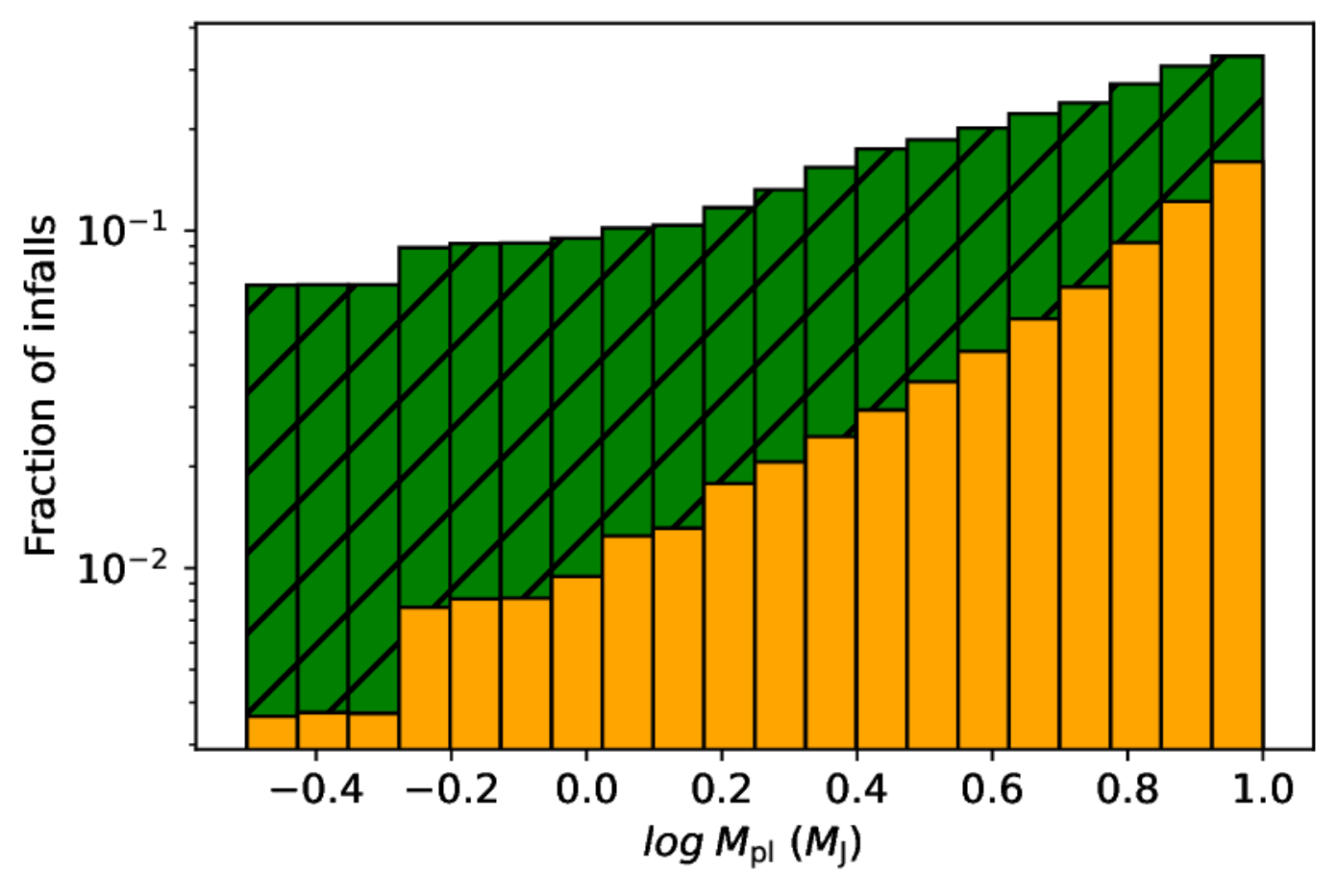}
    \caption{Engulfment ratio as a function of planetary mass. Green shaded bars represent the total number of infalls, the orange bars represent the number of infalls to date. Figure corresponds to the D1 distribution.}
    \label{fig13_F}
\end{figure}

\begin{figure*}
\begin{multicols}{2}
	\includegraphics[width=\linewidth]{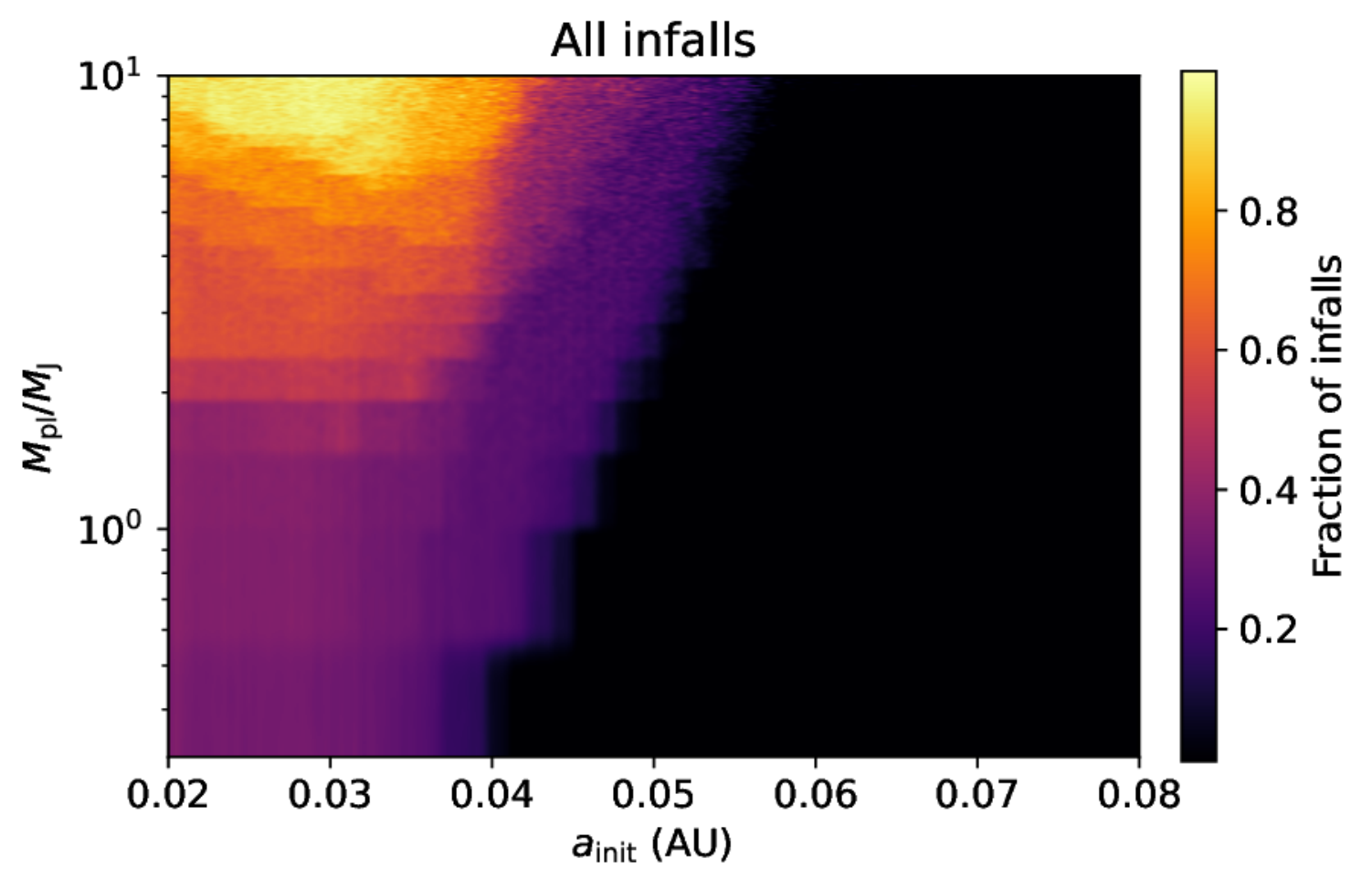}\par 
	\includegraphics[width=\linewidth]{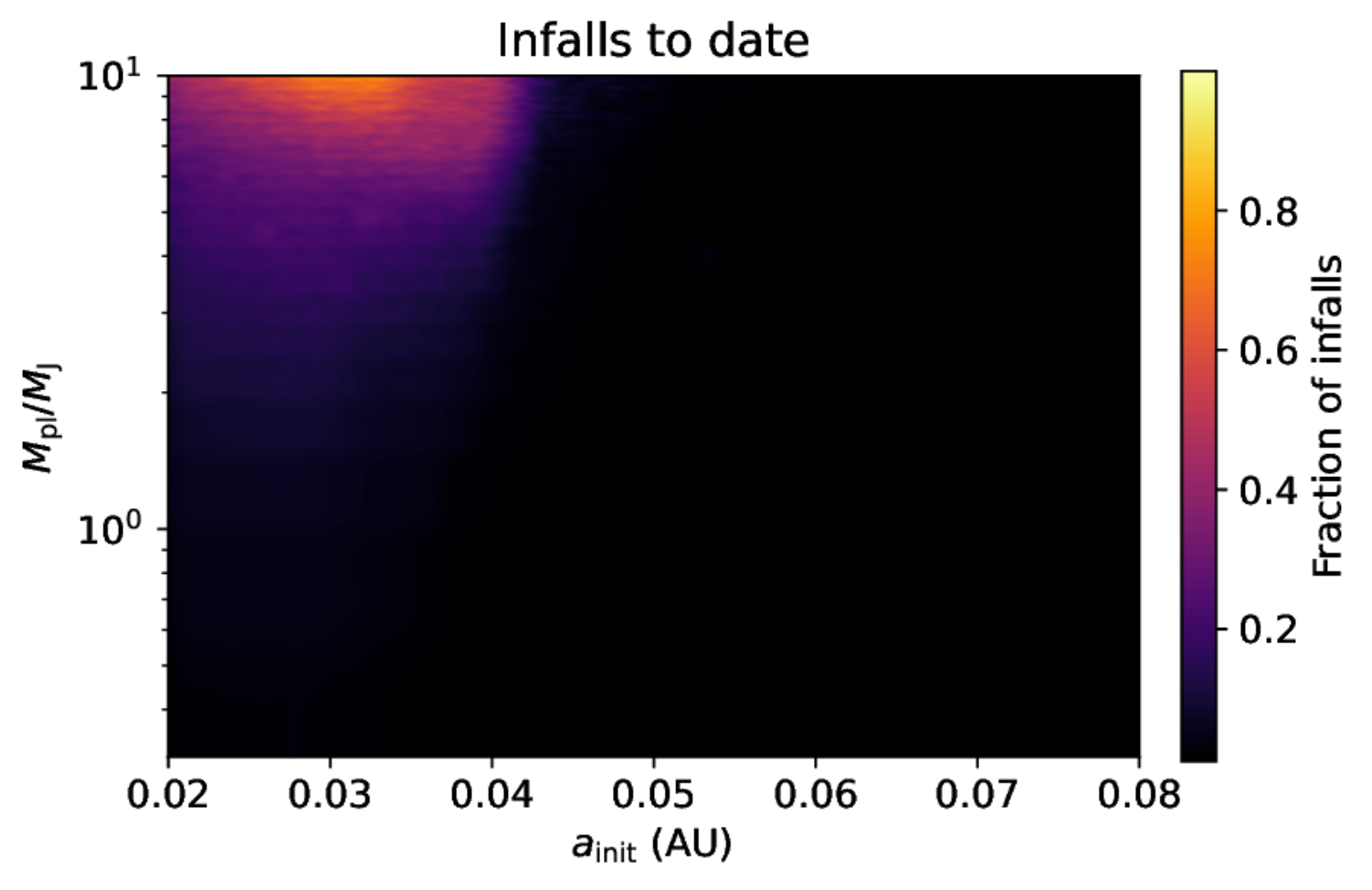}\par 
	    \end{multicols}
    \caption{Engulfment ratio distribution in the mass--separation diagram. Left panel corresponds to the total number of infalls (including mergers that have not happened yet). Right panel corresponds to the number of infalls that took place before the present epoch (based on the current system age).}
    \label{fig14_F}
\end{figure*}

The final results of the hot Jupiter population modeling are reported in Table~\ref{tab2}. One can see that switching from sharp (D1) to uniform (D2) log period distribution increases the total amount of infalls by the factor of two. Accordingly, the overall engulfment ratio ranges between 11 and 21\%. The green dashed histogram in Fig.~\ref{fig13_F} shows the engulfment ratio for every planetary mass bin in the case of the D1 distribution. As expected, more massive planets have a higher infall probability reaching 30\% for hot Jupiters with $M_\mathrm{pl} \simeq 10\;M_\mathrm{J}$. The shift toward massive planets is even more pronounced when we consider the completed mergers, depicted by the orange histogram. Only a negligible fraction of hot Jupiters, 1.5--3.0\%, is already engulfed by their host stars, meaning that the initial hot Jupiter occurrence rate must be close to the observed one. The same applies to the distribution of planetary mass when neglecting photoevaporation. Fig.~\ref{fig14_F} demonstrates the infall region in the mass--initial separation plane.  On the left panel, corresponding to the total number of events, all the unstable systems populate the distinct area restricted by 0.04 AU for low-mass hot Jupiters and 0.06 AU for massive hot Jupiters. We note that the colored map is not tied to any particular initial distribution of planetary parameters. Another interesting detail is the presence of a domain occupied by massive planets where the engulfment ratio is close to unity. Thereby, almost every hot Jupiter belonging to that area in the parameter space merges with the host FGK star, no matter what its mass or rotation rate is.

\begin{figure}
	\includegraphics[width=\columnwidth]{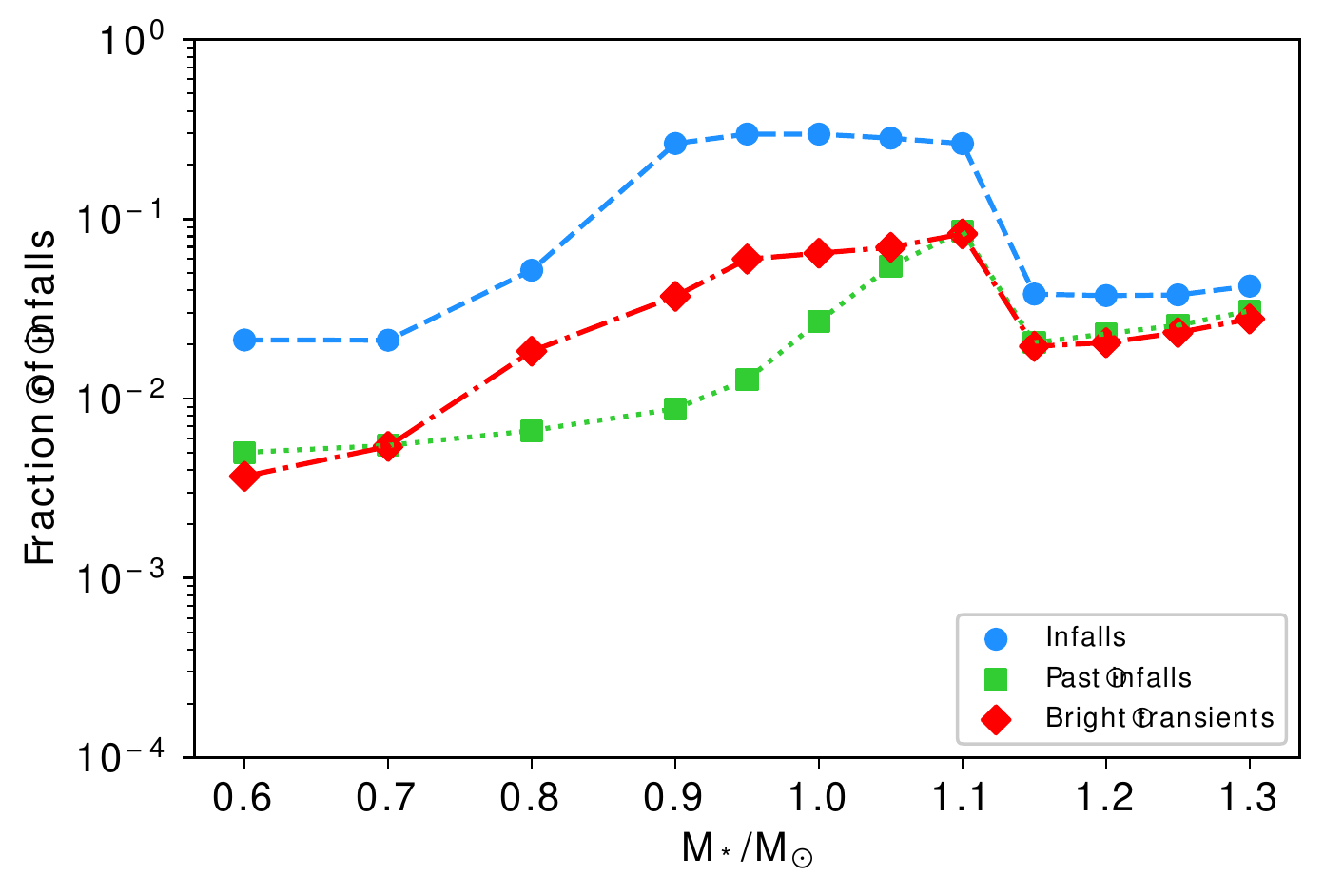}
    \caption{Engulfment ratio as a function of stellar mass. Blue circles denote the overall infalls, green squares denote the infalls occurred before the present epoch, red diamonds denote the infalls accompanied by transients. Figure corresponds to the D1 distribution.}
    \label{fig15_F}
\end{figure}

To clarify the degree of influence of stellar parameters on the outcome of our simulations, in Fig.~\ref{fig15_F}, we plot the infall probability for every stellar mass considered in the present work. The most prominent feature is the sharp decrease in the engulfment ratio corresponding to the onset of the convective core at 1.15 $M_{\odot}$. At the same time, the stars with mass above  1.15 $M_{\odot}$ have short TAMS ages, which is reflected in a relatively high fraction of the completed infalls demonstrated in green. Another noticeable drop of mergers characterizes the stars with mass below 0.9 $M_{\odot}$. This drop is related to the reduced contribution of gravity waves resulting from the MS lifetimes exceeding 14 Gyr. In addition, the coalescences with the low-mass stars very rarely produce bright transients, represented by red diamonds. Since the mean stellar density decreases during the MS stage, the condition $\rho_\mathrm{pl}/\rho_{*}>1$ required for tidal disruption or direct impact is more typical for the late ages of stellar evolution. However, with a decrease in the contribution of gravity waves, characteristic of stars with $M_* \leq 0.8 M_{\odot}$ (see subsection~\ref{subsec:stmass}), the average age of mergers is shifted toward lower values reducing the probability of a significant luminosity
enhancement.

After averaging the number of mergers over the Galactic lookback time from -50 Myr to 50 Myr, we obtain the infall rates varying between 340 and 650 events per million years, depending on the initial distribution of planetary separation. The majority of these events follow the stable accretion scenario that does not result in a significant luminosity enhancement. According to our estimates, only 107 -- 194 coalescences per million years are accompanied by a transient, half of which are powerful enough to be observed with the Large Synoptic Survey Telescope (LSST) having the limiting magnitude in g-band equal to $+24.8^{\mathrm{m}}$ for point sources \footnote{(\url{https://smtn-002.lsst.io/})}. Another important goal was to calculate the number of systems undergoing orbital decay intense enough to be detected. Given that the currently achieved precision of the best transit-timing observations following a single observing season is on the order of a few seconds (\citealt{Collier}), we select the infalls driven by the gravity waves, for which the cumulative shift in transit times $T_\mathrm{shift}$ exceeds 5 seconds over a 10-year baseline $T_\mathrm{dur}$. To find out which systems meet this criterion, we derive the expression, combining eqs. (55), (57), and (59) from B20:
\begin{equation}
T_\mathrm{shift} \approx 225\mathrm{s} \left( \frac{1 \,\mathrm{Myr}}{\tau_\mathrm{\alpha}}\right)\left( \frac{T_\mathrm{dur}}{10\,\mathrm{yr}}\right)^2,
    \label{eq:decay}
\end{equation}
with $\tau_\mathrm{\alpha}$ the orbital decay timescale. Substituting $T_\mathrm{dur} = 10 \, \rm yrs$ and $T_\mathrm{shift} = 5\, \rm s$ in eq. (\ref{eq:decay}) results in $\tau_\mathrm{\alpha} = 45 \,\rm Myr$. We purposely filter out the mergers induced by equilibrium tide dissipation as they are expected to proceed on much lower timescales. In addition, we choose only the decaying systems with stars brighter than $+12^{\mathrm{m}}$, $+13^{\mathrm{m}}$, and $+16^{\mathrm{m}}$, corresponding to the limiting magnitudes of TESS (\citealt{TESS}), PLATO (\citealt{PLATO}), and Kepler (\citealt{Kepler}), respectively. Our estimates are given in Table~\ref{tab2}. The number of decaying systems is computed by averaging over the Galactic lookback time between -200 Myr and 200 Myr.

\begin{table}
	\centering
	\caption{Statistics of planetary mergers \label{tab2}}

	\begin{tabular}{|c||c|c|} 
	    \hline
       Quantity 	 & \multicolumn{2}{c}{Value} \\
		\hline
		 Distribution & D1 &  D2 \\
		 \hline\hline
		Total number of hot Jupiters & $5.2 \times 10^7$ & $5.2 \times 10^7$ \\
		Total number of infalls & $5.7 \times 10^6$ (11\%) & $1.06 \times 10^7$ (21\%) \\
		Infalls to date & $8.0 \times 10^5$ (1.5\%) &  $1.6 \times 10^6$ (3.0\%)\\
        Stable accretion& $4.4 \times 10^6$ & $8.3 \times 10^6$\\
		Tidal disruption& $1.3 \times 10^6$ & $2.2 \times 10^6$  \\
		Direct impact& $6.7 \times 10^4$ & $1.0 \times 10^5$  \\
		Present-day infall rate & 340 $\pm$ 20 $\mathrm{Myr^{-1}}$ & 650 $\pm$ 30 $\mathrm{Myr^{-1}}$\\
		Rate of transients  & 107 $\pm$ 11 $\mathrm{Myr^{-1}}$ &  194 $\pm$ 15 $\mathrm{Myr^{-1}}$ \\
		Rate of transients with $m_\mathrm{app} < 24.8$   &  49 $\pm$ 8 $\mathrm{Myr^{-1}}$ & 89 $\pm$ 9 $\mathrm{Myr^{-1}}$  \\
		Decays\tablefootnote{Details are given in Section~\ref{sec:results}}  with $m_\mathrm{*, app} < 12$  & 0.8 $\pm$ 0.6 & 1.2 $\pm$ 0.8 \\
		Decays with $m_\mathrm{*, app} < 13$  & 2.4 $\pm$ 1.8 & 3.7 $\pm$ 1.2  \\
		Decays with $m_\mathrm{*, app} < 16$ & 22 $\pm$ 5 & 41 $\pm$ 7 \\
		\hline
	\end{tabular}
\end{table}


\section{Discussion} \label{sec:discussion}
As demonstrated in our work, gravity wave damping is the efficient mechanism driving the migration of close-in planets around the MS stars possessing radiative core. Its implementation results in a transformation of the present-day distribution of planets concerning the initial one. Thus, the conclusion from \cite{Heller} that the tidal migration is negligible compared with the migration inside the disc may only be valid for stars with the convective core.

Regarding the orbital evolution around fast rotators, our study is consistent with the findings of \cite{Ahuir} concerning the population of 'young migrators.' Specifically, we verified that the fate of massive planets around rapidly rotating stars is determined by the initial separation with respect to the corotation radius, as the planets located outside undergo intense outward migration, making them stable even after the initiation of gravity wave breaking. As predicted by the corresponding authors, the inclusion of gravity waves modifies the dynamics of 'old migrators,' providing the decrease in the occurrence of hot Jupiters with the decreasing orbital period in agreement with the observational statistics.

We showed in subsection~\ref{subsec:metallicity} that the effectiveness of the dynamical tide dissipation is enhanced in metal-rich stars, which is not consistent with \cite{BolmontGallet}. This discrepancy can be explained by the different tidal quality factor calculation techniques. We recall that \cite{BolmontGallet} simulated the secular evolution of a star-planet system relying on the homogeneous two-layer stellar model, according to which the structural part of the inertial wave dissipation depends on the stellar radius and mass aspect ratios, corresponding to the interface between the radiation and convection zones (\citealt{Mathis}). At the same time, our tidal quality factor estimates take into account both the relative sizes of layers and the physical conditions inside them.  Regarding the impact of metallicity, two main effects were found. Firstly, the infall probability correlates with metal abundance. Secondly, on average, metal-rich stars undergo coalescence with hot Jupiters closer to the end of the MS lifetime. These two features are caused by the increased contribution of gravity waves and the reduced dissipation of equilibrium tide, demonstrated in Fig.~\ref{fig10_F}. Combining the above two effects, we cannot support the idea, suggested by \cite{BolmontGallet}, that the observed trends in hot Jupiter occurrence with metallicity result from metal-poor stars engulfing planets more frequently than metal-rich stars. We are convinced that the predominance of metal-rich systems with hot Jupiter is entirely due to the effects associated either with the planetary formation or the migration inside the protoplanetary disc.

Based on our results, we cannot be confident about the role of tidal migration in the origin of the upper inner edge in the mass--separation diagram declared by \cite{Bailey,Collier}. The right panel of Fig.~\ref{fig14_F} illustrates the engulfment ratio related to the mergers completed before the present day. Only the population of hot Jupiters with $M_\mathrm{pl} \geq 3\;M_\mathrm{J}$ and $a_\mathrm{init} \leq 0.04$ AU is significantly affected by the infalls. However, it is not enough to make the boundary on the mass--separation plane as sharp and well-defined as observed today. One possible way to achieve this within our simulations is to take into account the correlation of the hot Jupiter occurrence rate with stellar mass. In future, we will address this problem in detail by tracing the evolution of the hot Jupiter population.

Our estimates of the present-day infall rate are several orders of magnitude lower than in \cite{Metzger}. One order of magnitude is supposed to be attributed to the impracticability of a steady-state rate of planetary mergers used in the above work. The remaining discrepancy is a product of the extrapolation approach applied by \cite{Metzger} in comparison with the present research.  We recall that, in this study, we do not treat the Galaxy as a whole. Instead, we focus on the inner thin disc where the mean stellar metallicity is higher, and so is the frequency of the hot Jupiter formation. In addition, we limit ourselves to studying stars in the mass range between 0.6 and 1.3 $M_{\odot}$ as the main contributors to the total number of coalescences due to gravity waves. The difference in the considered range of stellar and planetary parameters is also the main reason for the inconsistency between our findings and the values from \cite{Popkov}. It turned out that, within the same planetary and stellar mass intervals, our engulfment ratio is formally equal to their results obtained with $Q_* \sim 10^6$ (Popkov, A.V., private communication). The rate of transients, given in Table~\ref{tab2}, is two orders of magnitude lower than in \cite{Popkov}. We believe that this is, again, is associated with the choice of the explored range in the parameter space. 

Comparing the Galactic velocity dispersions of the hot Jupiter host and field star samples from Gaia Data Release 2, \cite{Hamer} inferred that the stars hosting hot Jupiters are systematically younger. Their evaluated constraint on the tidal quality factor ($Q_* < 10^7$) is in good agreement with the value mentioned in the previous paragraph, although the main conclusion about the destruction of the majority of the initial hot Jupiter population during the MS lifetime contradicts the findings of the present paper.  The origin of this discrepancy may be in  uncertainties in the initial occurrence rate and distribution of hot Jupiters in the mass--separation diagram. If we assume a trend in hot Jupiter occurrence rate with stellar mass in favor of more massive hosts within each of F, G, and K spectral types (discussed by \cite{Johnson}), the average MS lifetime of the hot Jupiter host sample will be significantly shorter compared to the field star sample. This potentially may lead to the same bias as observed by \cite{Hamer}. Another explanation may reside in the impact of the initial stellar rotation on the planetary formation. One of the manifestations of this effect may be the lack of planets around rapidly rotating stars, discovered by \cite{McQuillan}. If we increase the initial number of hot Jupiters orbiting slow and median rotators by the cost of (partially) ignoring rapid rotators, the fraction of infalls will increase. It will increase even further if we reduce the number of hot Jupiters with $P_\mathrm{orb, init} > 5$ in accordance with the distribution of planets detected by the Kepler telescope (\citealt{Santerne}).

The derived rate of transients makes a discovery of a bright event highly unlikely. To prove this, we calculated the size of the region within which the overall rate of transients is equal to one event per year, assuming the observability of every transient from any distance. After scaling the present-day SFR in the Galaxy with SFR density of 0.01915 $M_{\odot} \rm / yr / Mpc^3$ (\citealt{Brinchmann}), multiplied by the rate of the Galactic transients, we obtained the radius of 57 Mpc, exceeding the distance to the Virgo Cluster. Thus, the detection of a transient produced by hot Jupiter is beyond the capabilities of LSST. However, this does not negate the feasibility of a planetary infall detection. As discussed by \cite{Popkov}, the majority of transients correspond to the ingestions of close-in planets by the expanding massive MS stars, having non-tidal nature. Thus, if direct impact or tidal disruption is ever to be observed, most likely, it would be related to a merger of Earth or Neptune mass planet.

Nonetheless, the theory of tidal interaction in a star-planet system can be constrained via the  measurement of the orbital decay rate through precise transit timing. Our results indicate that the detection of the orbital decay with TESS is sufficiently challenging. At least a few decaying systems can be discovered within the first decade of the observation at PLATO. The second decade will increase this number by about four times. Finally, if the large-scale survey with the facility having the same capacities as Kepler is ever to be launched, the sample of the identified decaying systems will grow substantially.  We note that the estimates of the number of decays might be somewhat underestimated. Firstly, we do not take into account the infalls driven by equilibrium tide. This is largely justified because, as noted in Sec.~\ref{sec:results}, the observation of such decays is significantly more challenging, given the low dissipation rates of the non-wavelike tide (demonstrated in Fig.~\ref{fig2_F}). Secondly, the mergers with the evolved stars can potentially make a significant contribution. Besides, including the stars with mass outside the analyzed range will undoubtedly provide additional decays. We remind the reader that the only confirmed decaying planetary system to date, WASP-12, is composed of a star with mass lying above the adopted interval from 0.6 to 1.3 $M_{\odot}$. At last, applying a more realistic interstellar extinction will expand the observed stars' sample by those located above the Galactic plane.

\section{Summary and conclusions} \label{sec:conclusion}
In this paper, we study the orbital evolution of hot Jupiters around solar-mass stars following the tidal prescriptions from \cite{Barker}, describing the dissipation of equilibrium tide, inertial waves,
and gravity waves. These prescriptions are applied to the models of the rotating FGK stars computed with the MESA code to obtain the tidal dissipation rates, which are subsequently used to simulate the dynamics of a star-planet system until the host's main sequence termination. We vary stellar and planetary parameters to understand the principles underlying the evolution of the hot Jupiter population after the dissipation of the protoplanetary disc. Special attention is paid to the investigation of the possibility of a star-planet coalescence since the infall factor is the one that shapes the orbital architecture of Jovian planets in the most straightforward and observable way.  In particular, we find that:
\begin{itemize}
\item gravity waves substantially expand the infall region up to 0.06 AU by incorporating those planets that do not get captured on the $n = 2 \Omega_{*}$ limit;
\item the initiation of gravity wave breaking induced by low-mass planets is shifted toward late ages. As a result, the high fraction of systems composed of a hot Jupiter with $M_\mathrm{pl} \leq 3\;M_\mathrm{J}$  remains stable throughout the MS stage;
\item the contribution of gravity waves is determined by the host star's initial spin. For planets orbiting fast rotators, inertial wave dissipation is the key mechanism driving the migration of hot Jupiters. On the contrary, the coalescence with slow rotator occurs mainly due to the activity of gravity waves;

\item the transition between regimes of inertial wave domination and gravity wave domination depends on stellar metallicity. Metal-rich stars are favored in terms of tidal migration under the dissipation of gravity waves;
\item generally, fast rotators engulf their planets earlier than slow rotators;
\item the impact of gravity wave dissipation is especially noticeable against the background of stars possessing a convective core. In the present study, we assume that such stars do not undergo gravity wave dissipation before the TAMS. The latter reduces the infall probability;  
\item for stars with radiative core, the impact of gravity waves sharply decreases with decreasing stellar mass.
\end{itemize}

We are interested in improving our model by implementing more mechanisms essential in the context of the evolution of a star-planet system. Firstly, photoevaporation needs to be taken into account. As shown by \cite{Rao1}, mass loss by close-in planets with $M_\mathrm{pl} \leq 2.5\;M_\mathrm{J}$ results in the obliteration at early ages. Thus, low-mass hot Jupiters captured on the $n = 2 \Omega_{*}$ limit may be destroyed before the tidal interaction leads to their engulfment.

Secondly, we would like to explore the effect of magnetic interaction on the dynamics of a star-planet system. Depending on stellar and planetary magnetic field strengths and star-planet separation, this type of interaction occurs in two different regimes, namely unipolar and dipolar. \cite{Strugarek4} demonstrated that, when enabled, unipolar regime forces the effective drag reducing the migration timescale by several orders of magnitude. The combination of tidal interaction and dipolar magnetic interaction has been extensively studied by \cite{Ahuir}. It was found that even though the tidal effect dominates in systems with a high planetary mass, one should consider both mechanisms to explain the observed properties of the planetary population.

Another important step would be to investigate the probability of gravity wave breaking to take place on the edge of radiation zone in stars with convective cores (Ivanov, P.B., private communication). This would require the development of another wave-damping criterion, one of the promising areas of our research.

Finally, we are aimed to expand our model by considering the orbital evolution around the evolved stars. The latter is expected to significantly increase the estimated number of decaying systems as the efficiency of both equilibrium (\citealt{Mustill}) and dynamical tide (\citealt{Barker}) dissipation is enhanced in giant stars.

The above steps would allow us to reproduce the key statistical patterns of the observed population of Jovian planets from the initial distribution in the mass--separation diagram. In the present study, we focus on estimating the rate of hot Jupiter infalls in the Milky Way Galaxy. The considered range of stellar mass extends from 0.6 to 1.3 $M_{\odot}$. Adopting the distributions of stellar and planetary properties, mentioned in Sec.~\ref{sec:populaton}, we obtain the following results: 
\begin{itemize}
\item 11 -- 21\% of the initial hot Jupiter population merge with the host star within the main sequence lifetime (or 14 Gyr);
\item 1.5 -- 3.0\% of the initial hot Jupiter population is already engulfed;
\item the present-day infall rate in the Galaxy is 340 -- 650 events per million years. The frequency of transients is 107 -- 194 events per million years, which makes it unlikely to observe transients involving hot Jupiter in the Virgo Cluster;
\item  the number of decaying systems composed of the MS star with $m_\mathrm{app}$ < 12, 13, 16 (resp. TESS, PLATO, and Kepler limiting magnitudes) and hot Jupiter for which the predicted cumulative shift in transit times is greater than 5 s after 10 years is 0.8 -- 1.2, 2.4 -- 3.7, and 22 -- 41, respectively.

\end{itemize}

The main uncertainty of our approach arises from the initial distribution of star-planet separation. In this study, we employed the power law from \cite{Petigura} based on the hot Jupiter observations and the commonly used uniform distribution in log period, although it is not necessarily the best option.  We note that the current population of Jovian planets is modified regarding the initial one due to star-planet interaction (including tides). However, we showed that the most significant migration falls on the phase of gravity wave dissipation close to the end of the MS, while the majority of the observed planetary systems currently do not exhibit gravity wave breaking (\citealt{Barker}). Anyway, the detailed study of the orbital evolution needs to be coupled with the finding of the planet population synthesis approach. We hope that our research draws special attention to the population synthesis of hot Jupiters, encouraging more efforts in this field. Collaboration between different studies will pave a path toward the comprehensive picture of exoplanetary systems.

\section*{Acknowledgements}
The work on the orbital evolution code is supported by the Theoretical Physics and Mathematics Advancement Foundation ``BASIS''. We acknowledge the support provided by the Ministry of Science and Higher Education of the Russian Federation grant 075-15-2020-780 (N13.1902.21.0039) in the implementation of the hot Jupiter population modeling.

Special thanks to Prof. Sergei Popov for coordinating the work and comments on the manuscript. I am grateful to Drs. Pavel Ivanov and Louis Amard for the fruitful discussion. I would also like to thank Prof. Bill Paxton and the MESA community for making this work possible. Finally, we acknowledge Dr. Adrian Barker for tidal interaction prescriptions, and Dr. Seth Gossage for MESA extension.  

\textit{Software}: \textbf{MESA} r11701 \citealt{MESA1,MESA2,MESA3,MESA4,MESA5}
\section*{Data Availability}

The data underlying this article will be shared on reasonable request to the corresponding author




\bibliographystyle{mnras}

\bibliography{example} 



\appendix

\section{Infall diagrams}

\begin{figure*}
\begin{multicols}{2}
    \includegraphics[width=\linewidth,height=5.5cm]{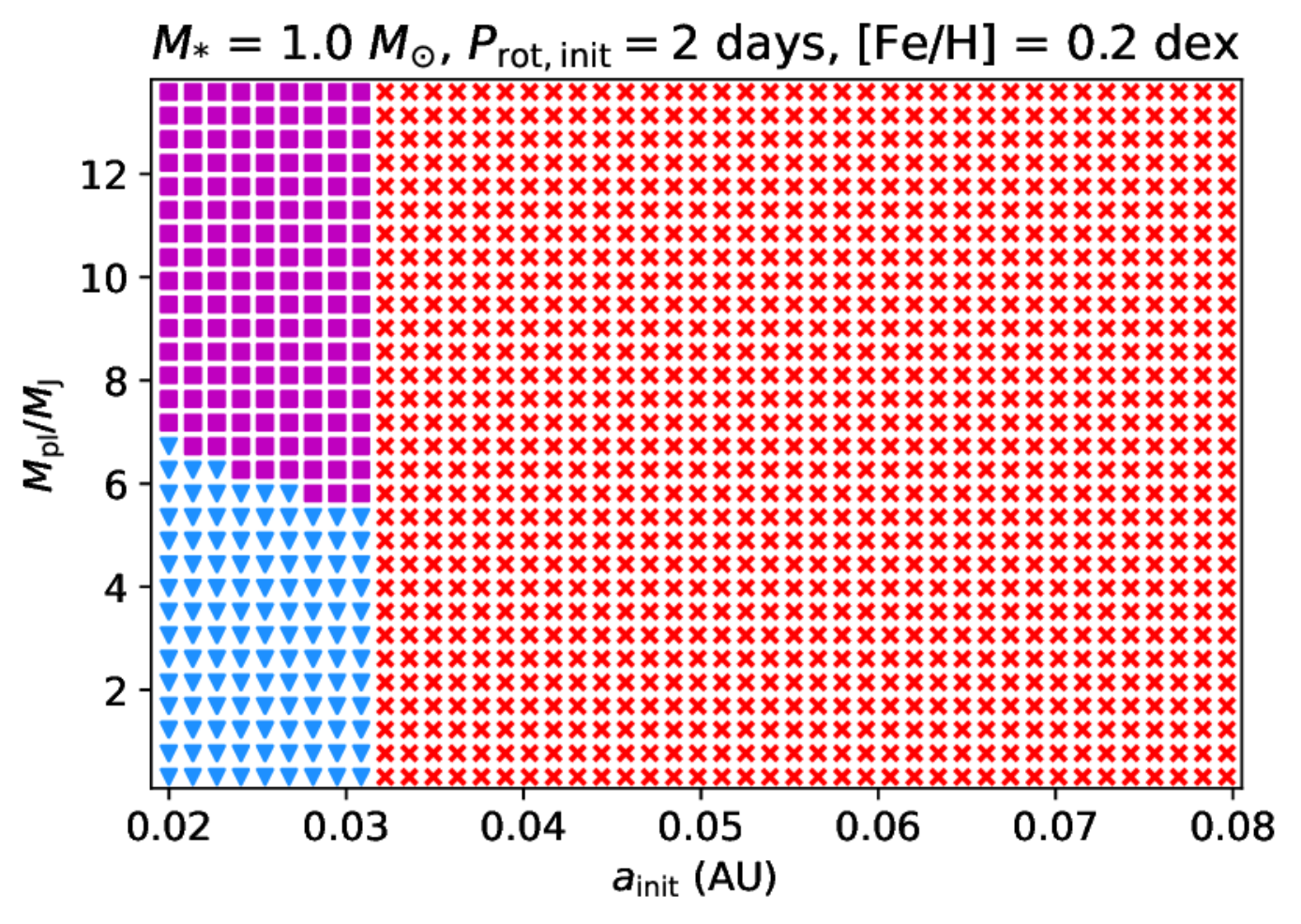}\par 
    \includegraphics[width=\linewidth,height=5.5cm]{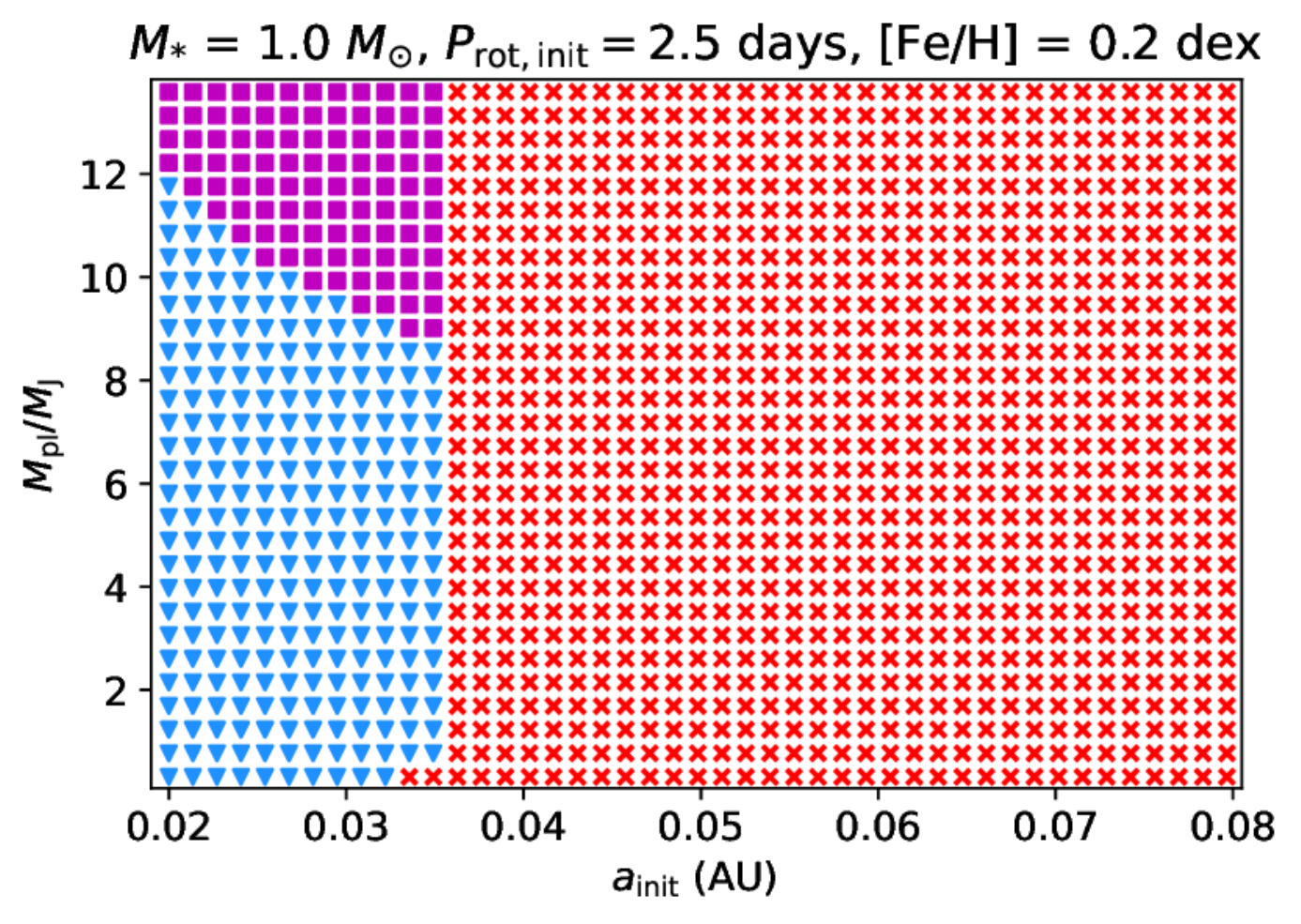}\par
    \includegraphics[width=\linewidth,height=5.5cm]{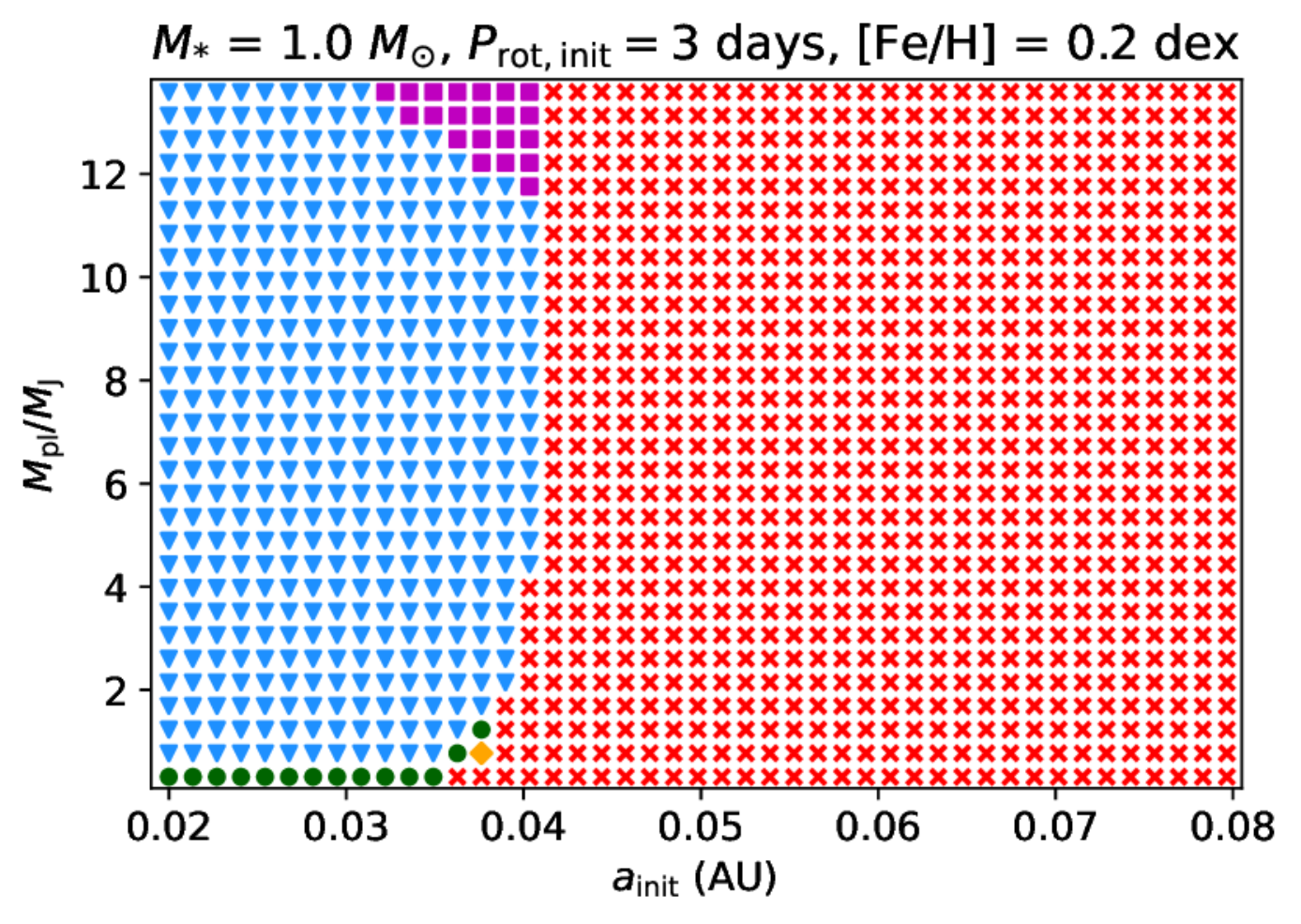}\par 
    \includegraphics[width=\linewidth,height=5.5cm]{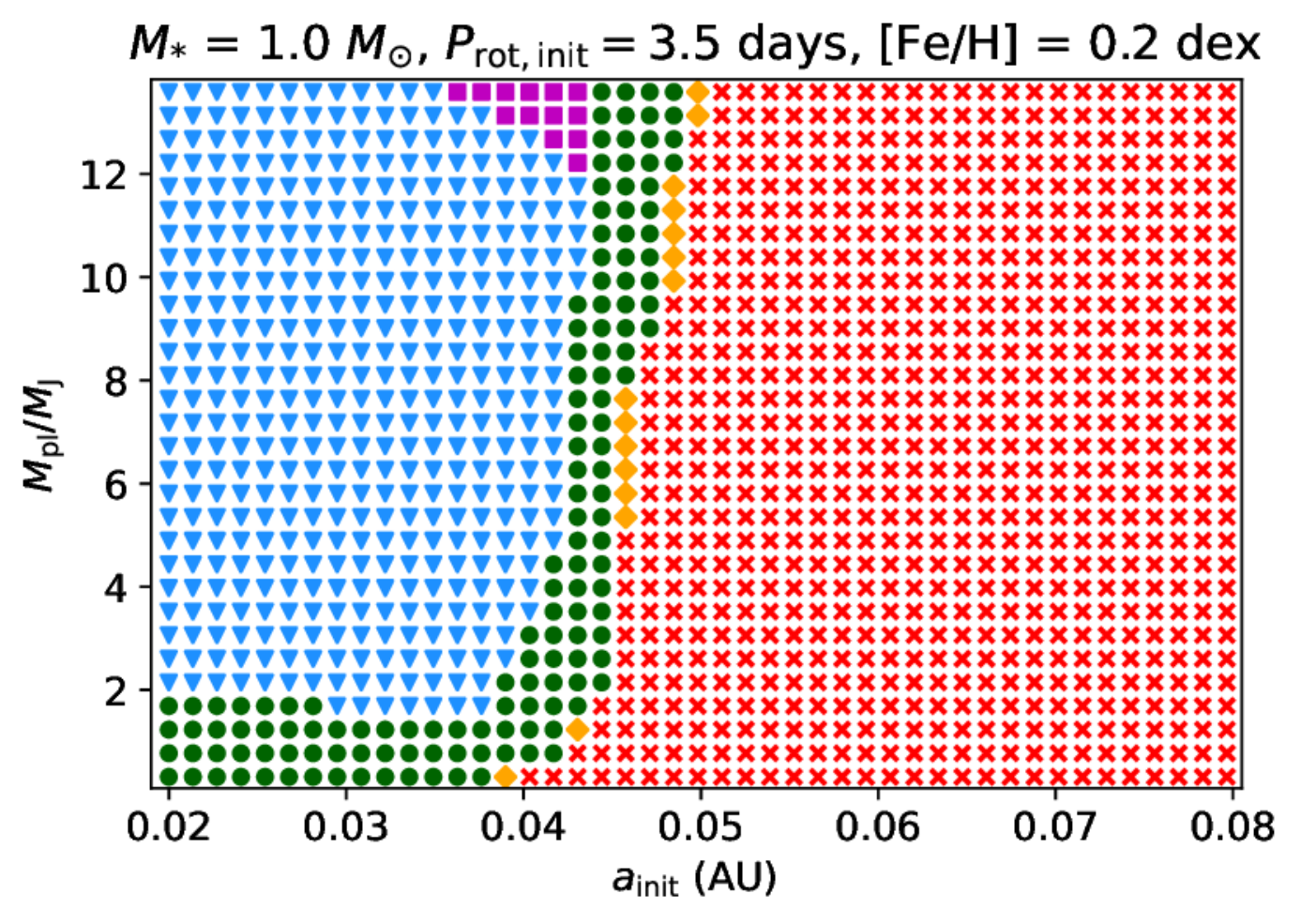}\par
    \includegraphics[width=\linewidth,height=5.5cm]{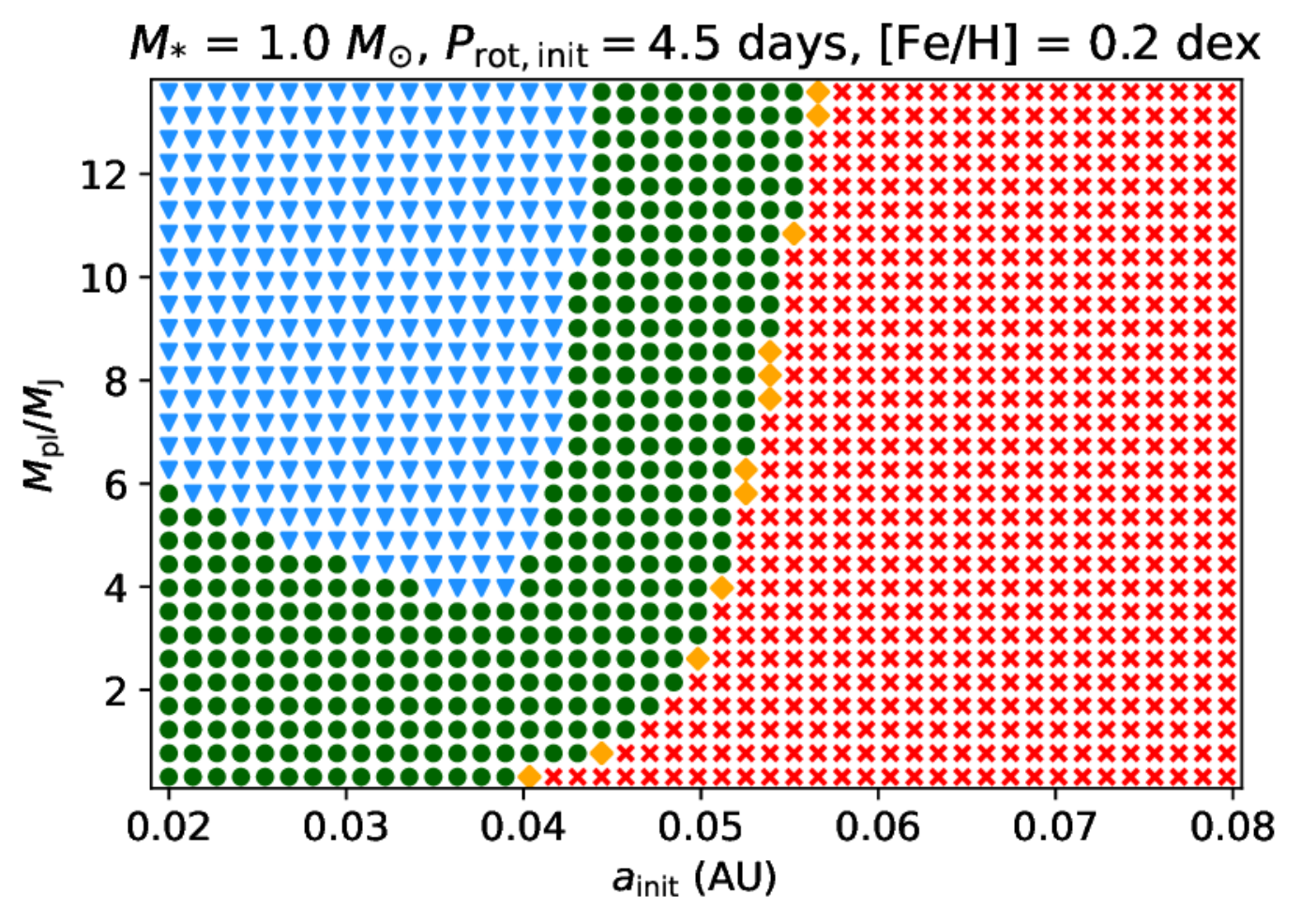}\par 
    \includegraphics[width=\linewidth,height=5.5cm]{M10R55.pdf}\par
    \includegraphics[width=\linewidth,height=5.5cm]{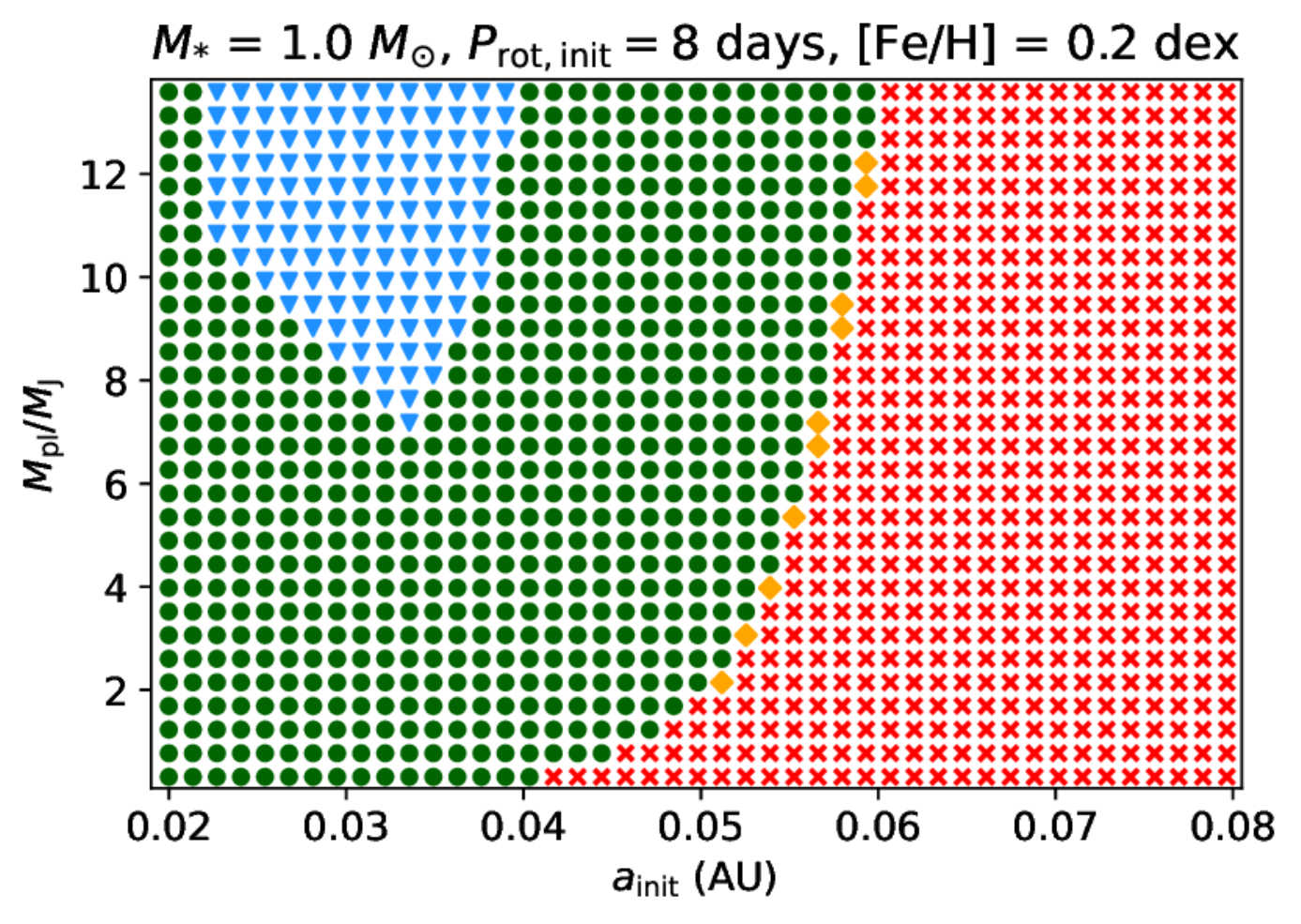}\par 
    \includegraphics[width=\linewidth,height=5.5cm]{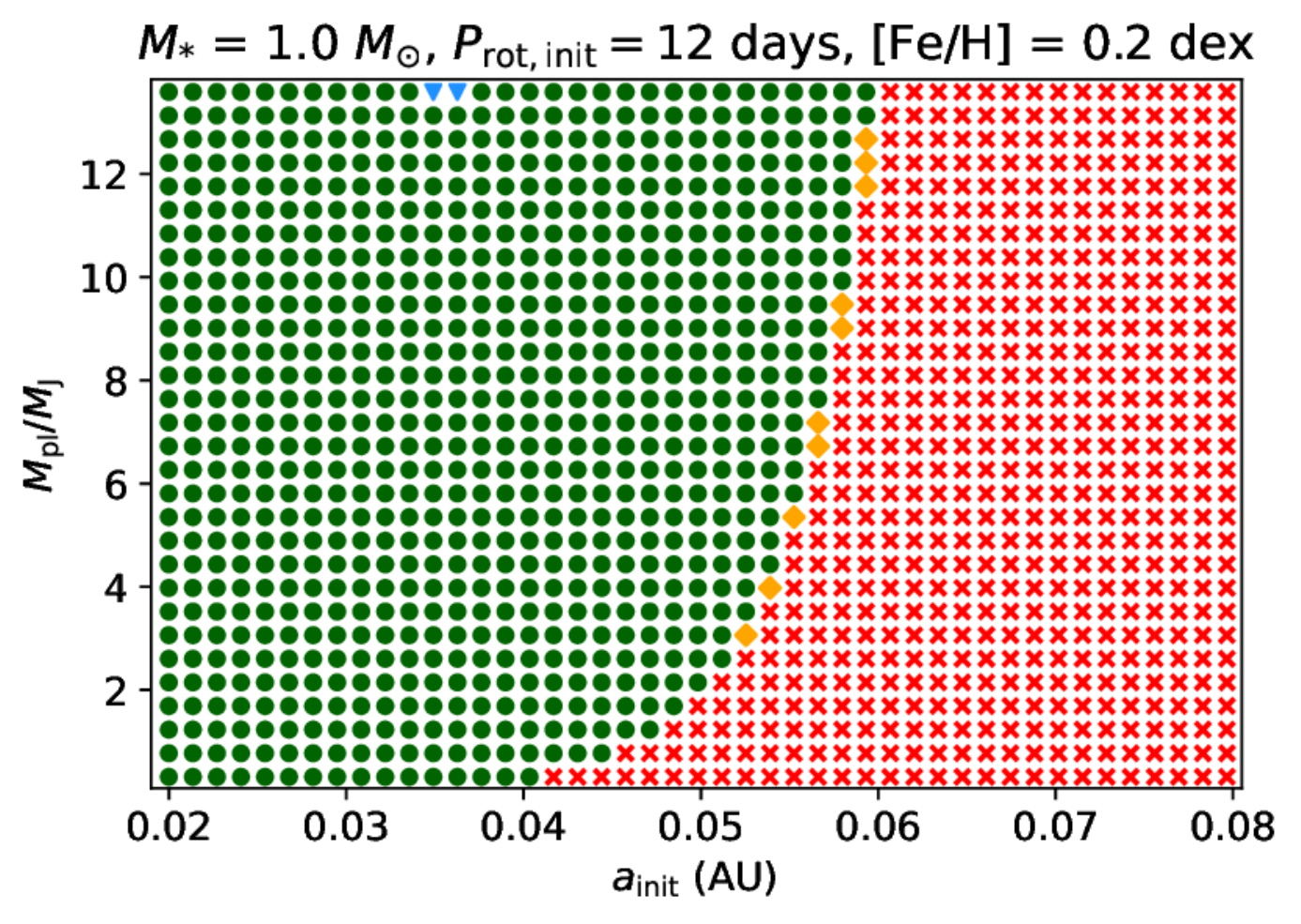}\par
    \end{multicols}
\caption{Infall diagram for 1 $M_\mathrm{\odot}$ star. Designations are the same as those in Fig.~\ref{fig9_F}.}
\label{ap1}

\end{figure*}

\begin{figure*}
\begin{multicols}{2}
    \includegraphics[width=\linewidth,height=5.5cm]{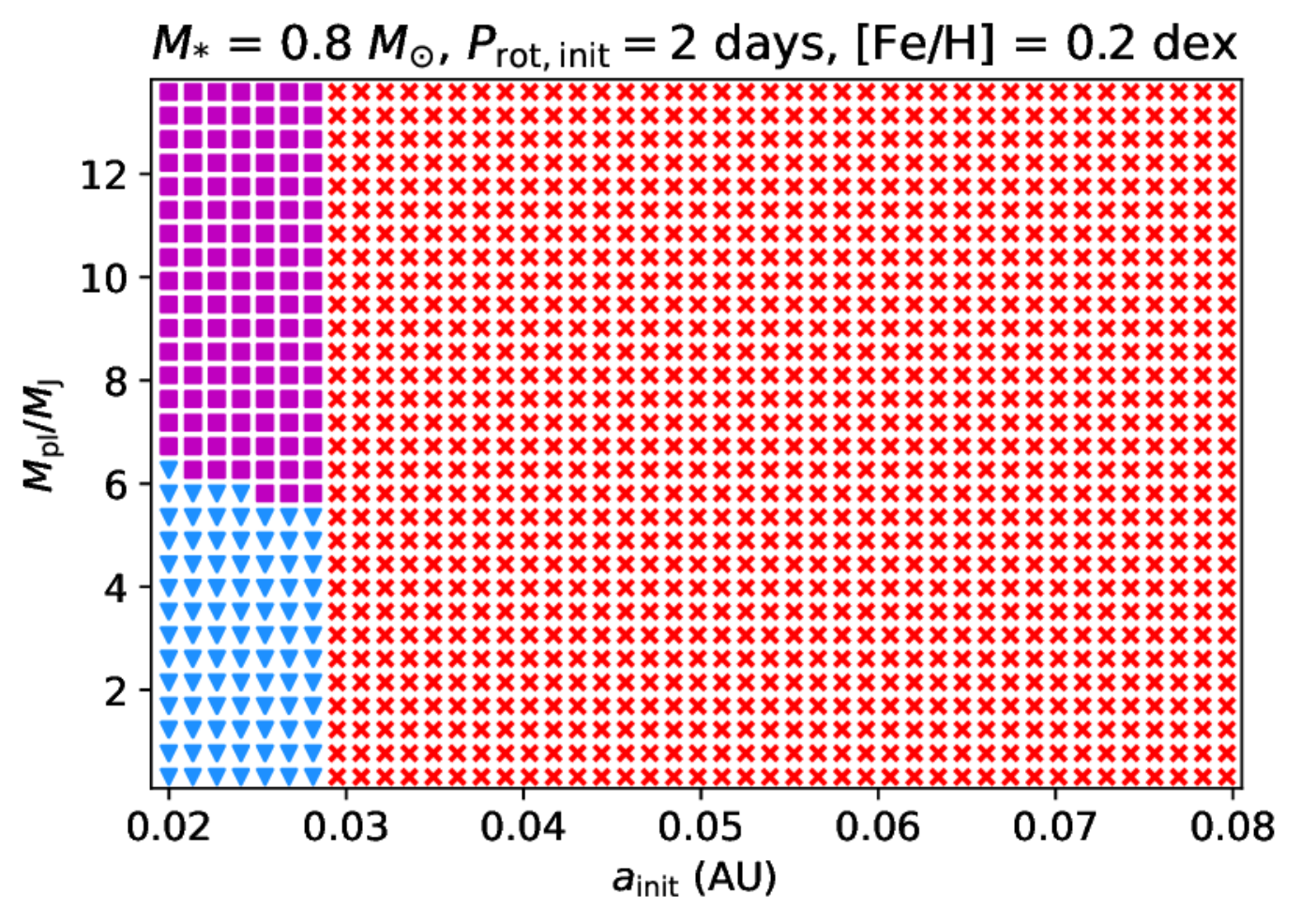}\par 
    \includegraphics[width=\linewidth,height=5.5cm]{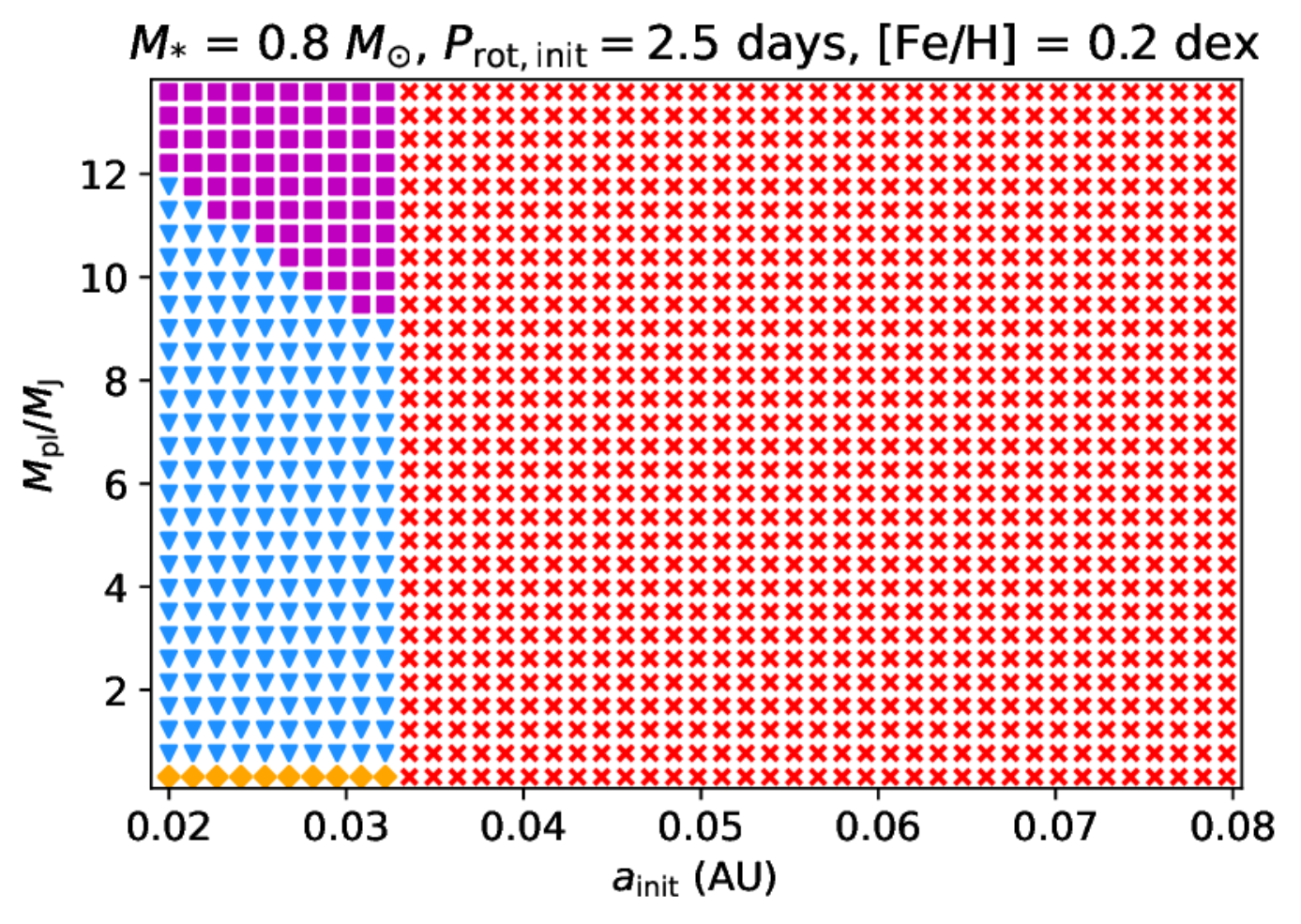}\par
    \includegraphics[width=\linewidth,height=5.5cm]{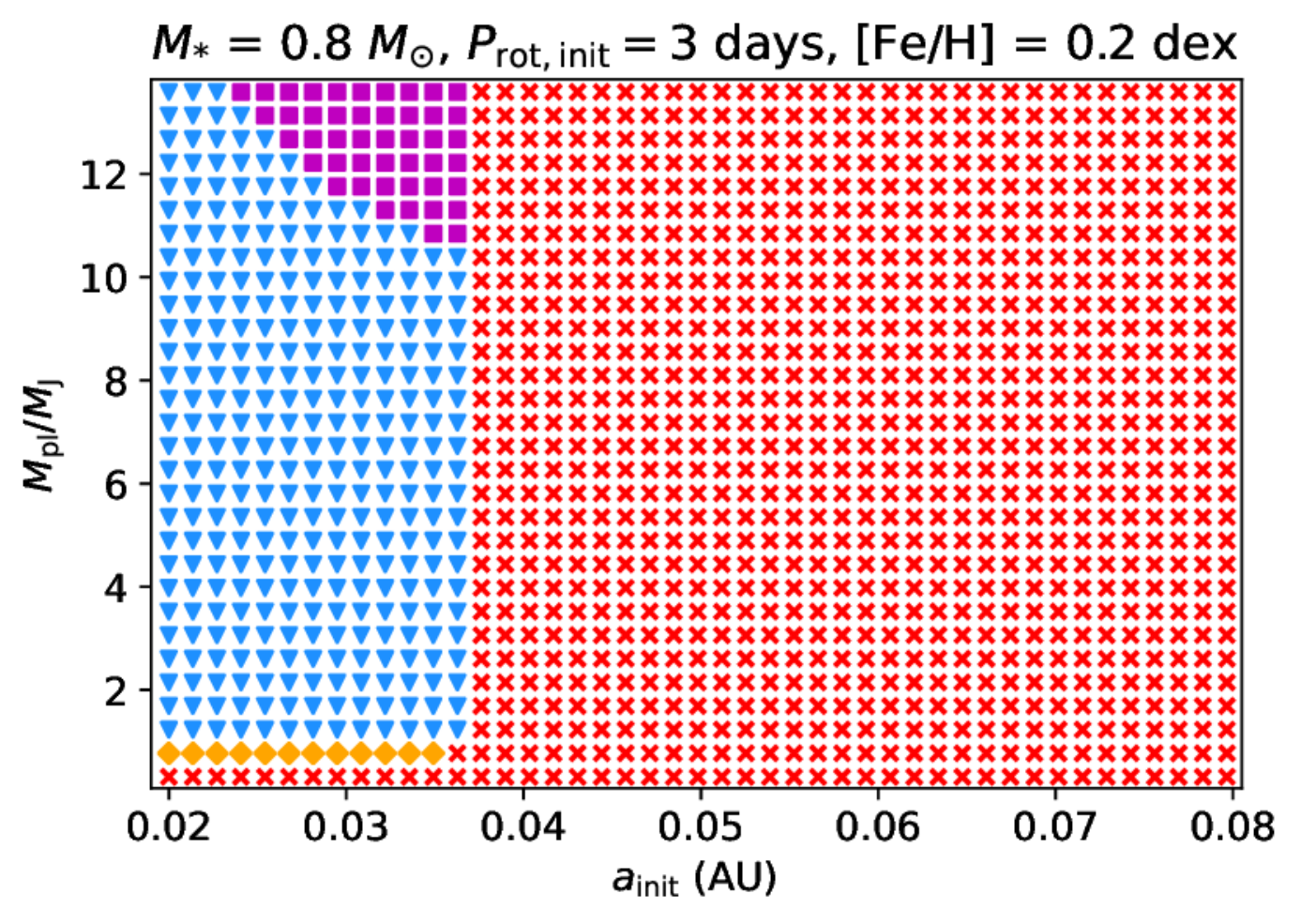}\par 
    \includegraphics[width=\linewidth,height=5.5cm]{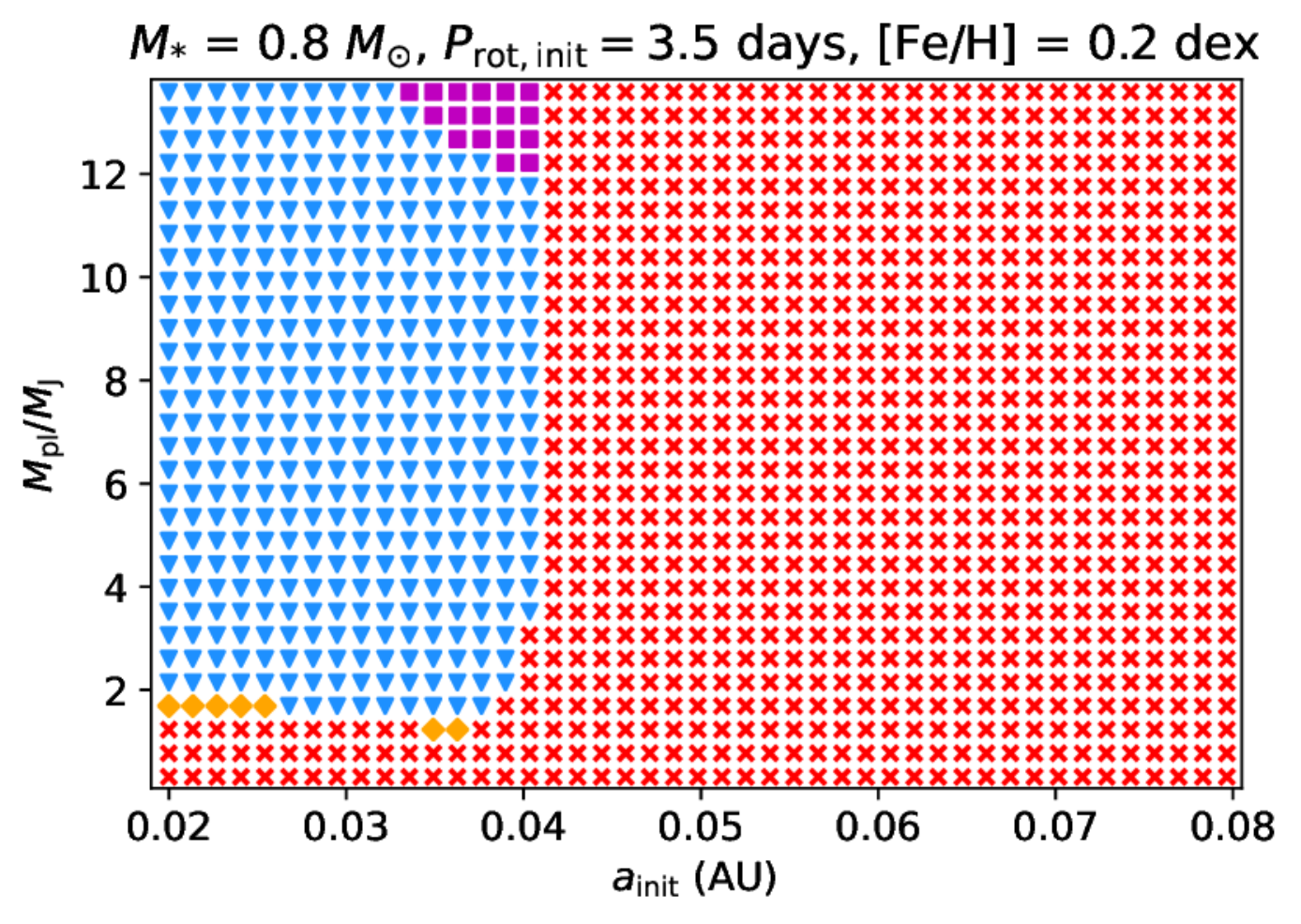}\par
    \includegraphics[width=\linewidth,height=5.5cm]{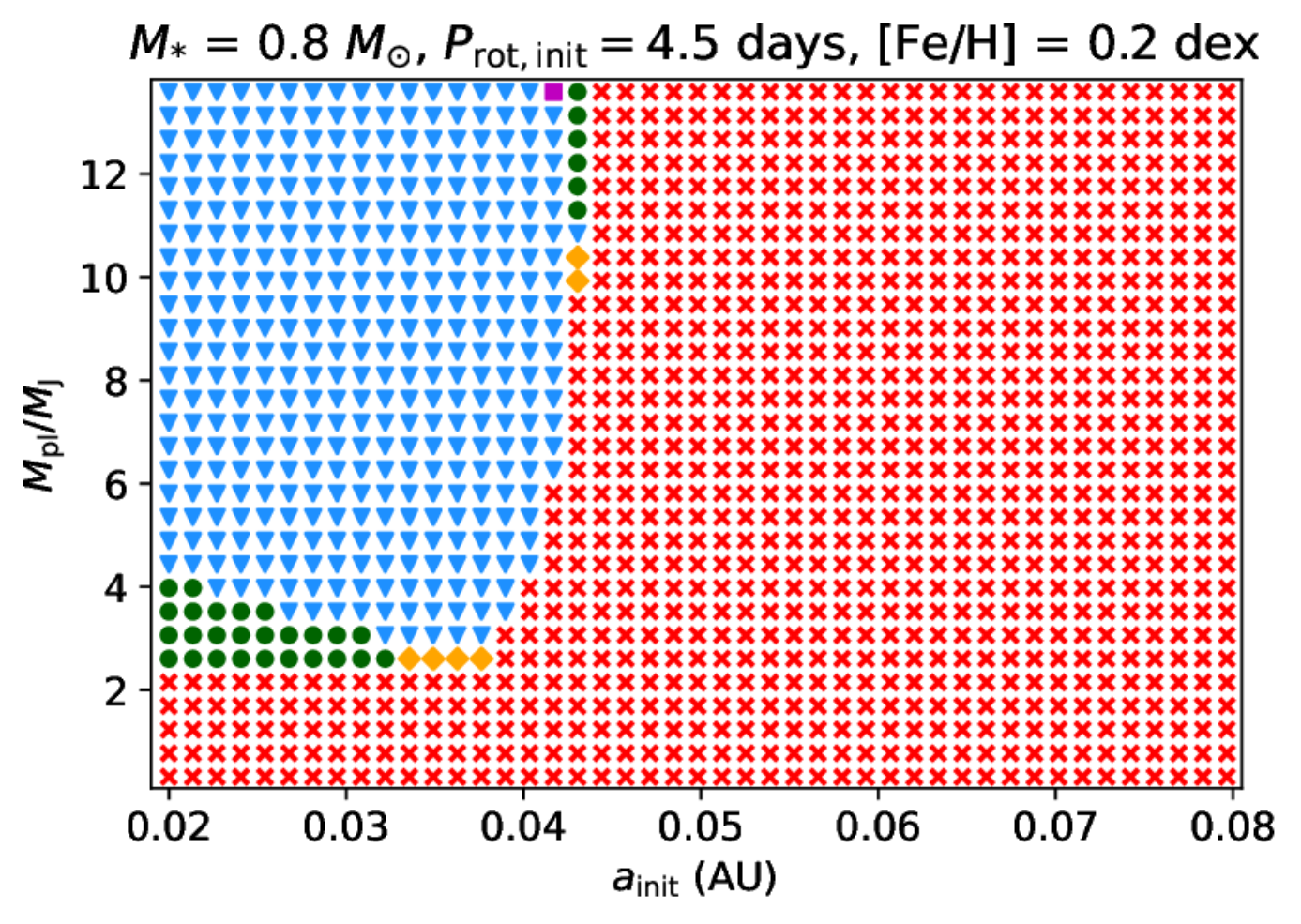}\par 
    \includegraphics[width=\linewidth,height=5.5cm]{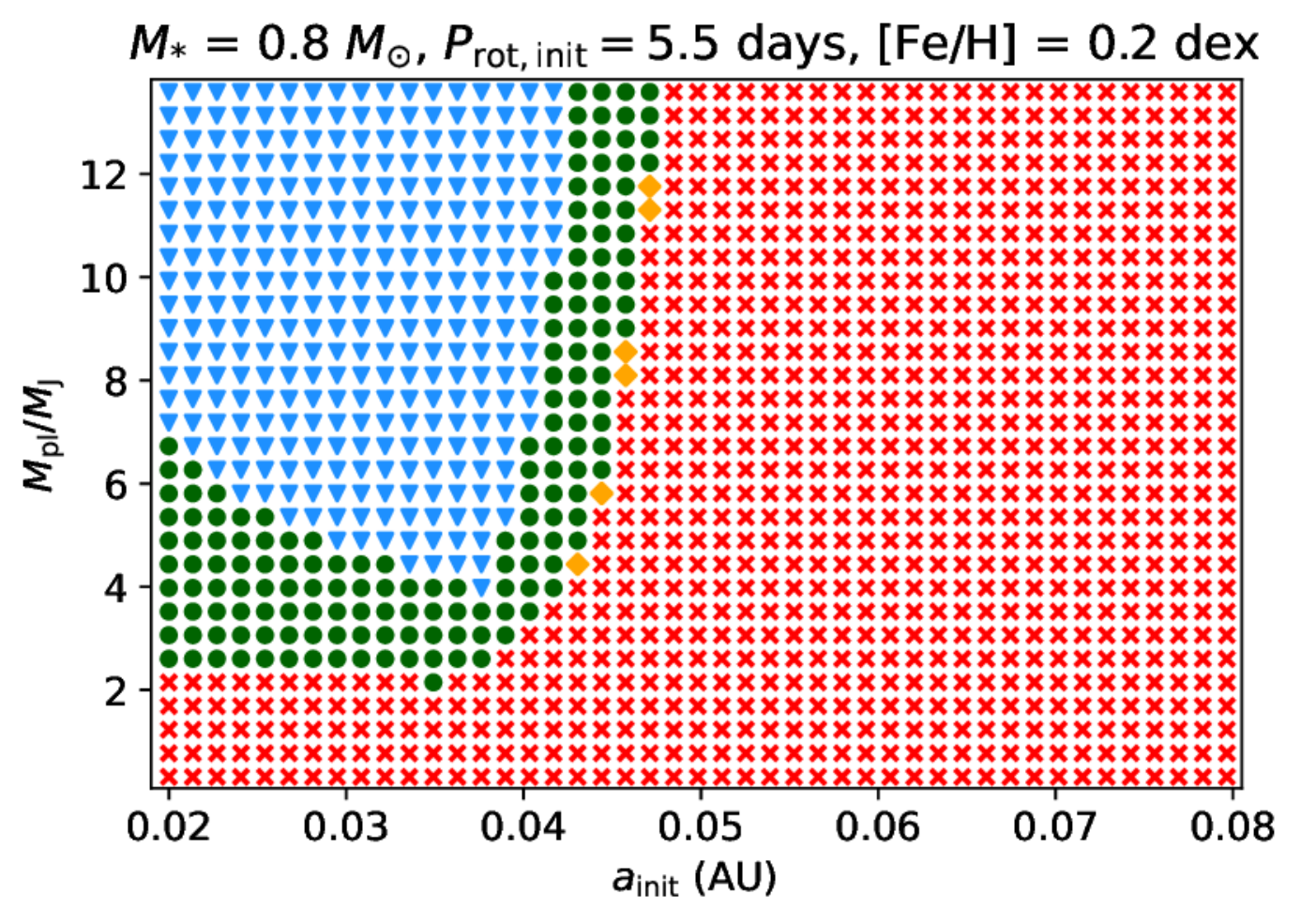}\par
    \includegraphics[width=\linewidth,height=5.5cm]{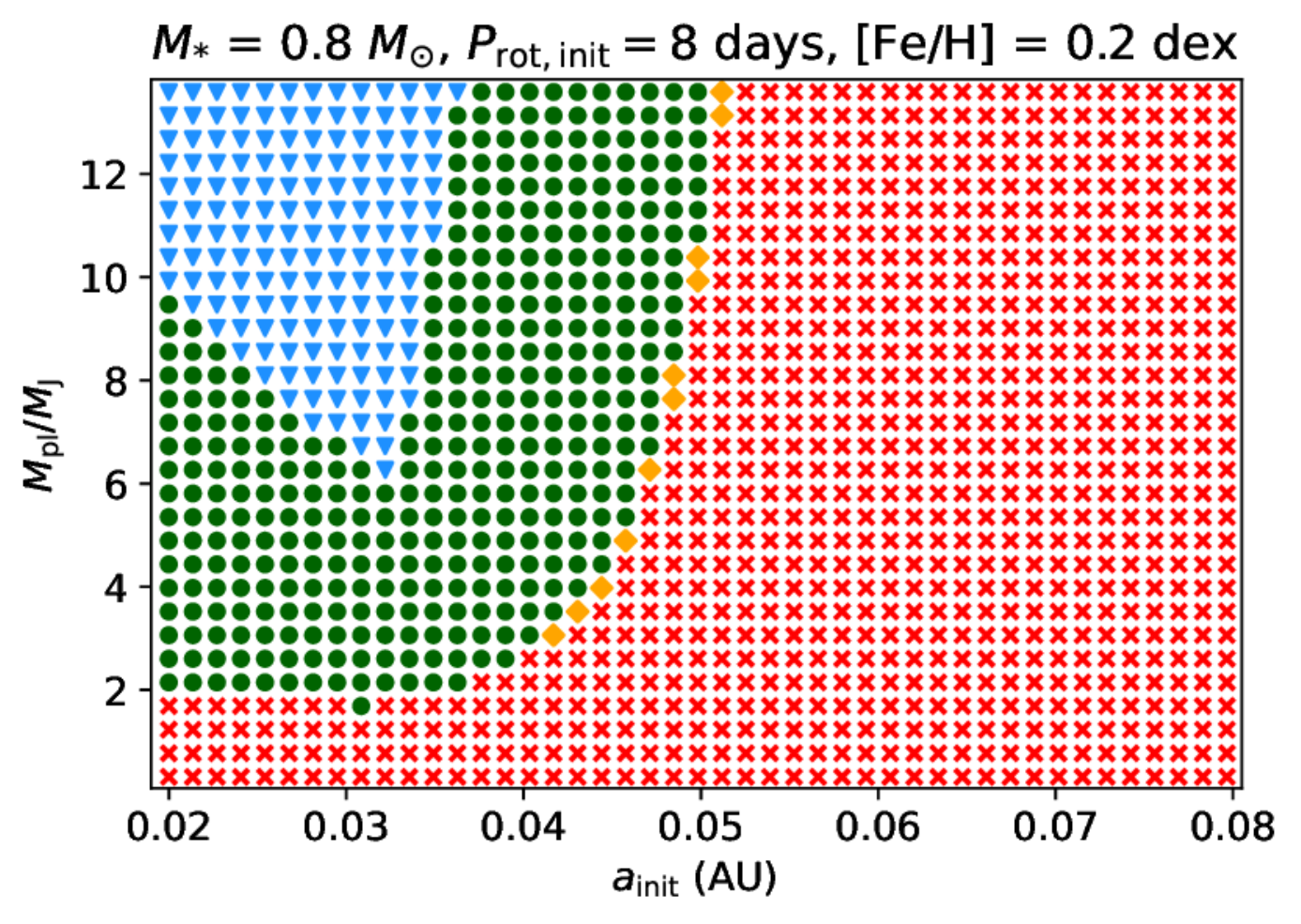}\par 
    \includegraphics[width=\linewidth,height=5.5cm]{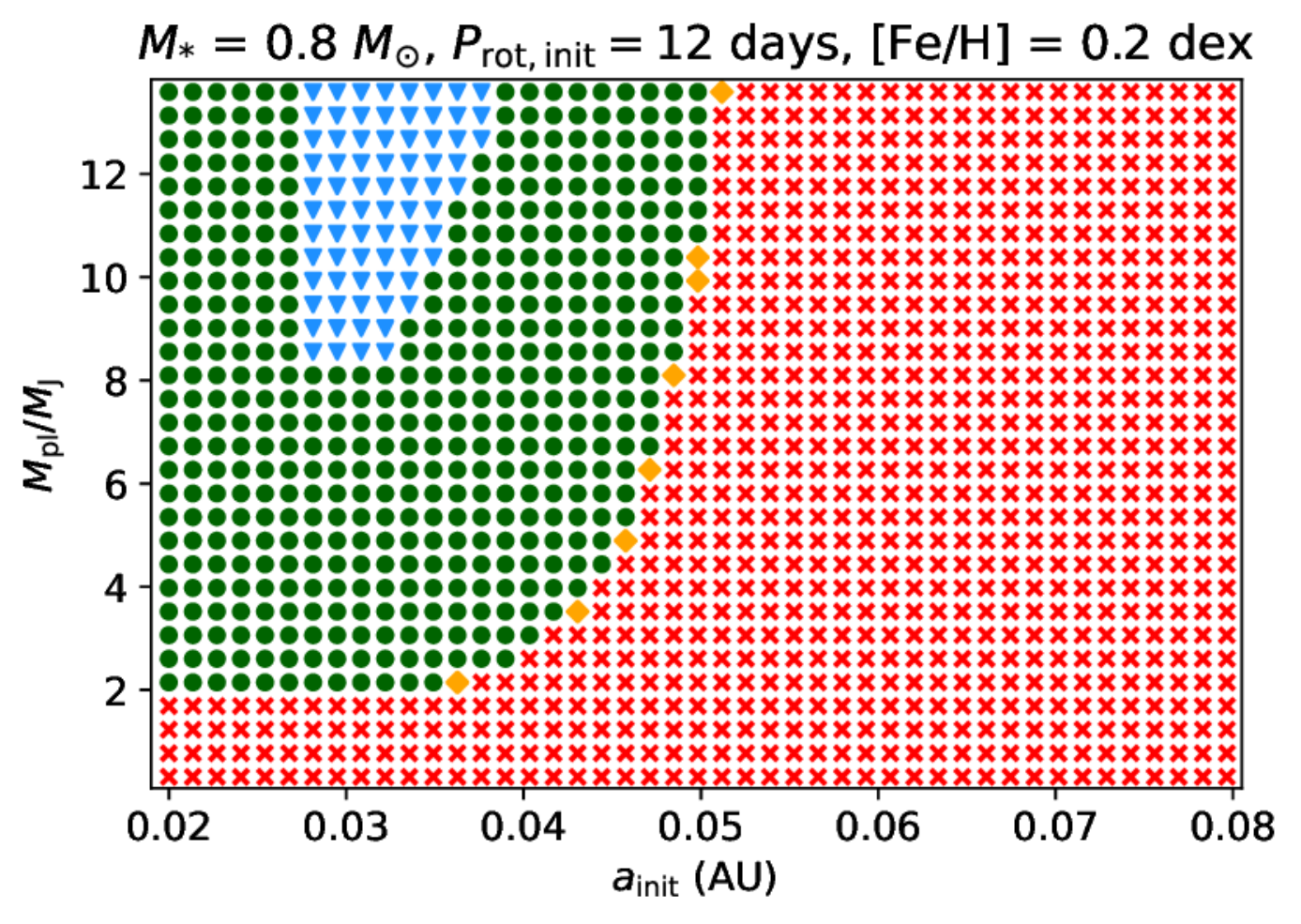}\par
    \end{multicols}
\caption{Infall diagram for 0.8 $M_\mathrm{\odot}$ star. Designations are the same as those in Fig.~\ref{fig9_F}, except we applied new definition for intermediate outcome when gravity waves do not dissipate (no infall, but the planet reduces the semi-major axis by more than 20 \% during the MS stage). }
\label{ap2}
\end{figure*}

\begin{figure*}
\begin{multicols}{2}
    \includegraphics[width=\linewidth,height=5.5cm]{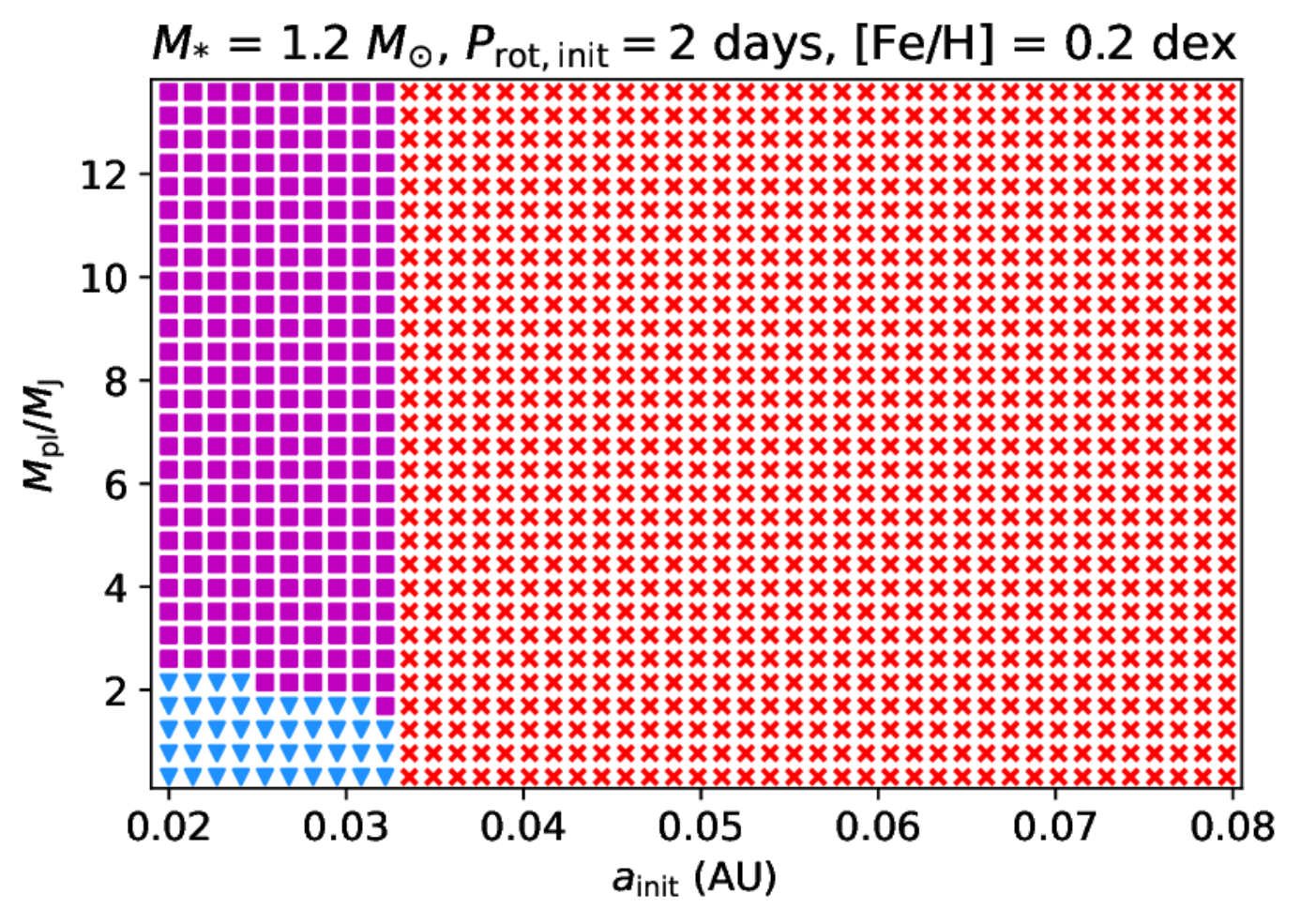}\par 
    \includegraphics[width=\linewidth,height=5.5cm]{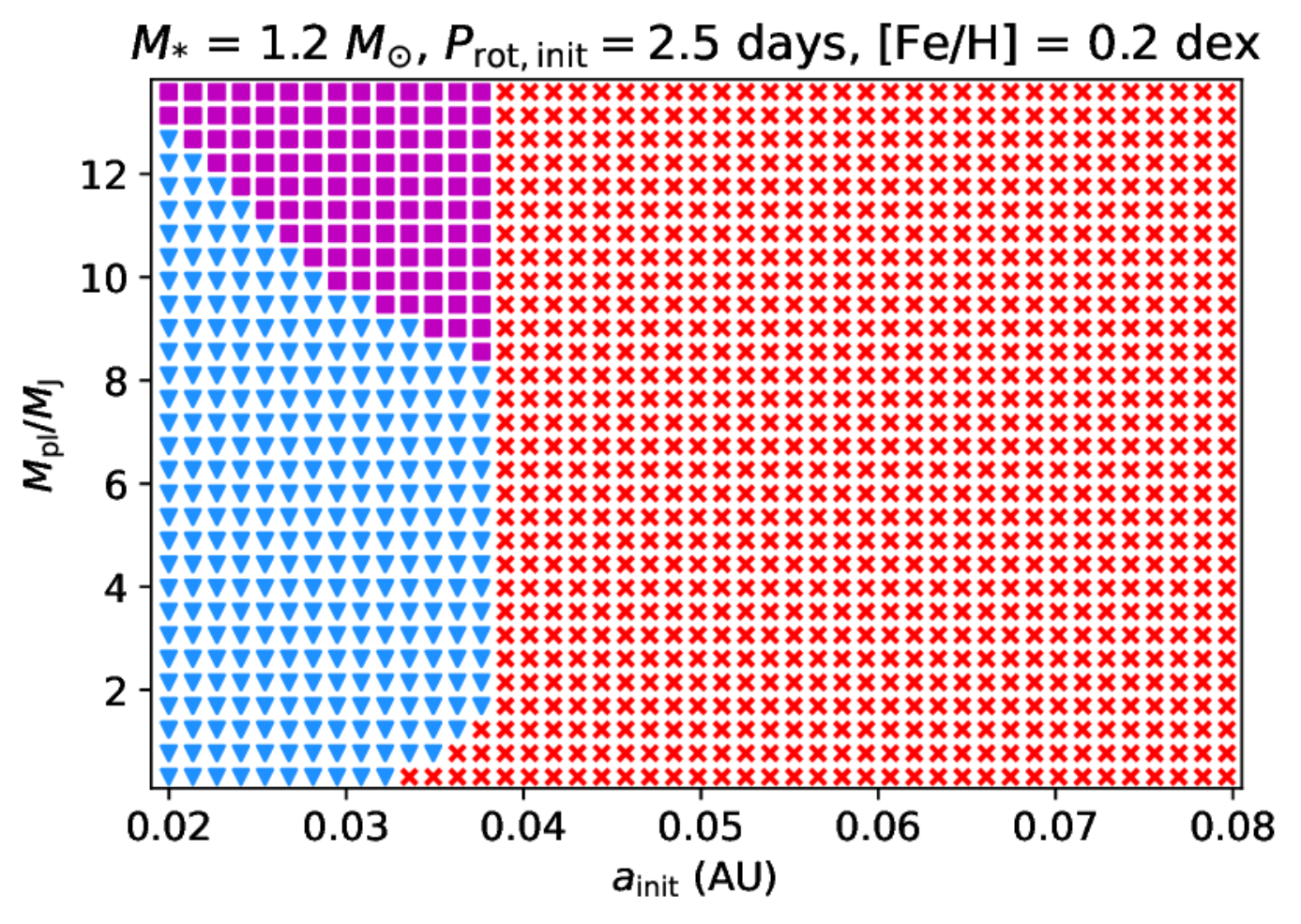}\par
    \includegraphics[width=\linewidth,height=5.5cm]{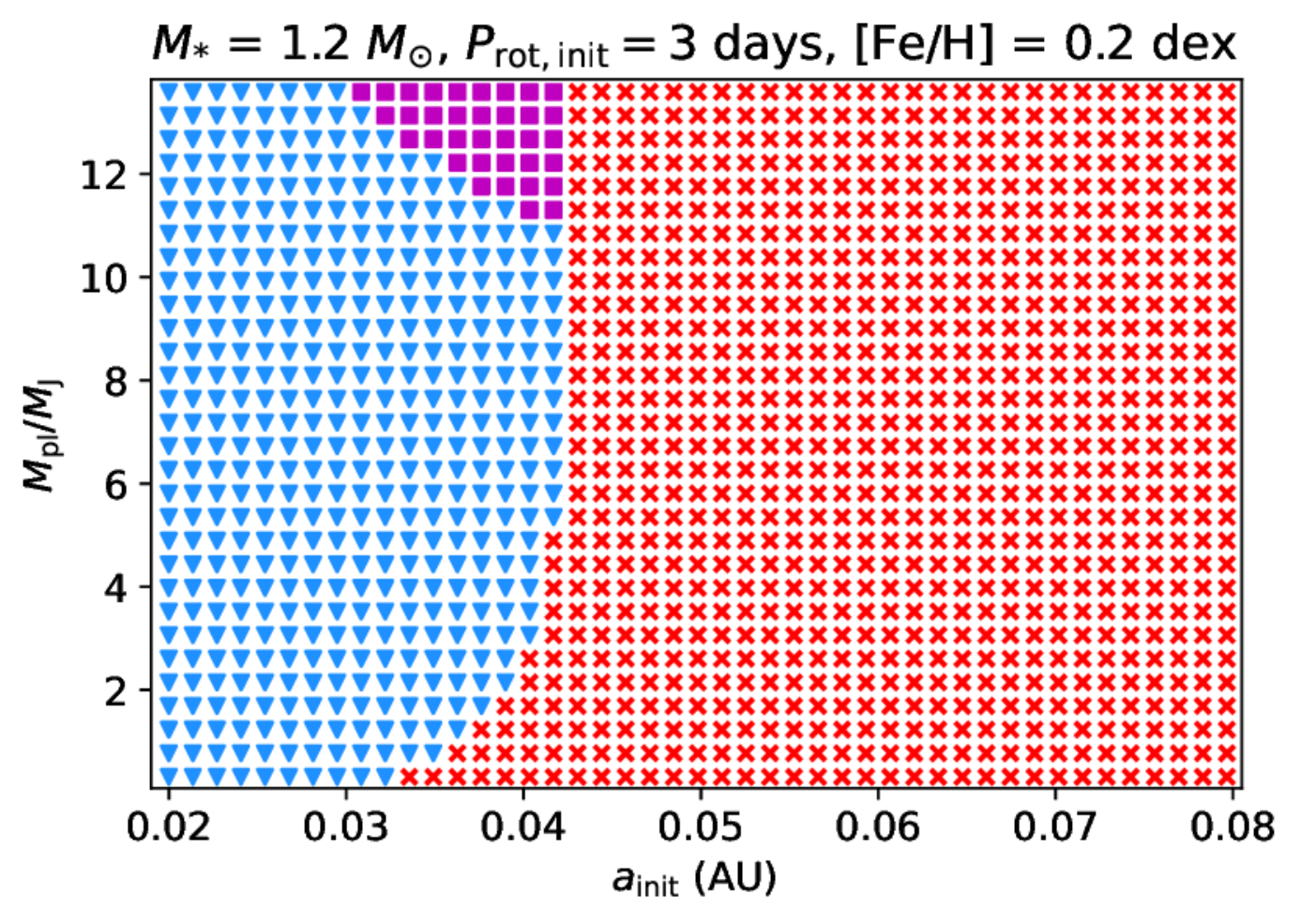}\par 
    \includegraphics[width=\linewidth,height=5.5cm]{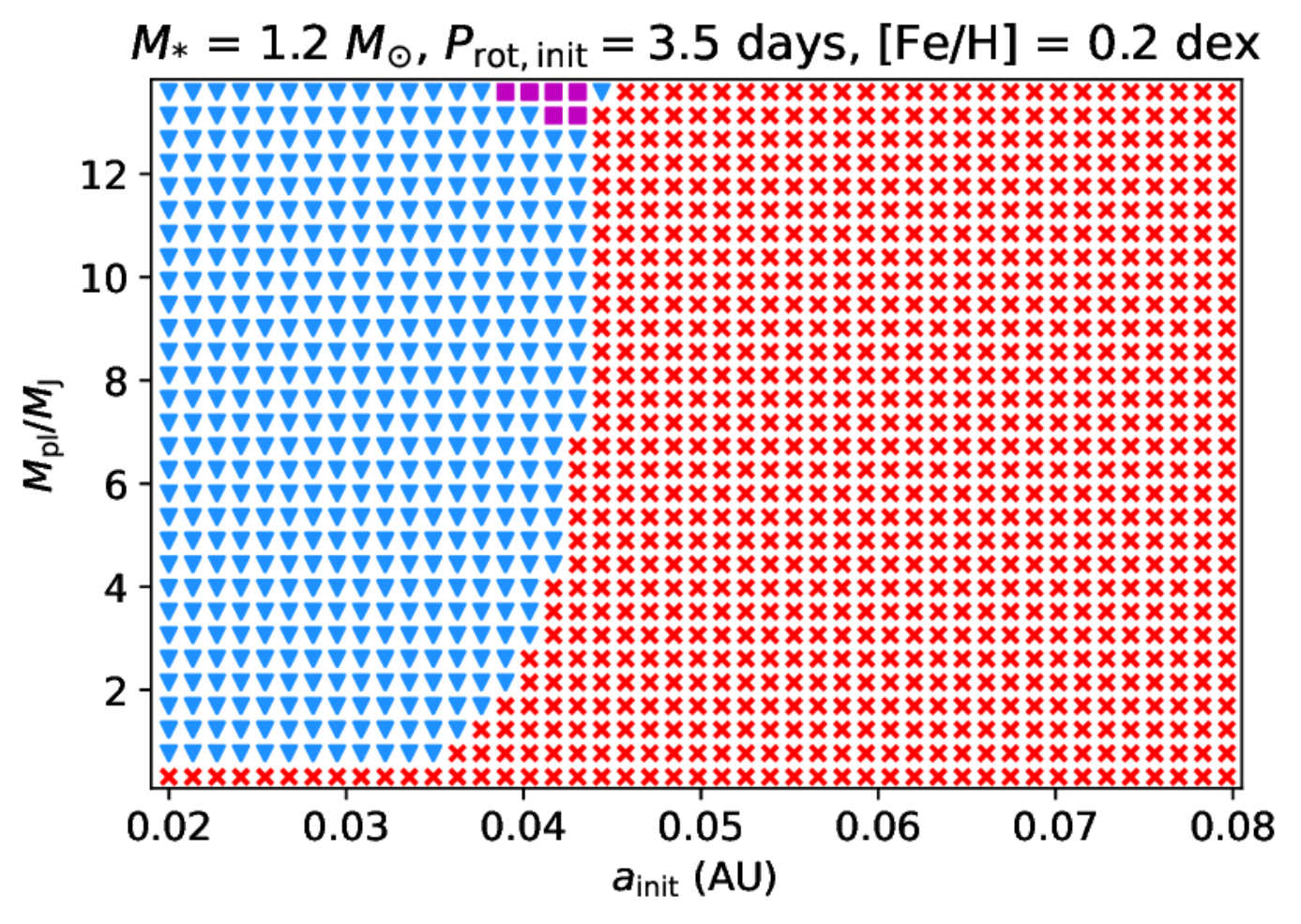}\par
    \includegraphics[width=\linewidth,height=5.5cm]{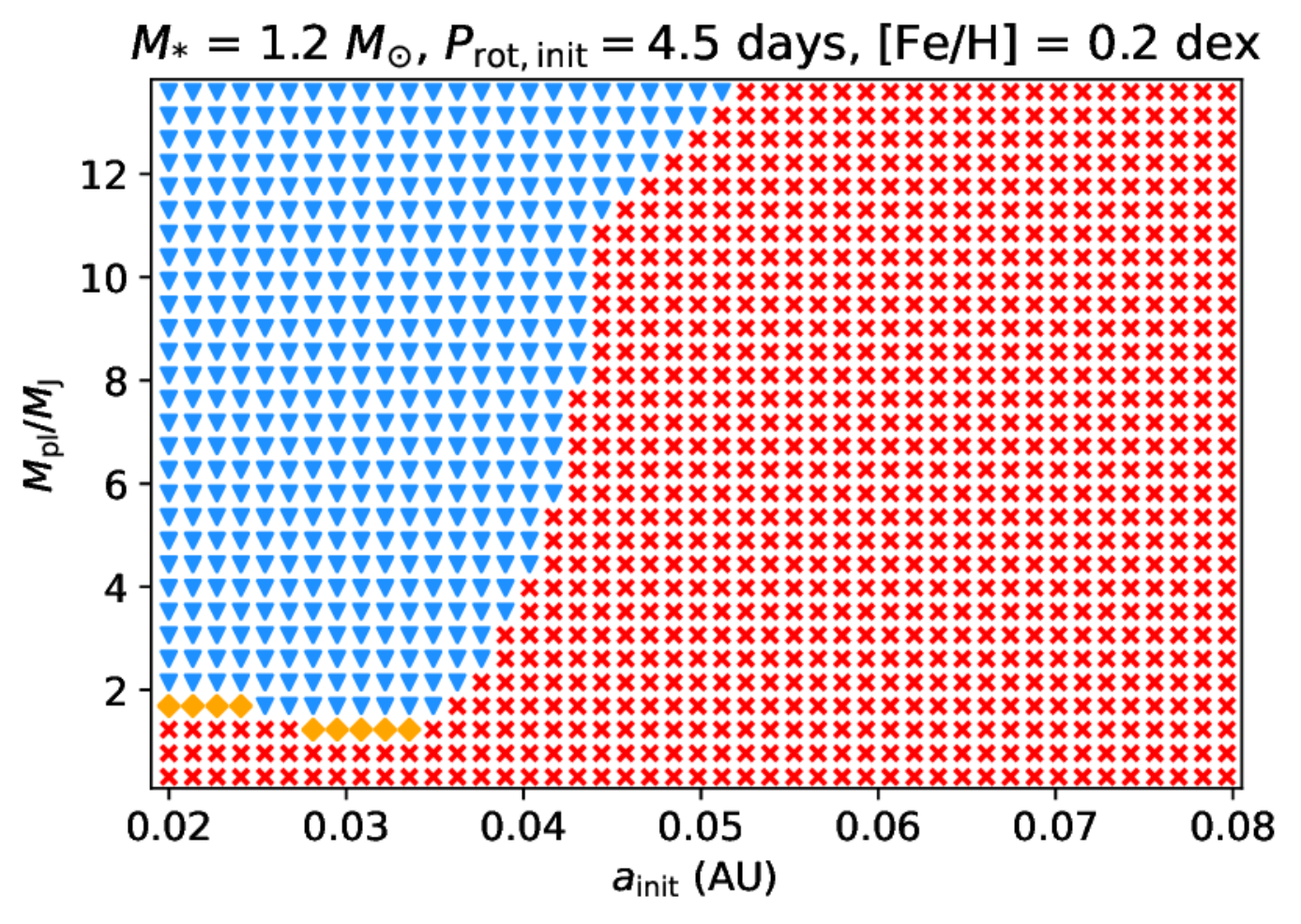}\par 
    \includegraphics[width=\linewidth,height=5.5cm]{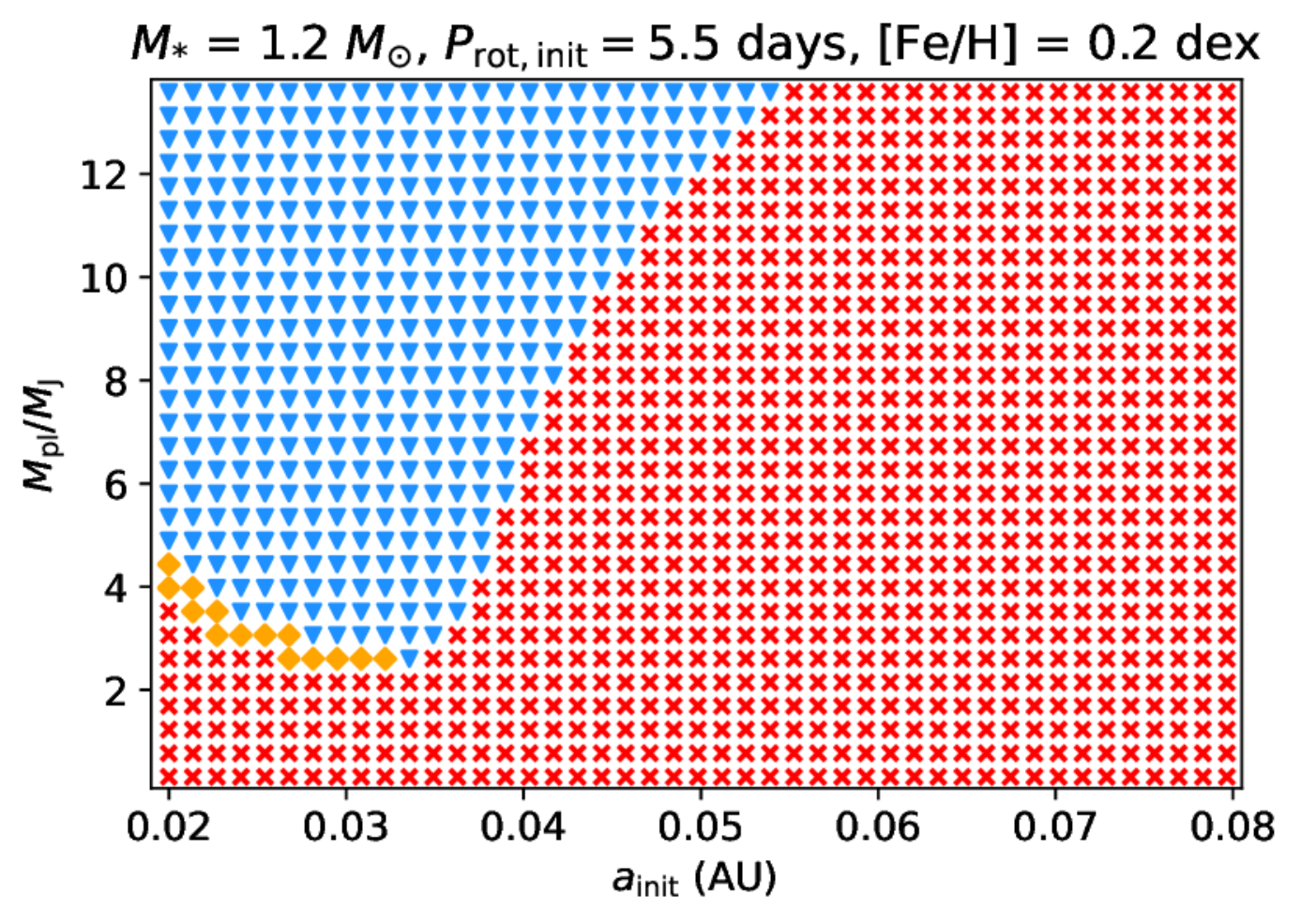}\par
    \includegraphics[width=\linewidth,height=5.5cm]{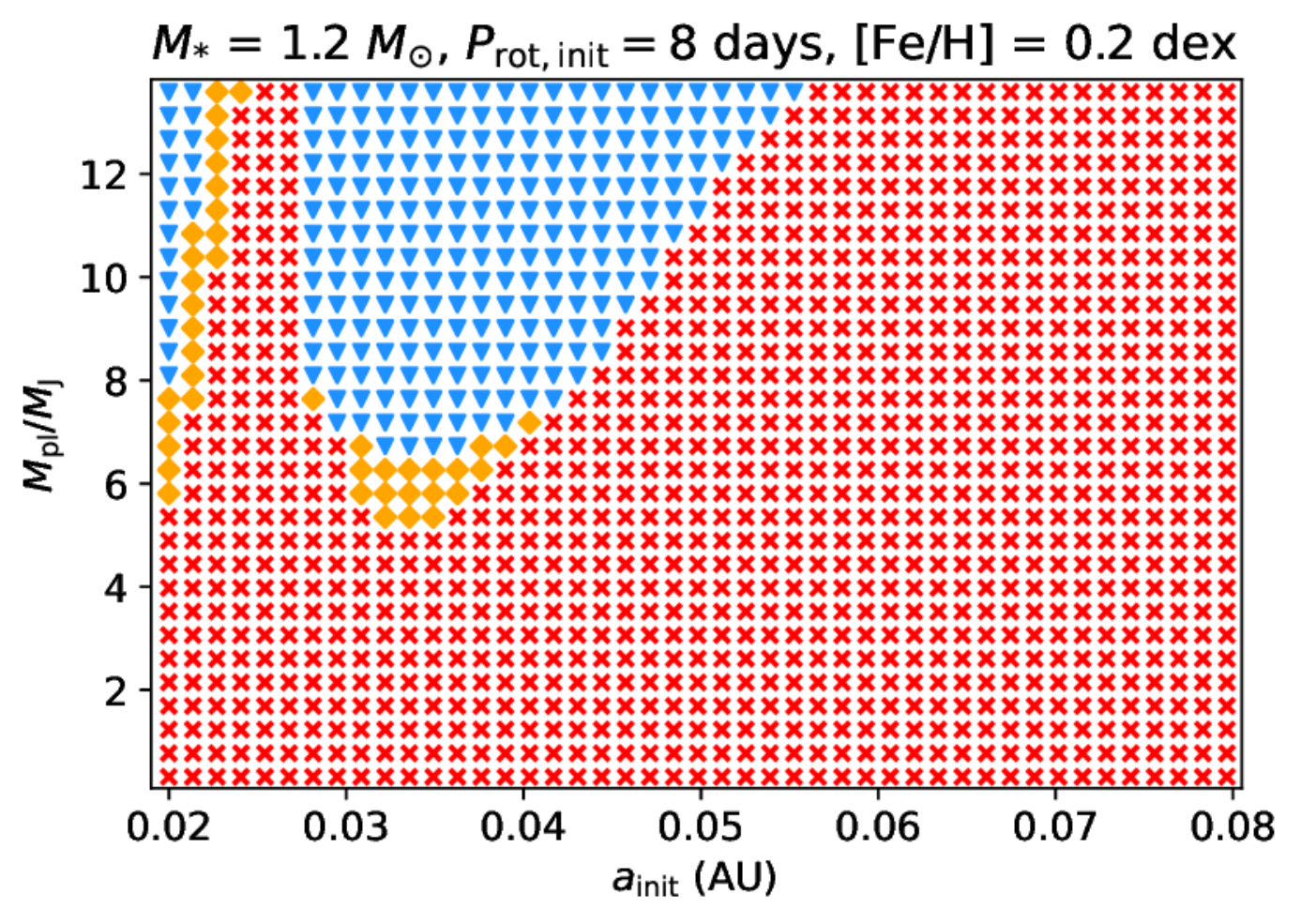}\par 
    \includegraphics[width=\linewidth,height=5.5cm]{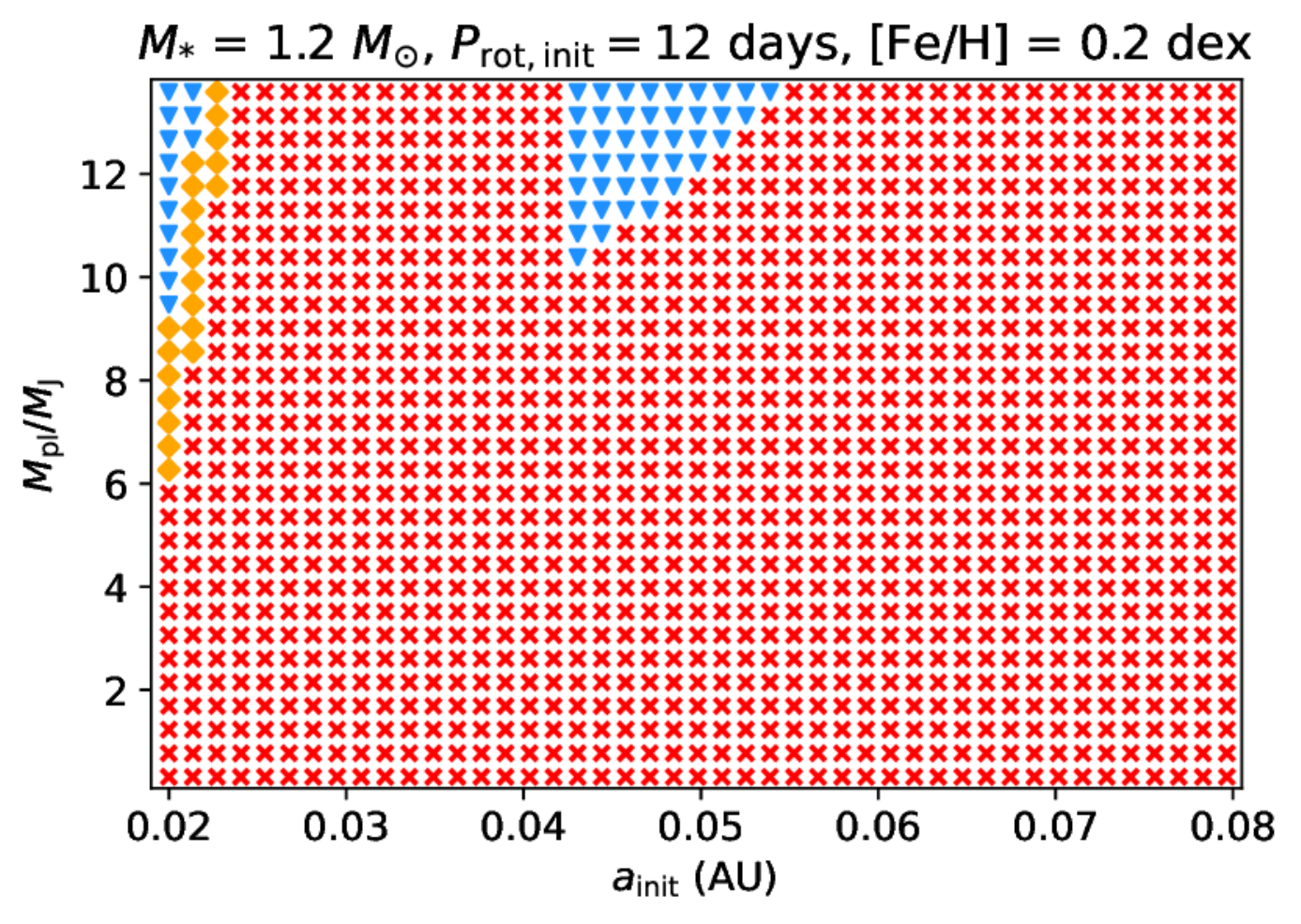}\par
    \end{multicols}
\caption{Infall diagram for 1.2 $M_\mathrm{\odot}$ star. Designations are the same as those in Fig.~\ref{ap2}.}
\label{ap3}
\end{figure*}


\bsp	
\label{lastpage}
\end{document}